\documentclass[final,3p,times]{elsarticle}

\usepackage{lineno}
\usepackage{comment}
\excludecomment{omitext}

\biboptions{comma,square,sort}

\usepackage{times,colordvi,epsfig,graphicx,float,color,multicol,bbm}
\usepackage{pstricks,pst-node,pst-text,pst-3d}
\usepackage{tikz}
\usetikzlibrary{patterns}
\usepackage{subfig}
\usepackage[latin1]{inputenc}
\usepackage[fleqn]{amsmath}
\usepackage{amsfonts,amssymb,amsbsy,amscd,amsthm}
\usepackage[mathscr]{eucal}
\usepackage{times}
\usepackage{psfrag}
\usepackage{graphics}
\usepackage{wrapfig,booktabs}
\usepackage{tabularx,multirow}
    \newcolumntype{L}{>{\raggedright\arraybackslash}X}
    \newcolumntype{C}{>{\centering\arraybackslash}X}
\usepackage{color}
\usepackage{mathrsfs}
\usepackage{array}%,supertabular}
\usepackage{verbatim}
\usepackage{comment}
\usepackage{enumitem}
\usepackage[linesnumbered,ruled,lined,boxed]{algorithm2e}
\RestyleAlgo{boxruled}
\SetKwRepeat{Do}{do}{while} %--- macro to create a do-while construct for algorithm2e

\usepackage{listings}
\usepackage[off]{auto-pst-pdf}
\usepackage{ifthen}
\usepackage{mathtools}
\DeclarePairedDelimiter{\nm}{\lVert}{\rVert}
\DeclarePairedDelimiter{\abs}{\lvert}{\rvert}

\usepackage[colorlinks=true,urlcolor=blue,citecolor=black,linkcolor=black]{hyperref}

\mathchardef\breakingcomma\mathcode`\,
{\catcode`,=\active
  \gdef,{\breakingcomma\discretionary{}{}{}}
}
\newcommand{\mathlist}[1]{$\mathcode`\,=\string"8000 #1$}

\DeclareMathOperator*{\argmax}{argmax} % thin space, limits underneath in displays

\renewcommand{\vec}[1]{{\ensuremath{\boldsymbol{\mathrm #1}}}}
\newcommand{\vdot}{{\ifnum\thedotStyle=0{\ensuremath\cdot}\else
    {\boldsymbol{\mathsf{\ensuremath\cdot}}}\fi}}

\newcommand{\bx}{\mathbf{x}}

\newcommand{\bv}{\mathbf{v}}

\newcommand{\bw}{\mathbf{w}}
\newcommand{\bz}{\mathbf{z}}

\newcommand{\btx}{\tilde{\mathbf{x}}}

\newcommand{\bhat}[1]{\widehat{\vec {#1}}}

\newcommand{\pd}[2]{\frac{\partial{#1}}{\partial{#2}}}
\newcommand{\pdt}[1]{\pd{#1}t}
\newcommand{\ddt}[1]{\frac{d \,{#1}}{dt}}

\newcommand{\R}{\ensuremath{\mathbb{R}}}
\newcommand{\RR}{\ensuremath{\mathcal{R}}}

\newcommand{\lrp}[1]{\left( #1 \right)}

\newcommand{\lrb}[1]{\left[ #1 \right]}

\newcommand{\lrB}[1]{\left\{ #1 \right\}}

\newcommand{\etal}{\textit{et al}.\ }
\newcommand{\ie}{\textit{i}.\textit{e}.\ }
\newcommand{\eg}{\textit{e}.\textit{g}.\ }

\newcommand{\dif}{\, \mathrm{d}}
\newcommand{\dx}{\dif x}
\newcommand{\ds}{\dif s}

\newcommand{\strongRes}{\boldsymbol{\mathfrak{R}}}

\newtheorem{theorem}{Theorem}[section]

\newtheorem{lemma}[theorem]{Lemma}

\theoremstyle{definition}
\newtheorem{definition}{Definition}[section]

\journal{Journal of Computational Physics}

\begin{document}

\begin{frontmatter}


\title{A greedy non-intrusive reduced order model for shallow water equations}


\author[erdc]{Sourav Dutta}
\ead{sourav.dutta@erdc.dren.mil}
\author[erdc]{Matthew W. Farthing}
\ead{matthew.w.farthing@erdc.dren.mil}
\author[unig]{Emma Perracchione}
%\ead{perracchione@dima.unige.it}
\author[erdc]{Gaurav Savant}
\author[unipd]{Mario Putti}

\address[erdc]{U.S. Army Engineer Research and Development Center, Coastal and Hydraulics Laboratory (ERDC-CHL), Vicksburg, MS 39180, USA}
\address[unig]{Department of Mathematics, University of Genoa, Genova, Italy}
\address[unipd]{Department of Mathematics ``Tullio Levi-Civita", University of Padova, Padova, Italy}

\begin{abstract}
In this work, we develop Non-Intrusive Reduced Order Models (NIROMs) that combine Proper Orthogonal Decomposition (POD) with a Radial Basis Function (RBF) interpolation method to construct efficient reduced order models for time-dependent problems arising in large scale environmental flow applications. The performance of the POD-RBF NIROM is compared with a traditional nonlinear POD (NPOD) model by evaluating the accuracy and robustness for test problems representative of riverine flows. Different greedy algorithms are studied in order to determine a near-optimal distribution of interpolation points for the RBF approximation. A new power-scaled residual greedy (\textit{psr-greedy}) algorithm is proposed to address some of the primary drawbacks of the existing greedy approaches. The relative performances of these greedy algorithms are studied with numerical experiments using realistic two-dimensional (2D) shallow water flow applications involving coastal and riverine dynamics.
\end{abstract}


\end{frontmatter}

%\linenumbers

\section{Introduction}\label{introduction}\footnote{Please cite this article as: S. Dutta, M.W. Farthing, E. Perracchione \etal., A greedy non-intrusive reduced order model for shallow water equations, Journal of Computational Physics, vol. 439, p. 110378, 2021. \url{https://doi.org/10.1016/j.jcp.2021.110378}}
The shallow water equations (SWE) are used to model a wide variety of free-surface problems found across science and engineering, ranging from dam breaks \cite{SBMT2011} and riverine hydrodynamics to hurricane storm surge \cite{Westerink_Luettich_etal_08} and atmospheric processes \cite{V1992}. Despite the trend of hardware improvements and significant gains in the algorithmic efficiency of standard discretization procedures, \textit{high-fidelity} numerical resolution of shallow water models can still be very computationally intensive, due to the large amount of degrees of freedom (DOFs) needed to solve the PDE {\cite{Quarteroni_Manzoni_etal_16,WRCR:WRCR13404}}. The resulting computational expense poses a barrier to the inclusion of fully resolved two-dimensional shallow water models in many-query and real-time applications, particularly when the analysis involves optimal design, parameter estimation, risk assessment, and/or uncertainty quantification. \textit{Reduced order models} (ROMs) offer a valuable alternative way to simulate such dynamical systems with considerably reduced computational cost in comparison to the high-fidelity model (HFM).
The objective of these approaches is to replace the high-fidelity model by one with significantly reduced dimensions, thus trading in computational burden for a controlled loss of accuracy \cite{Benner_Gugercin_etal_15}.

\textit{Reduced basis} (RB) methods
\cite{Quarteroni_Manzoni_etal_16} constitute a family of widely popular ROM techniques which are usually implemented with an offline-online decomposition paradigm. The \textit{offline} stage
involves the construction of a solution-dependent basis spanned by a set of RB ``modes'', which are extracted from a collection of high-fidelity solutions, also called \textit{snapshots}. The RB
``modes'' can be thought of as a set of global basis functions that can approximate the dynamics of the high-fidelity model. The most well known method to extract the reduced basis is called \textit{proper orthogonal decomposition} (POD) \cite{Sirovich_87,BHL1993,Liang_etal_2002},
which is particularly effective when the coherent structures of the flow can be hierarchically ranked in terms of their energy content. In this method, a truncated Singular Value Decomposition (SVD) of the snapshot matrix produces a low rank global basis of the most significant empirical modes, that are optimal with respect to the Euclidean $2$-norm \cite{Quarteroni_Manzoni_etal_16}. The POD method has been successfully applied in various fields and is often referred to by different names. For instance, it is termed principal component analysis (PCA) method in statistics \cite{J1986}, and the Karhunen-Loeve decomposition in signal analysis and pattern recognition \cite{DM2008}. The POD technique has also been applied to ocean models
\cite{VH2006}, air pollution models \cite{FZPPBN2014}, convective Boussinesq flows \cite{SB2015}, and SWE models {\cite{DMD,SSN2014,LFKG2016,LFK2017}}. In the last decade, Koopman mode theory \cite{Koopman1931} has also provided a rigorous theoretical background for an efficient modal decomposition in problems describing oscillations and other nonlinear dynamics using a technique called dynamic mode decomposition (DMD) \cite{Rowley2009}. Several variants of the DMD algorithm have been successfully applied as a modal decomposition tool in nonlinear dynamics \cite{Schmid2010,ABBN2016} and have also been adopted as an alternative for POD in the determining the most optimal global basis modes for nonlinear problems
\cite{DMD,Bistrian2017}.

In the \textit{online} stage, a linear combination of the reduced order RB modes is used to approximate the truth solution (high-fidelity numerical solution) for a new configuration of flow
parameters. The procedure adopted to compute the expansion coefficients leads to the classification of RB methods into two broad categories: \textit{intrusive} and \textit{non-intrusive}. In an
intrusive RB method, the expansion coefficients are determined by the solution of a reduced order system of equations, which is typically obtained via a Galerkin or Petrov-Galerkin projection of the high-fidelity (full-order) system onto the RB space \cite{LFK2017}. For linear systems, the POD/Galerkin projection is the most popular choice. However, in the presence of nonlinearities, an affine expansion of the nonlinear (or non-affine) differential operator must be recovered in order to make the evaluation of the projection-based reduced model independent of the number of DOFs of the high-fidelity solution. Several different techniques, collectively referred to as hyper-reduction methods \cite{AZCF2015}, have been proposed to address this problem.  Barrault \etal \cite{EIM2004} proposed the empirical interpolation method (EIM) that constructs an approximation of the non-affine parametrized function. Chaturantabut and Sorensen \cite{CS2010} proposed a discrete empirical interpolation method (DEIM) in which, using a collocation-based strategy, the reduced
approximation is enforced to match the nonlinear function at a specific number of sampling points. The ``gappy POD" method \cite{W2006} similarly minimizes the least squares error on the set of
sampled points to seek the most optimal reduced approximation. Several other methods have also been proposed, namely a coefficient-function approximation consisting of a linear combination of precomputed basis functions \cite{NP2008}, and a combination of a quadratic expansion method and the DEIM called the residual DEIM method \cite{XFBPNDH2014}. Moreover, in complex nonlinear problems some of the intrinsic structures present in the high-fidelity model may be lost during order reduction using the projection-based POD/Galerkin approaches. This can result in qualitatively wrong solutions or instability issues \cite{BMQR2015,AF2012}. As a remedy, Petrov-Galerkin projection based approaches have been proposed {\cite{XFDPNBEH2013,FPNEDX2013}}.  Alternatively, introducing a
diffusion term into the reduced model \cite{CBF2011}, has also been shown to improve the stability of ROM results.

A valuable alternative family of methods to address the issues of instability and loss of efficiency in the presence of more general, non-affine differential operators is represented by
\textit{non-intrusive} reduced order models (NIROMs). The primary advantage of this class of methods is that complex modifications to the source code describing the physical model can be avoided, thus
making it easier to develop reduced models when the source code is not available or easily modifiable, which can often be the case for legacy and commercial codes. In these methods, instead of a Galerkin-type projection, the expansion coefficients for the reduced solution are obtained via interpolation on the space of a reduced basis extracted from snapshot data. However, since the reduced bases generally belong to nonlinear, matrix manifolds, a variety of interpolation techniques have been proposed that are capable of enforcing the constraints characterizing those manifolds. Regression-based non-intrusive methods have been proposed that, among others, use artificial neural networks (ANNs), in particular multi-layer perceptrons \cite{HU2018}, neural ODEs \cite{DRF2021}, as well as Gaussian process regression (GPR) \cite{GH2019} to perform the interpolation.

Radial Basis Function (RBF) interpolation is another effective tool 
for interpolation of multidimensional scattered data \cite{NW2004} and has been demonstrated to be flexible, convenient and accurate in various research areas \cite{FF2015,WWMGBPA2010,SNPP2019}. Adopting RBF interpolation for extracting the coefficients of the reduced basis has been shown to be quite successful for nonlinear, time dependent PDEs \cite{XFPH2015}, nonlinear, parametrized PDEs
\cite{ADN2013,XFPN2017,Chen_etal_2018}, and aerodynamic shape
optimization \cite{IQ2013}, to name a few. 
However in most of the current non-intrusive ROMs, a random or uniform sampling of the snapshot space is employed while computing the coefficients of the reduced basis which may lead to prohibitive computational cost in the offline stage and stability issues in the online computations. 
To address this issue, Xiao \etal \cite{XFPN2017} presented a Smolyak sparse grid collocation approach, while an adaptive greedy sampling approach was proposed by Chen \etal \cite{Chen_etal_2018} to select optimal interpolation points in the parameter space.

In this work, we present a greedy non-intrusive reduced order model for the SWE in \textit{fast replay} applications. The high-fidelity numerical solutions of the SWE are obtained using the 2D depth-averaged module of the Adaptive Hydraulics (AdH) finite element suite, which is a U.S. Army Corps of Engineers (USACE) high-fidelity, finite element resource for 2D and 3D dynamics and has a wide user base, including USACE districts, universities and consulting agencies \cite{MSSS2013}. The suite engine is based on a conservative, implicit, continuous Galerkin (CG) finite element (FE) approximation with residual-based stabilization techniques \cite{BC1996,HM1986}. All the models within the AdH suite are both spatially and temporally adaptive \citep{Trahan2018}. It should, however, be noted that, by design, the RBF-POD NIROM outlined in this work is easily applicable to other high-fidelity numerical solvers. Unlike the approach presented in \cite{XFPH2015}, this RBF-POD NIROM model employs RBF interpolation to approximate the time derivative of the coefficients in the POD-basis space, thus adding the ability to represent 
time-dependent problems with greater flexibility. In this approach, the high-fidelity snapshots are represented in the reduced space of optimal POD modes. The RBF interpolation method is adopted to approximate the dynamics of the reduced system of equations.

The novelty of the current work lies in the choice of interpolating the temporal derivative of the POD-projection coefficients of the high-fidelity snapshots and the greedy selection of snapshot information to generate the RBF interpolant itself. The goal is to provide improved flexibility and efficiency in approximating online temporal dynamics. Specifically, three different greedy
algorithms are studied for the selection of an optimal subset of the collected snapshots as the centers for the radial kernel in order to improve the efficiency of the RBF NIROM. The \textit{p-greedy} \cite{DSW2005} algorithm is designed to optimize the selection of centers based on an iterative minimization of the error introduced by the kernel properties and measured by the power function. The \textit{f-greedy} \cite{SW2000,WH2013} algorithm performs a greedy selection of centers to minimize the residual error in the RBF approximation, which has been applied here by taking $l^2$-norms of the interpolated vector-valued functions. Finally, a novel power-scaled greedy or \textit{psr-greedy} algorithm is introduced that minimizes the scaled residual error where the power function is adopted as the variable scaling factor. The greedily
obtained set of centers is used to construct a RBF interpolant for approximating the time derivative of the projection coefficients of the high-fidelity snapshots. In the online stage, this RBF interpolant is employed to compute reduced solutions for any new time configuration queried by the fast-replay application.

The paper is organized as follows. In Section \ref{sec:hfm}, we present the general SWE in two dimensions for depth-averaged, free-surface flows and then summarize the SUPG-stabilized, CG
high-fidelity numerical method. In Section \ref{sec:rom}, the traditional nonlinear POD (NPOD) global model reduction technique via SVD-basis collection and subsequent Galerkin projection is presented, and relevant details for its application to the discrete, stabilized high-fidelity SWE model are provided. Section \ref{sec:rbf} provides a preliminary introduction to the theory of kernel-based approximation, while in Section \ref{sec:rbf-pod}, the POD-RBF NIROM is introduced, that uses the projected HFM snapshots to construct the RBF interpolant for online evaluations of the reduced model. In Section \ref{sec:greedy} three different adaptive greedy sampling strategies are proposed for an improved selection of near-optimal interpolation points to be used in constructing the RBF interpolant. Numerical tests have been performed for tidal flows with temporally varying boundary forcing, as well as riverine flows and the results are provided in Section \ref{sec:results}, with a careful examination of both the accuracy and the relative efficiency. 
Section \ref{sec:discussions} provides a brief review of some of the salient features of the NIROM framework. Finally, in Section \ref{sec:conclusions}, the concluding remarks are presented.

%%%---------------------------------------------------------------
%%%---------------------------------------------------------------
\section{High-fidelity formulation for the shallow water equations}\label{sec:hfm}
\subsection{Continuous formulation}
For our high-fidelity model, we consider a standard depth-averaged SWE formulation written as \cite{SBMT2011}

\begin{equation}\label{eq:conservative_compact}
\strongRes \equiv \pd{\vec q}{t} + \pd{\vec{p}_x}{x} + \pd{\vec{p}_y}{y} + \vec r = 0,
\end{equation}
with $\vec {q} = [q_1,q_2,q_3]^T$ the unknown conservative variable consisting of the flow depth, $q_1=h$, and discharges in the $x$ and $y$ directions given by $q_2=u_x h$ and $q_3=u_y h$, respectively. Here $u_x$ is the velocity in the $x$ direction and $u_y$ is the velocity in the $y$ direction. In addition to the conservative variable $\vec q$, we will also make direct reference to the primitive state variable $\vec u = [u_1,u_2,u_3]^T = [h,u_x,u_y]^T$ in some cases below as well. The flux vectors in the lateral directions are

\begin{equation}\label{eq:swe-flux-x}
  \vec p_x = \left\{\begin{array}{c}
  u_x h \\
  u_x^2h + (1/2)gh^2 - h\left(\sigma_{xx}/\rho\right) \\
  u_x u_y h - h\left(\sigma_{yx}/\rho\right)
  \end{array}
  \right\},
\end{equation}

\begin{equation}\label{eq:swe-flux-y}
  \vec p_y = \left\{\begin{array}{c}
  u_y h \\
  u_x u_y h - h\left(\sigma_{xy}/\rho\right) \\
  u_y^2h + (1/2)gh^2 - h\left(\sigma_{yy}/\rho\right) 
  \end{array}
  \right\},
\end{equation}
and

\begin{equation}\label{eq:swe-source}
  \vec r = \left\{\begin{array}{c}
  0 \\
  gh\pd{h_b}{x} + gh\lrb{\lrp{n^2_{mn}u_x\sqrt{u_x^2+u_y^2}}/h^{4/3}} - f_c h u_y\\
  gh\pd{h_b}{y} + gh\lrb{\lrp{n^2_{mn}u_y\sqrt{u_x^2+u_y^2}}/h^{4/3}} + f_c h u_x\\
  \end{array}
  \right\},
\end{equation}
where we have written eqs.~\eqref{eq:swe-flux-x}-(\ref{eq:swe-source}) in terms
of the primitive variables for convenience. Here, $\rho$ is the fluid
density, $g$ is the gravitational acceleration, and $h_b$
is the elevation of the bottom surface. $f_c$ is the Coriolis
coefficient, and a standard Manning's parameterization for bottom
roughness is used in eq.~\eqref{eq:swe-source} with coefficient $n_{mn}$. 
$\sigma_{xx,xy,yx,yy}$ are Reynolds stresses due to turbulence 
which are approximated using the Boussinesq approach for the 
gradient in the mean currents

\begin{align}\label{eq:reynolds-stress-def}
  \sigma_{xx}&=2\rho\nu_t\pd{u_x}{x}, \\
  \sigma_{yy}&=2\rho\nu_t\pd{u_y}{y}, \mbox{ and } \\
  \sigma_{xy}&=\sigma_{yx}=2\rho\nu_t\lrp{\pd{u_x}{y}+\pd{u_y}{x}},
\end{align}
where $\nu_t$ is the kinematic eddy viscosity \citep{SBMT2011}.

%%%%------------------------------------------------------
\subsection{Stabilized finite element scheme for the semi-discrete weak formulation}
Given a spatial domain $\Omega \subset \R^2$, temporal domain $[0,T]$, and a suitable test space $\mathbf{W}$ defined on $\Omega$, the classical weak form
for eq.~\eqref{eq:conservative_compact} can be written as

\begin{equation}\label{eq:Galerkin}
  \int_{\Omega}\pd{\vec q}{t}\bw \dx - \int_{\Omega}\lrp{\pd{\bw}{x}\vec p_x + \pd{\bw}{y}\vec p_y}\dx 
  + \int_{\partial\Omega}\lrp{\vec p_x n_x  + \vec p_y n_y}\bw\ds 
  +\int_{\Omega}\vec r\bw \dx = 0,
\end{equation}
where $\bw$ is an arbitrary test function in $\mathbf{W}$ and $\vec n = \lrb{n_x,n_y}^T$ is the outer unit normal for $\Omega$. Note that we have integrated the divergence terms in eq.~\eqref{eq:conservative_compact} by parts and assume that eq.~\eqref{eq:Galerkin} holds for $t\in [0,T]$. 
If we denote ${\xi}_{s}$ as the $s$-th basis function from a standard piece-wise linear CG approximation space, then the discrete solution $q_{h,i}$ $(i=1,2,3)$, to the weak problem given by eq.~\eqref{eq:Galerkin} can be represented as $q_{h,i} = \sum_{s=1}^{N}{\xi}_{s} q_{i,s}$, where $\bhat{ q_i} =[q_{i,1},\ldots, q_{i,N}]^T$, $(i=1,2,3)$, are the $N$ dimensional coefficient vectors for each solution component. Similarly, the primitive variables can be represented in terms of $\xi_s$ as $u_{h,i} = \sum_{s=1}^{N}\xi_s u_{i,s}$, with $N$-dimensional coefficient vectors $\bhat{ u_i}=[u_{i,1},\ldots, u_{i,N}]^T, \, (i=1,2,3)$. Unfortunately, straightforward Galerkin approximations to eq.~\eqref{eq:Galerkin} fail to produce accurate solutions in a number of flow regimes \citep{Aizinger_Dawson_02}. In particular, advection-dominated conditions with shocks or near-shocks are particularly challenging. For this reason, we use residual-based stabilization from \citep{Berger_Stockstill_95} which is an extension of the original SUPG method \cite{HM1986,HFMQ1998}. Letting $\bw_h$ be an arbitrary test function in $\mathbf{W}_h$, we write

\begin{align}\label{eq:gridscale_final}
  \int_{\Omega}\pd{\vec q_h}{t}\bw_h \dx - \int_{\Omega}\lrp{\pd{\bw_h}{x}\vec p_x + \pd{\bw_h}{y}\vec p_y}\dx
  + \int_{\partial\Omega}\lrp{\vec p_x n_x  + \vec p_y n_y}\bw_h\ds \nonumber \\
  +\int_{\Omega}\vec r\bw_h \dx + \sum_{e}\int_{\Omega_e}\strongRes_h\lrb{\lrp{\mathbf{J}_x\vec \tau_x}^T\pd{\bw_h}{x} +
    \lrp{\mathbf{J}_y\vec \tau_y}^T\pd{\bw_h}{y}}\dx = 0.
\end{align}
Here, $\mathbf{J}_{x,y} = \vec p_{x,y}^{\prime}$ are the advective flux Jacobians, $\lrB{\Omega_e}_1^{N_e}$ is the collection of elements in a simplicial triangulation of $\Omega$, and $\strongRes_h$ is an approximation of the strong residual $\strongRes$. The specific form of the intrinsic time scale parameters, $\vec \tau_{x,y}$, can be found in \citep{Trahan2018}.

There are a range of boundary conditions that can be incorporated into eq.~\eqref{eq:gridscale_final} in practice, and the test space $\mathbf{W}$ is chosen accordingly. For example, at riverine inflow boundaries we consider normal flux conditions on the total discharge,

\begin{equation}
q_2 n_x+q_3 n_y = (u_x h)n_x + (u_y h)n_y = q_b. 
\end{equation}
At land boundaries, a no-flow condition $u_x n_x + u_y n_y=0$ holds, while at sea boundaries, the free surface elevation $\eta=h+h_b = \eta_b$, is specified. Radiation boundary conditions may also be specified as well \cite{Trahan2018}.

\subsection{Time discretization}
Eq.~\eqref{eq:gridscale_final} together with the appropriate boundary conditions can be written formally as a semi-discrete nonlinear system 
\begin{align}\label{eq:gridscale_semidiscrete}
    \vec R\left(\pd{\vec q_h}{t},\vec q_h,t\right)=0.
\end{align}

To address the temporal discretization of eq.~\eqref{eq:gridscale_semidiscrete}, we consider a second-order backward Euler (BDF-2) approximation in which the nonlinear terms from eq.~\eqref{eq:gridscale_final} are extrapolated in time. The time derivatives in the continuity and momentum equations are expressed as

\begin{align}\label{eq:time-disc}
  \begin{split}
    \left( \pdt{q}\right)_i^{n+1} = \alpha \left[ \frac{\left(3/2 q_i^{n+1} - 1/2q_i^n \right) - \left(3/2 q_i^n - 1/2 q_i^{n-1} \right)}{\Delta t}\right] \\
    + (1-\alpha) \left[ \frac{q_i^{n+1} - q_i^n}{\Delta t}\right],
  \end{split}
\end{align}
where $0 \leq \alpha \leq 1$ is a factor that determines the order of time stepping, $\Delta t$ is the time step size and the superscript $n$ indicates the $n^{th}$ time step, so that the actual time is $t^n= n \Delta t$.

After introducing the time discretization \eqref{eq:time-disc} in eq.~(\ref{eq:gridscale_final}), we can recover a fully discrete nonlinear system in the primitive variables at time step $t^{n+1}$ as

\begin{align*}%\label{eq:discrete-prim-residual}
    \vec R^{n+1}_u(\vec u_h^{n+1}; \vec u_h^{n},\vec u_h^{n-1}) = 0,
\end{align*}
where $\vec u_h = [\bhat{ u_1}^T,\bhat{ u_2}^T,\bhat{ u_3}^T]^T$. For simplicity, in the rest of the article we suppress the explicit dependence of the discrete residual on $t$ and on the values of the solution at $t^n$ and $t^{n-1}$ and write,

\begin{align}\label{eq:discrete-prim-residual}
    \vec R^{n+1}_u(\vec u_h^{n+1}) = 0,
\end{align}

In terms of the conservative variables, the discrete nonlinear system can be written similarly as $\vec R^{n+1}_q(\vec q_h^{n+1}) = 0$
where $\vec q_h = [\bhat{ q_1}^T,\bhat{ q_2}^T,\bhat{ q_3}^T]^T$ .

%%%---------------------------------------------------------------
%%%---------------------------------------------------------------
\section{Global basis reduced order model}\label{sec:rom}
One of the main advantages of a projection-based reduced model
framework is the systematic splitting of the computational procedure
into a resource intensive offline stage and an efficient online stage,
either in a real-time or many-query context. The
offline stage primarily includes all simulations carried out with the 
high-fidelity numerical model introduced in Section 2. It is assumed
that the temporal and spatial resolutions adopted are sufficient to
extract all the relevant dynamics and key features of the flow. In the
online stage, due to lack of computational resources and/or time
constraints, a more efficient model (\ie ROM) with possible relaxed
accuracy is required and an efficient ROM that leverages the
fine-scale information collected in the offline stage is introduced.

The non-intrusive ROM approach proposed in this work involves two 
additional steps in the offline computations. First, the high-dimensional 
solution to the dynamical system is represented in a low-dimensional subspace 
that is constructed by choosing the most significant (in terms of energy) 
global empirical basis modes obtained by POD. Second, a multivariate RBF 
interpolant is adopted to model the time evolution of the projection coefficients of the high-dimensional snapshots on to the lower-dimensional 
space of POD modes. This approach does not involve the Galerkin projection 
of the governing equations to the space of the POD modes, which 
circumvents the need for additional approximation to account for the 
nonlinearity of the governing equations.

\subsection{Proper Orthogonal Decomposition}
POD is a widely used projection-based model reduction
method and has been successfully applied to a host of large-scale
dynamical systems \cite{AS2001}. We use a generic algebraic
formulation of dynamical systems to present a brief overview of the procedure.

Consider a generic vector of $3N$ degrees of freedom, $\vec v =
[\vec v^T_1,\vec v^T_2,\vec v^T_3]^T$, which can represent either the
discrete conservative variables, $[\bhat{ q_1}^T,\bhat{ q_2}^T,\bhat{ q_2}^T]^T$ or the discrete primitive variables $[\bhat{u_1}^T,\bhat{u_2}^T,\bhat{u_3}^T]^T$.  Then let
$\widetilde{\vec{S}}_i = [\bv_i^1, \ldots \bv_i^M]$ be a $N\times M$
matrix of solution snapshots, obtained from the high-fidelity offline computations from time $t=0$ to $t=T$ for each of the three state variables, using potentially variable time steps. It is assumed that the set $\widetilde{\vec{S}}_i$ captures all the key features of the flow phenomena. Following the usual practice \cite{Lassila_Manzoni_etal_14}, a new ``normalized" set of snapshots $\vec S_i$ is generated by subtracting the time averaged value or the column means from each row of the snapshots in $\widetilde{\vec{S}}_i$ such that $\vec S_i = (\bv_i^1 - \bar{\bv}_i, \ldots, \bv_i^M - \bar{\bv}_i)$ where $\bar{\bv_i} = \sum_{n=1}^M \bv_i^n/M$. In some cases, the initial condition is used to offset the set of snapshots \cite{CFCA2013,CBF2011}, while the non-normalized
solution may also be directly used to build the basis representative in
order to simplify the analysis of the procedure \cite{LFKG2016}. The updated snapshot matrices for each of the variables $\vec S_1, \vec S_2, \vec S_3$ are treated separately, but in an identical manner. Thus, the subscripts are omitted and the details are provided for a general snapshot matrix $\vec S$.

A ``thin" singular value decomposition (SVD) of the snapshot matrix $\vec S$ is performed

\begin{align}\label{eq:thin_SVD}
    \vec S = \widetilde{\vec \Theta} \widetilde{\boldsymbol{\Sigma}} \widetilde{\vec \Psi}^T,
\end{align}
where $\widetilde{\boldsymbol{\Sigma}} = \text{diag}(\sigma_1,\ldots,\sigma_\RR)$ is a $\RR\times \RR$ diagonal matrix containing the singular values arranged in decreasing order of magnitude, $\sigma_1 \geq \sigma_2 \ldots \geq \sigma_\RR$ and $\RR < \min \{N,M\}$ is the rank of $\vec S$. $\widetilde{\vec \Theta}$ and $\widetilde{\vec \Psi}$ are $N\times \RR$ and $M \times \RR$ matrices respectively, whose columns are the orthonormal left and right singular vectors of $\vec S$ such that $\widetilde{\vec \Theta}^T \widetilde{\vec \Theta} = \vec I_{\RR} = \widetilde{\vec \Psi}^T \widetilde{\vec \Psi}$, and that are defined as

\begin{align*}
    \vec S \vec S^T \vec \theta_n = (\sigma_n)^2 \, \vec \theta_n, \quad \vec S^T \vec S \vec \psi_n = (\sigma_n)^2 \, \vec \psi_n, \qquad  1 \leq n \leq M.
\end{align*}
The columns $\vec \theta_n$ of the matrix $\widetilde{\vec \Theta}$ are also
ordered corresponding to the singular values $\sigma_n$ and these provide the desired basis vectors (or empirical modes) for the solution snapshot vector $\bv$. This follows directly from eq.~(\ref{eq:thin_SVD}), as every column of the snapshot matrix $\vec S$ lies in the range space of $\widetilde{\vec \Theta}$, \ie every solution snapshot vector can be represented as $\bv^n = \bar{\bv} + \widetilde{\vec \Theta} \vec c$ for some vector $\vec c$. As a consequence of this property of the SVD, a lower dimensional (reduced order) approximation can be obtained by choosing a small number, $m \ll M$, of the leading empirical modes as the set of basis vectors. Let ${\vec \Theta}$ denote the matrix of the first $m$ columns of $\widetilde{\vec \Theta}$, $\vec \Psi$ be the matrix containing the first $m$ rows of $\widetilde{\vec \Psi}$, and ${\boldsymbol{\Sigma}}$ be a diagonal matrix containing the first $m$ singular values from $\widetilde{\boldsymbol{\Sigma}}$, then the high-fidelity solution $\bv^n$ at time $t^n$ has the reduced order representation $\bz^n$ when projected onto the space of POD modes,

\begin{align}\label{eq:pod-basis}
    \bv^n \approx \bar{\bv} + {\vec \Theta} \bz^n = \bar{\bv} + \sum_{i=1}^m z_i^n \vec \theta_i.
\end{align}
Although the POD basis provides an optimal rank-$m$ approximation $\widehat{\vec S} = {\vec \Theta} {\boldsymbol{\Sigma}} {\vec \Psi}^T$ of the snapshot matrix $\vec S$, some information is lost due to the inexactness of eq.~(\ref{eq:pod-basis}). However, a desired level of accuracy, $\tau_{POD}$ can be obtained by choosing the truncation number $m$ such that: $\lrp{\nm{\vec S - \widehat{\vec S}}_F^2/ \nm{\vec S}_F^2} = \lrp{{\sum_{i=m+1}^M \sigma_i^2}/{\sum_{i=1}^M \sigma_i^2}} \leq \tau_{POD}$, where $\nm{\cdot}_F$ denotes the Frobenius norm.

In traditional Galerkin/POD approach, a reduced order model for the
time evolution of the state vectors is obtained by a Galerkin projection of the system of high-dimensional equations onto the reduced space spanned by the
POD empirical modes. Returning to the high-fidelity SWE model in
eq.~(\ref{eq:gridscale_final}), let $\bz = (\bz_1^T, \bz_2^T, \bz_3^T)^T \in \R^{m_1 + m_2 + m_3}$ denote the projection coefficients of the high-dimensional snapshots $\bv = (\bv_1^T, \bv_2^T, \bv_3^T)^T$ where $m_1, m_2, m_3$ are the POD truncation levels for the three state vectors. Also, define a global basis $\vec \Theta = \text{diag}({\vec \Theta}_1, {\vec  \Theta}_2, {\vec \Theta}_3)$ as a block-diagonal matrix. Using the representation given by eq.~(\ref{eq:pod-basis}) in the high-fidelity, fully discrete, nonlinear system 
(\ref{eq:discrete-prim-residual}) and performing a standard Galerkin
projection yields a reduced system of equations 

\begin{align}\label{eq:dynamic_system_reduced}
    \vec \Theta^T \vec R^{n+1}(\bar{\bv} + \vec \Theta \bz^{n+1}) = 0.
\end{align}

\noindent
The system of $m$ equations for the evolution of the POD coefficients $\bz$, given by eq.~(\ref{eq:dynamic_system_reduced}), is a reduced order approximation of the high-fidelity system of $3N$ equations, given by eq.~(\ref{eq:discrete-prim-residual}), and the full order solution can be recovered using eq.~(\ref{eq:pod-basis}).
In the numerical experiments below, eq.~(\ref{eq:dynamic_system_reduced}) will be referred to as the NPOD (Nonlinear POD) ROM. Due to the presence of nonlinearities, the NPOD model may still be slow, since the evaluation procedure scales like the fine dimension. Several \textit{hyper-reduction} (approximation of the nonlinearity in a reduced space) strategies have been proposed to recover the lost efficiency \citep{CFCA2013,W2006}.
Note, that unlike the alternating direction implicit (ADI) scheme considered in \citep{SSN2014}, our fine-scale approximation involves non-polynomial nonlinear terms due to the bottom roughness parametrization and stabilization.

\section{RBF-POD reduced order model formulation}\label{sec:nirom}

In this section, we introduce the fundamental framework for approximating the time evolution of the coefficients of the POD expansion via kernel-based approximation schemes. The striking feature underlying this approach is that the construction of the kernel-based interpolation framework is independent of the POD-based reduced basis representation of the snapshot space.

\subsection{Remarks on reproducing kernel Hilbert spaces}\label{sec:rbf}
The problem of multivariate scattered data interpolation is stated as: given a set of $N_d$ distinct points $X=\{\bx_i \in \mathbb{R}^d, i=1,2,\ldots,N_d\}$, and a set of $N_d$ real numbers, $\{f_i, i=1,2,\ldots,N_d \}$, find a continuous function $F(\bx)$ that satisfies $F(\bx_i) = f_i, \, \forall \, i=1,2,\ldots,N_d $. We employ radial basis function (RBF) interpolation for determining the function $F( \bx)$.

A function $\Psi:\mathbb{R}^d \rightarrow \mathbb{R}$ is called radial if for each $\bx \in \mathbb{R}^d$, $\Psi(\bx) = \phi(\nm{\bx})$, where $\phi: [0,\infty) \rightarrow \mathbb{R}$ is an univariate function, generally referred to as the radial basis function.
Given the set of scattered centers $X = \{\bx_i\}_{i=1}^{N_d}$, a RBF interpolant is defined as the linear combination of $N_d$ instances of a chosen radial basis function $\phi$, that are translated about the centers and takes the form

\begin{align}\label{eq:interp_def}
    F(\bx) = \sum_{j=1}^{N_d} \alpha_j \phi (\nm{\bx - \bx_j}),
\end{align}
where $\nm{\bx - \bx_j}$ is the Euclidean distance between the observation point $\bx$ and the center $\bx_j$. The unknown coefficients $\alpha_j, j = 1,2,\ldots,N_d$ are determined by solving a system of linear equations of order $N_d$ generated by the interpolation conditions,

\begin{align}
    f_i = F(\bx_i) \equiv \sum_{j=1}^{N_d} \alpha_j \phi(\nm{\bx_i - \bx_j}).
\end{align}
The linear system may be written in the matrix form $\vec A \vec \alpha = \vec \gamma$, where $\vec \alpha = [\alpha_1, \alpha_2, \ldots, \alpha_{N_d}]^T $, $\vec \gamma = [f_1, f_2, \ldots, f_{N_d}]^T$ and

\begin{align}\label{eq:rbf_interp_matrix}
    \vec A = [A_{ij}] = [\phi(\nm{\bx_i - \bx_j})].
\end{align}

Such a system, admits a unique solution if the kernel is strictly positive definite. Otherwise, for conditionally positive definite functions, a polynomial term needs to be added to make the problem well-posed \cite{C1991,F2007}. 
We now introduce some basic notions of reproducing kernel Hilbert spaces (RKHS) which will be useful later for studying greedy algorithms and their convergence results \cite{SW2000,WH2013}. 

Let us assume $\Phi: \Omega \times \Omega \rightarrow \R$ denotes a continuous, symmetric, and strictly positive definite kernel on a compact set, $\Omega \subset \R^d$. Let $V_{\Omega} = \text{span}\{\Phi(\cdot,\bx): \bx \in \Omega) \}$ denote the vector space spanned by all functions $\Phi(\cdot, \bx)$, which can be equipped with the natural inner product

\begin{align}\label{eq:rkhs_in_prod}
    \left( \sum_{j=1}^{N_m} \alpha_j \Phi(\cdot, \bx_j), \sum_{k=1}^{N_p} \beta_k \Phi(\cdot, \btx_k) \right)_\Phi := \sum_{j=1}^{N_m} \sum_{k=1}^{N_p} \alpha_j \beta_k \Phi(\bx_j, \btx_k).
\end{align}
It can be seen that $\Phi$ is the reproducing kernel of $V_\Omega$ with respect to the inner product $(\cdot,\cdot)_\Phi$ \ie for each symmetric, positive definite $\Phi$ there is a unique such space with the reproducing property

\begin{align}\label{eq:rkhs_prop}
    \lrp{f, \Phi(\cdot,\bx)}_\Phi = f(\bx), \quad \forall f \in V_\Omega, \, \bx \in \Omega.
\end{align}
The closure of $V_\Omega$ yields a Hilbert space, induced by the reproducing kernel $\Phi$ over $\Omega$, which is known as the reproducing kernel Hilbert space (RKHS) or the native Hilbert space to $\Phi$ and will be denoted by $\mathcal{N}_\Phi(\Omega)$.

Let $X = \{\bx_1, \ldots, \bx_{N_d} \} \subset \Omega$ be the finite discrete set defined before and $\eta_j \in V_X := \text{span}\{\Phi(\cdot,\bx): \bx \in X \subset \Omega\}$ denote the cardinal functions, also known as the Lagrangian basis, \ie $\eta_j$ satisfies $\eta_j(\bx_k) = \delta_{jk}$, $j,k=1,2,\ldots, N_d$. Then the interpolant given by eq.~(\ref{eq:interp_def}) can also be represented by

\begin{align}\label{eq:interp_cardinal}
    F(\bx) = \sum_{j=1}^{N_d} f(\bx_j) \eta_j(\bx).
\end{align}
Therefore, using the reproducing property of eq.~(\ref{eq:rkhs_prop}) and the cardinal representation in eq.~(\ref{eq:interp_cardinal}), the interpolation error for a function $f \in \mathcal{N}_\Phi(\Omega)$ can be expressed as

\begin{align}\label{eq:interp-err-power}
    f(\bx) - F(\bx) = \left( f, \Phi(\cdot,\bx) - \sum_{j=1}^{N_d} \eta_j(\bx) \Phi(\cdot, \bx_j)\right)_\Phi.
\end{align}
Applying the Cauchy-Schwarz inequality to (\ref{eq:interp-err-power}) leads to

\begin{align}\label{eq:power-err-est}
    \abs*{f(\bx) - F(\bx)} \leq P_X(\bx) \nm{f}_\Phi,
\end{align}
where $P_X(\bx)$ denotes the power function (see \cite{W2005}, 11.2) which takes the explicit form

\begin{subequations}
\begin{align}
    P^2_X(\bx) &:= \nm*{\Phi(\cdot,\bx) - \sum_{j=1}^{N_d} \eta_j(\bx) \Phi(\cdot, \bx_j)}_{\Phi}^2 \label{eq:pow-func}\\
    &= \Phi(\bx,\bx) - 2 \sum_{j=1}^{N_d} \eta_j(\bx) \Phi(\bx, \bx_j) + \sum_{j,k=1}^{N_d} \eta_j(\bx)\eta_k(\bx) \Phi(\bx_j, \bx_k) \label{eq:pow-func-Lag}\\
    &= \Phi(\bx,\bx) - \vec b(\bx)^T \vec A^{-1} \,\vec b(\bx). \label{eq:pow-func-num}
\end{align}
\end{subequations}
As can be seen in eq.~(\ref{eq:pow-func}), the power function is essentially the norm of the pointwise error functional, and it can be computed numerically as a quadratic form given by eq.~(\ref{eq:pow-func-Lag}) using the Lagrange basis. Alternatively, the power function can also be evaluated using the representation given by eq.~(\ref{eq:pow-func-num}), where $\vec A$ is the interpolation matrix as defined in eq.~(\ref{eq:rbf_interp_matrix}) and $\vec b(\bx) = [\Phi(\bx,\bx_{1}), \ldots, \Phi(\bx,\bx_{N_d}) ]^T$ (see \cite{F2007} for further details).
It can be readily observed from eq.~(\ref{eq:pow-func}) that whenever $\Phi$ is a strictly positive definite kernel, \ie, $\vec A$ is a positive definite matrix, then the power function satisfies the bounds

\begin{align}\label{eq:pow-bounds}
    0 \leq P_X(\bx) \leq \sqrt{\Phi(\bx,\bx)}.
\end{align}

Moreover, if $X \subseteq Y$ are two point sets in $\Omega$, then the associated power functions satisfy the following necessary monotonicity property

\begin{align}\label{eq:pow-min}
    P_X(\bx) \geq P_Y(\bx), \quad \bx \in \Omega.
\end{align}

The remarks above are meaningful for introducing greedy methods for our problem, as done in the next subsections.

\subsection{RBF approximation of the reduced coefficients}\label{sec:rbf-pod}
%%%%%%--------------------------------------------------------------
In this work, the temporal dynamics of the governing system of equations is approximated using RBF interpolation over the set of projection coefficients of the high-fidelity snapshots. For simplicity, it is assumed that the time evolution of the coefficients $\bz$ can be represented as a semi-discrete dynamical system,

\begin{align}\label{eq:POD-semi-discrete}
\dot{\bz} = \vec f(\bz,t), 
\end{align}
where all the information about the temporal dynamics including the SUPG stabilization and other nonlinear terms are embedded in $\vec f(\bz,t)$. 
Introducing a first-order time discretization, the reduced solution at time level $n+1$ can be obtained by

\begin{align}\label{eq:POD-time-deriv}
z_j^{n+1} = z_j^n + \Delta t^n  f_j(\bz^{n}), \quad n \in \{0,1,\ldots M-1\}, j \in \{1,2,\ldots, m\},
\end{align}
subject to the initial condition

\begin{align}
  \vec z^0 = \vec {\Theta}^T\lrp{\vec v^0 - \bar{\vec v}}.
\end{align}
In general, the NPOD reduced model given by eq.~(\ref{eq:dynamic_system_reduced}) can be recast in a similar discrete dynamical system form that is based on a Galerkin projection framework. In the POD-RBF NIROM framework proposed here, the Galerkin projection step is avoided and the time derivatives $f_j (j=1,\ldots,m)$ in eq.~(\ref{eq:POD-time-deriv}) are approximated using RBF interpolation. This approach is a generalization of the strategy adopted in \cite{XFPH2015}, where RBF interpolation was employed to approximate the evolution of the projection coefficients using the direct iteration scheme $\bz^{n+1} = \widetilde{\vec f}(\bz^n) $.
The approach in eq.~(\ref{eq:POD-time-deriv}) allows us to isolate the error in the discrete approximation of the time derivative of the reduced solution from the overall error of the reduced order model. The numerical experiments in this paper have been obtained with a first-order time discretization, as shown in eq.~(\ref{eq:POD-time-deriv}), but the application of higher order discretization schemes would follow similarly.

There are many kernels available in RBF literature which differ in terms of smoothness. In the context of the shallow water applications, kernels with limited regularity may be preferable, and this motivates the choice of the strictly positive definite Mat\'{e}rn $C^0$ kernel, given by $\phi(r) = e^{-cr}$, where $r$ denotes the Euclidean distance.
The constant $c$ is referred to as the shape factor of the corresponding RBF and can affect the accuracy of the fit. For further details about its tuning, we refer the reader to \cite{F2007}. 

Let $F_j$ denote a RBF interpolant approximating the time derivative function $f_j$ for a single POD coefficient $z_j^{n+1}$ at time level $(n+1)$, defined by a linear combination of $N_i$ instances of a radial basis function $\phi$. Then it assumes the form,

\begin{equation}\label{eq:nirom-rbf-interpolant}
F_j (\vec z) = \sum_{k=1}^{N_i} \alpha_{j,k} \, \phi \lrp{\nm{\vec z - \widehat{\vec z}_k}}, \quad j=1,\ldots,m,
\end{equation}
where $\{ \widehat{\vec z}_k \, | \, k = 1,\ldots, N_i \}$ denotes the set of ``centers" or trial points and $\alpha_{j,k}$ $(k = 1,\ldots, N_i)$ is the unknown interpolation coefficient corresponding to the $k^{th}$ center for the $j^{th}$ projection coefficient.
These interpolation coefficients are computed by enforcing the interpolation function $F_j$ to exactly match the time derivative of the projection coefficients at $N_e$ test points ($N_e \geq N_i$), that is,

\begin{align}\label{eq:nirom-interp-cond}
    g_{j,n} \equiv \frac{z_j^{n+1} - z_j^n}{\Delta t^{n+1}} = \sum_{k=1}^{N_i} \alpha_{j,k} \, \phi \lrp{\nm{\vec z^n - \widehat{\vec z}_k} }, \quad n = 1, \ldots N_e; \, j=1,\ldots,m.
\end{align}

In this work, the set of centers and the test points have been identically chosen from the set of snapshot projection coefficients as $\{\vec z^l \,|\, l = 0, \ldots, M-1\}$ such that $N_i = N_e = M$. The time derivative functions are assumed to be independent of the time step which leads to a symmetric system of $M$ equations to obtain the unknown interpolation coefficients, $\alpha_{j,k}$ for $k \in 0,\ldots M-1$ and $j=1,\ldots,m$. Thus, for $j=1,\ldots,m$,  the problem reduces to solving a system of $M$ linear equations  

\begin{gather}\label{eq:interp-cond}
 \vec A \vec \alpha^j=\boldsymbol{g}^j, 
\end{gather}
where 

\begin{align*}%\label{eq:rbf_interp_matrix2}
    [A_{n,k}] = [\phi\lrp{\nm{\vec z^n - \vec z^k}}],  \quad \quad  n,k = 0, \ldots M-1,\\
    \vec \alpha^j = [\alpha_{j,0}, \alpha_{j,1}, \ldots, \alpha_{j,M-1}]^T , \quad  \boldsymbol{g}^j = [g_{j,0}, g_{j,1}, \ldots, g_{j,M-1}]^T.
\end{align*}

The coefficients $\vec \alpha^j$ define a unique RBF interpolant which can then be used to approximate eq.~(\ref{eq:POD-time-deriv}) and generate a non-intrusive model for the evolution of the reduced solution as

\begin{align}\label{eq:RBF-time-deriv}
\tilde{z}_j^{n+1} = \tilde{z}_j^n + \widetilde{\Delta t}^n  F_j \lrp{\widetilde{\bz}^{n}}, \quad n \in \{0,\ldots M-1\}; j \in \{1,\ldots,m\},
\end{align}
with an appropriate initial condition $\tilde{\bz}(\cdot,0) = \tilde{\bz}^0$.
We conclude by pointing out that the RBF methodology is easy to implement in any dimension, but the computational need may scale up with the number of interpolation centers. Thus, we also present three algorithms dedicated to selecting an optimal (minimal) number of centers and their locations.

\subsection{Optimal distribution of RBF interpolation points}\label{sec:greedy}
A key aspect of finding an efficient sparse approximation using a radial basis kernel is the concept of $m$-term approximation \cite{D1998}, which is basically a measure of how accurately a function from a given function space can be approximated by the linear combination of $m$ functions belonging to a subset of the same space. This leads to the challenge of finding methods and algorithms that determine the best or near-best $m$-term approximations. In the context of RBF interpolation several adaptive schemes have been proposed for the selection of an optimal set of centers like thinning algorithms \cite{FI1998}, greedy algorithms, and k-mean clustering methods \cite{LZ2008,LH2009,SBNK2003}. In this work, we consider methods belonging to the family of greedy algorithms, which have been demonstrated to yield near-optimal $m$-term approximations under various conditions \cite{TGS2006,T2008,SH2017,SNPP2019}. The ``greedy" aspect of these algorithms has its foundation in a greedy step, which determines the next center to be added to the existing set of chosen centers according to certain minimizing criteria involving residuals or power functions. Some approximation and convergence results have been established for greedy algorithms in general spaces, \eg Hilbert \cite{D1998} or Banach spaces \cite{LT2005}. The greedy algorithms used in this work are primarily driven by the goal of ensuring approximation accuracy while improving efficiency, and hence any resulting gains in sparsity of representation are obtained as a bonus.

\subsubsection{\textit{p-greedy} algorithm}\label{sec:p-greedy}
It can be observed from eq.~(\ref{eq:power-err-est}), that the power function helps us to estimate the interpolation error by allowing us to decouple the effects due to the values of the data function $f$ from the effects of the kernel $\Phi$ and the location of the centers $\{\bx_j\}_{j=1}^{N_d}$. This crucial observation, along with the error bounds (\ref{eq:pow-bounds}), and the minimization property (\ref{eq:pow-min}) were the key ingredients used by De Marchi \etal \cite{DSW2005} to develop the {\textit{p-greedy}} algorithm as a method to iteratively obtain a near-optimal set of center locations that are independent of the data values.

Let $X = \{ \bz_i \,\lvert \, 0\leq i \leq M\}$ be the set of projection coefficients or centers for the radial basis kernel. The first selected center is given by $\vec z_1 = \argmax_{\vec z_i \in X} \{\Phi(\vec z_i, \vec z_i) \}$. Let us assume that after $k$ greedy iterations, the set of selected centers is given by $X_k$ such that $|X_k| = k$. Also, let the $k$-term RBF interpolation matrix be denoted by $\vec A^k$ and the corresponding RBF interpolant function be denoted by $\vec F^k$. 
In the $(k+1)^{th}$ iteration, the power function is evaluated at each of the remaining centers in $X \setminus X_k$, and the worst approximated center $\vec z_{k+1}$ \ie the center at which the power function attains the maximum value, is selected to enrich the existing set of centers. This process continues until the maximum value of the power function drops below a chosen tolerance $\tau_p$. The \textit{p-greedy} algorithm has been summarized in Algorithm {\ref{alg:p-greedy}}.

\begin{algorithm}[H]
  \DontPrintSemicolon
  \SetAlgoLined
	\KwResult{ $\widetilde{X}$ : Optimal set of RBF centers }
  \SetKwInOut{Input}{Input}\SetKwInOut{Output}{Output}
  \Input{$X = \{\vec z_1,\vec z_2, \ldots, \vec z_{M}\}$, $M > N_{max}$ \;
         }
  \BlankLine
	Initial center: $\vec z_1 = \argmax\limits_{\vec z_i \in X} \{\Phi(\vec z_i, \vec z_i) \}$ \;
	\While{max$\{P_{X_k}(\vec z_i)\}> \tau_p$ \& $k < N_{max}$}
	{
		Compute $P_{X_k}(\vec z_i), \quad \forall \vec z_i \in X\setminus X_k $ \;
		Let $\vec z_{k+1} = \argmax\limits_{\vec z_i \in X\setminus X_k} \{P_{X_k}(\vec z_i) \}$\;
		$X_{k+1} \leftarrow X_k \cup \{ \vec z_{k+1}\}$\;
		 Re-compute RBF system: $\vec F^{k+1}, \vec A^{k+1}$\;
	}
	\caption{\textit{p-greedy} algorithm for selecting optimal set of RBF centers}\label{alg:p-greedy}
\end{algorithm}

\textbf{Remark.} The main advantage of the \textit{p-greedy} approach is that the set of reduced centers is computed only one time, independently of the number of modes. The main drawback is that this approach is sub-optimal, indeed it only minimizes one term in the upper bound given by eq.~(\ref{eq:power-err-est}). To partially overcome this, we also take into account the function values and we thus drive our attention towards the $f$-greedy schemes.

\subsubsection{\textit{f-greedy} algorithm}\label{f-greedy}
The {\textit{p-greedy}} algorithm is designed to primarily capture the effect of the radial kernel and the location of the center points on the interpolation error, and this is expected to be helpful when the variation in the values of the interpolated function is relatively regular. In the context of RBF-based NIROMs, the underlying function is the time derivative of the projection coefficients. Depending on the characteristics of the shallow water flow problem being studied, the solution snapshots and, in turn, the projection coefficients can have highly nonlinear temporal evolution patterns. In order to capture the trend in the time derivative function, we consider a residual-based {\textit{f-greedy}} algorithm.

The idea of iteration on residuals leading to the scalar {\textit{f-greedy}} algorithm, was introduced by Schaback and Wendland in \cite{SW2000}, where they also provided a convergence proof using the orthogonality relation,

\begin{align}\label{eq:orthogonality}
  \left(F, f-F\right)_{\Phi} = 0, \quad \forall f \in \mathcal{N}_{\Phi}.
\end{align}

The orthogonality property is a direct consequence of the fact that the RBF interpolant $F$ has the minimal norm among all functions in the native space $\mathcal{N}_\Phi$ that interpolate $f$ on $X$. The orthogonality property allows the following Pythagorean splitting of the ``energy'' of the function $f$ using the native norm

\begin{align}\label{eq:pythagorean}
  \nm{f}^2_\Phi = \nm{f-F}_\Phi^2 + \nm{F}_\Phi^2.
\end{align}
A recursive application of (\ref{eq:pythagorean}) easily shows that $\lim_{k \rightarrow \infty} \nm{f-F^k}_\Phi = 0$, where $F^k$ is the $k$-term interpolant, and this forms the key ingredient of the scalar \textit{f-greedy} algorithm. In the $(k+1)^{th}$ iteration, the absolute value of the $k^{th}$ function residual \ie the difference between the data value $f$ and the $k$-term interpolant function $F^k$, is computed for each center remaining in the pool of unused centers $X \setminus X_k $. Then, the center ${\vec z_{k+1}}$ at which the maximum error of the $k^{th}$ function residual occurs, is added to the existing set of centers \ie $X_{k+1} = X_k \cup \{\vec z_{k+1} \}$.

In the context of our interpolation problem, the function residual is vector valued, \ie $\abs*{\vec f- \vec{F}_k} \in \R^m$ where $m = m_1 + m_2 + m_3$ is the total number of POD modes for all the solution components. However, the interpolation conditions (\ref{eq:nirom-interp-cond}) are applied individually for each mode to generate independent modal interpolants. Hence, transferring the scalar \textit{f-greedy} algorithm directly to the vectorial case by a mode-wise application results in $m$ near-optimal sets of centers $X^j_k (j=1,\ldots ,m)$.
In the absence of a mechanism to enforce a coordinated choice over all $m$ modes the resulting sets of centers $X^j_k$ might only have very few points in common or even be pairwise disjoint. Then, assuming that the sets of centers have approximately the same number of elements (\ie $|X^j| \approx |X^i|, j\neq i$), the kernel has to be evaluated at roughly $m|X^j|$ different centers and the cost of evaluation in the online computation scales with a factor of $m$. Moreover, the offline computational cost also scales with a factor of $m$ in the worst case, which can lead to substantially longer times for large training sets of snapshots.

To reduce the computational burden, the \textit{f-greedy} approach is applied to approximate the $l^2$-norm of the function $g(\bz) = \nm*{\vec{f}(\bz)}_2$ by the interpolant $\tilde{F}(\bz)$, and the center at which the scalar function residual is maximum is selected in each greedy iteration. The procedure is summarized in Algorithm {\ref{alg:f-greedy}}.

\begin{algorithm}[H]
  \DontPrintSemicolon
  \SetAlgoLined
	\KwResult{ $\widetilde{X}$ : Optimal set of RBF centers }
  \SetKwInOut{Input}{Input}\SetKwInOut{Output}{Output}
  \Input{$X = \{\vec z_1,\vec z_2, \ldots, \vec z_{M}\}$, $M > N_{max}$ }
  \BlankLine
	Initial center: $\vec z_1 = \argmax\limits_{\vec z_i \in X} \{ \abs*{g(\vec z_i)} \equiv \nm*{\vec{f}(\bz_i)}_2\}$ \;
  
    \Do{$\abs*{\max\{\xi_{X_k}\} } > \tau_f $ \& $k < N_{max}$}
    {
      Compute $\xi_{X_k}(\vec z_{i}) = g(\vec z_{i}) -  \tilde{F}^{k}(\vec z_{i}), \quad \forall \vec z_i \in X\setminus X_k $ \;
      Set $\vec z_{k+1} = \argmax\limits_{\vec z_i \in X\setminus X_k} \{\abs*{\xi_{X_k}(\vec z_i)} \}$\;
      $X_{k+1} \leftarrow X_k \cup \{ \vec z_{k+1}\}$\;
      Re-compute RBF system: $\tilde{F}^{k+1}, \vec A^{k+1}$\;
    }

	\caption{\textit{f-greedy} algorithm for selecting optimal set of RBF centers}\label{alg:f-greedy}
\end{algorithm}

\subsubsection{\textit{psr-greedy} algorithm}\label{sec:psr-greedy}
In this work, a new strategy called the power-scaled greedy or \textit{psr-greedy} algorithm has been developed to construct NIROMs for problems where both the variation in data values as well as the dependence on the kernel functions are considered equally significant. 

A scalar \textit{fp-greedy} algorithm was proposed in \cite{SW2006}, where a ratio of the function residual and the power function was adopted as the selection metric, thus minimizing the native RKHS norm $\nm{f - F^k}_\Phi $ in each greedy step. Essentially this scalar variant weighs the residual error at the center being examined with how effectively the considered center is already contained in the current set of centers. A direct application of the scalar algorithm to the vectorial setting suffers from the same computational disadvantages that were discussed in the context of the \textit{f-greedy} algorithm. A vectorized strategy named the Vectorial Kernel Orthogonal Greedy algorithm (VKOGA, \cite{WH2013}) was suggested which attempts to extend the scalar orthogonality property of RBF interpolants (see (\ref{eq:orthogonality})) by introducing the concept of a vectorial gain function. Then a greedy maximization of the gain function leads to an iterative addition of the largest possible ``gain'' in approximation with respect to the native space norm.

The novelty of the \textit{psr-greedy} algorithm proposed here is the use of the product of the power function and the function residual in order to greedily select the worst approximated centers. Thus it aims to simultaneously eliminate the error introduced by both the kernel as well as the function variations. 
In order to efficiently approximate the vector function residual, an alternative strategy is proposed to identify a subset of modes ($\ll m $) corresponding to each solution component for which a unified set of centers $X$ can be determined. First, the ``energy'' content of individual modes is computed as

\begin{align}\label{eq:greedy-modes-energy}
    \hat{e}_j^i = \sum_{k=0}^{M-1} \lrp{\left. \ddt{z_{k,j}}\right|_i}^2,
\end{align}
where $z_{k,j}\restriction_i$ is the $j^{th}$ mode of the $k^{th}$ center, $\bz_k = [z_{k,0},z_{k,1}, \ldots, z_{k,m}]^T$ for the $i^{th}$ solution component. The modes are arranged in the descending order of energy content and then, the set of the most significant modes for each component, $L^i = \{j \,| \, j \leq m_i^j \ll m_i\}$ is identified by selecting a desired fraction $\tau_{greedy}$ such that

\begin{align}\label{eq:greedy-modes-energy-selection}
\lrp{\sum_{j=0}^{j=m_i^j} \hat{e}^i_j} / \lrp{\sum_{j=0}^{j=m_i} \hat{e}^i_j} \leq \tau_{greedy}. 
\end{align}

A combined set of modes $L = L^1 \cup L^2 \cup L^3$ is defined by sequentially adding the modes from each subset $L_i$ and generating a continuous linear ordering of the modes. The unified set $L$ is used to iteratively apply the \textit{psr-greedy} algorithm such that, for a particular mode in each greedy iteration, the center with the maximum value of the scalar product of the residual error and the power function is chosen to enrich the existing set of centers. This process continues until the absolute maximum of the scalar product becomes less than a chosen tolerance $\tau_{psr}$. The initial center is defined by $\vec z_1 = \argmax_{\vec z_i \in X} \{ \Phi(\vec z_i, \vec z_i)\, \abs{f_j(\vec z_i)} \}$ where $j$ denotes the index of the first mode in $L$, and $f_j(\vec z_i)$ denotes the corresponding scalar component of the data function evaluated at the center point, $\vec z_i$. The greedy iterative process is then repeated for every mode in the set of selected significant modes $L$ and the entire procedure is presented in Algorithm {\ref{alg:psr-greedy}}. 

\textbf{Remark 1.} It is worth mentioning that instead of taking the $l^2$-norm of the function residual, this modal approach is equivalent to computing a weighted norm of the function residual where the weights are determined by the relative magnitudes (or ``energy") of the function components.
The modal components of the data function have different scales of variation, and weighting them equally using the $l^2$ norm may translate to underestimating oscillations in the leading dominant modes. Also, even though in comparison to the $l^2$-norm approach, this modal approach requires the optimization of one additional hyperparameter, $\tau_{greedy}$, the gain in overall efficiency and approximation accuracy, as observed in our numerical experiments, makes it a superior choice.

\textbf{Remark 2.} The main motivation for studying the \textit{psr-greedy} scheme instead of the classical \textit{fp-greedy} comes from stability issues. Indeed, from eq. (\ref{eq:power-err-est}) it is evident that by minimizing the ratio of the scalar function residual to the power function, the \textit{fp-greedy} algorithm, in a way,  seeks to minimize the native space norm of the function $\nm*{f}_{\Phi}$. However, dividing any quantity by the power function, $P_X(\bx) \ll 1$, might be extremely unstable. Indeed, the power function vanishes at each center and therefore, for a point $\bx$ that is close to an already selected center we have $P_X(\bx)  \approx 0$. On the other hand, we observe that multiplying (\ref{eq:power-err-est}) by the power function and taking into account (\ref{eq:pow-bounds}), we have

\begin{align}\label{eq:fp-err-est}
    P_X(\bx) \abs*{f(\bx) - F(\bx)} \leq P^2_X(\bx) \nm{f}_\Phi \leq \nm{f}_\Phi,
\end{align} 
provided that $\Phi(\bx,\bx)=1$, which is true for the Mat\'{e}rn $C^0$ kernel (as well as for some other strictly positive definite kernels like the Gaussian, multiquadric, inverse quadratic etc.). Eq. (\ref{eq:fp-err-est}) points to the fact that similar to the classical \textit{fp-greedy} approach, the \textit{psr-greedy} algorithm also seeks to essentially minimize the native space norm of the function while avoiding potential computational instabilities, and thus provides an empirical motivation for the approach.

Moreover, from a computational standpoint, the centers selected using the \textit{f-greedy} algorithm may exhibit a strong clustering pattern. This means that where the function has high variations owing to nonlinearities, the \textit{f-greedy} may suffer from oversampling, whereas the centers where steep gradients may still be present but the nonlinearities are less dominant, may be under represented. The \textit{psr-greedy} is able to overcome this sub-optimal behavior by the application of the power function as a variable scaling factor and by using a suitable tolerance $\tau_{psr}$. The power function increases the relative importance of centers in under-sampled regions, while the tolerance factor ensures that all of the relevant nonlinear features are suitably represented. This may explain the redistribution of the clustering patterns using the \textit{psr-greedy} algorithm and also provides an intuitive understanding of the approach.

\begin{algorithm}[H]
  \DontPrintSemicolon
  \SetAlgoLined
	\KwResult{ $\widetilde{X}$ : Optimal set of RBF centers }
  \SetKwInOut{Input}{Input}\SetKwInOut{Output}{Output}
  \Input{$X = \{\vec z_1,\vec z_2, \ldots, \vec z_{M}\}$, $M > N_{max}$ \;
        $L = \{j \,|\,  j \leq m_i^j\}, \, |L| \ll m$ : List of selected significant modes }
  \BlankLine
	Initial center: $\vec z_1 = \argmax_{\vec z_i \in X} \{ \Phi(\vec z_i, \vec z_i)\, \abs*{f_j(\vec z_i)} : j \text{ is the first mode in } L\}$ \;
  \For{$j$ in $L$}
  {
    \Do{$\abs*{\max\{\xi^j_{X_k}\} } > \tau_{psr} $ \& $k < N_{max}$}
    {
      Compute $\xi^j_{X_k}(\vec z_{i}) = P_{X_k}(\vec z_i) \, \abs*{f_j(\vec z_{i}) - F^k_j(\vec z_{i})}, \quad \forall \vec z_i \in X\setminus X_k $ \;
      Set $\vec z_{k+1} = \argmax\limits_{\vec z_i \in X\setminus X_k} \{\xi^j_{X_k}(\vec z_i) \}$\;
      $X_{k+1} \leftarrow X_k \cup \{ \vec z_{k+1}\}$\;
      Re-compute RBF system: $\vec F^{k+1}, \vec A^{k+1}$\;
    }
  }
	\caption{\textit{psr-greedy} algorithm for selecting optimal set of RBF centers}\label{alg:psr-greedy}
\end{algorithm}

The straightforward implementation of the greedy algorithms using the standard basis of translates $\{\Phi(\vec z_1,\cdot),\ldots,\Phi(\vec z_M, \cdot)\}$ or using the Lagrange basis functions $\eta_j, j = 1,\ldots, M$ with the property $\eta_j(\vec z_i) = \delta_{ij}$, may pose severe computational challenges. This is because every greedy iteration involves potentially ill-conditioned kernel matrices and requires recomputing the entire set of RBF coefficients. In this work, we have adopted the fairly general Newton basis approach formulated in \citep{PS2011} for the \textit{p-greedy} iterations (we refer the reader to the Appendix for further details on the topic). This approach was extended to vector-valued functions for VKOGA in \citep{WHM2013}, and has been applied here, with suitable modifications, for the modal residual computations required in the \textit{psr-greedy} iterations. 

\subsection{Combined RBF-POD reduced order model}\label{sec:algorithm}
The key steps in the RBF-POD NIROM procedure are outlined below.
\begin{enumerate}[label=(\roman*)]
    \item The high-fidelity snapshots $\bv_i^1, \ldots \bv_i^M$ for each component $u_i$ are obtained by solving eq.~(\ref{eq:discrete-prim-residual}), then normalized and stored in $\vec S_i$.
    \item The truncated set of POD basis vectors $\vec \Theta_i$ are obtained by performing a SVD of $\vec S_i$ as given by eq.~(\ref{eq:thin_SVD}).
    \item The normalized high-fidelity snapshots are projected onto the space spanned by the POD basis to obtain the corresponding projection coefficients $\bz_i^1, \ldots \bz_i^M$ as defined by eq.~(\ref{eq:pod-basis}).
    \item A subset of the projection coefficients is selected using an appropriate greedy algorithm (Algorithms \ref{alg:p-greedy}, \ref{alg:f-greedy} or \ref{alg:psr-greedy}) to define a near-optimal set of centers $\widetilde{X}$ for the RBF interpolant.
    \item The RBF interpolants $F_j(\bz)$ for the modal time derivatives of the projection coefficients $\ddt{\bz_j}$ are obtained by solving the linear system of interpolation conditions, given by eq.~(\ref{eq:interp-cond}), on the set of near-optimal centers $\widetilde{X}$.
    \item These interpolants $F_j(\bz)$ are used in eq.~(\ref{eq:RBF-time-deriv}), to advance the reduced solution in time for any new configuration queried by the application. 
\end{enumerate}

%%%---------------------------------------------------------------
%%%---------------------------------------------------------------
\section{Numerical results}\label{sec:results}

Numerical tests have been conducted with different types of shallow water flow problems using AdH \cite{Trahan2018} as the high-fidelity numerical solver. To compare the performance of the POD-RBF NIROM with
the nonlinear POD (NPOD) model, a riverine flow problem with moderately large degrees of freedom has been studied numerically, as discussed in
Section \ref{sec:results1}. In Section \ref{sec:results2}, we consider two
different flow regimes to evaluate the performance of the greedy
algorithms. The first example represents tidal flow conditions in an urbanized bay, while the second example considers a different riverine flow with variable inflow boundary conditions. The bathymetry data in all of the examples were obtained from USACE Hydrographic surveys. All the bed topography measurements
and water surface elevations were referenced with respect to the
bathymetry data, following the NAVD88 convention.

In all of the numerical experiments, first a sufficiently well-resolved high-fidelity simulation was conducted. For the purpose of demonstrating the efficiency of the model reduction methodology, a set of temporal snapshots was selected as the offline training data by uniformly skipping over a fixed number of available high-fidelity snapshots. These offline snapshots were used to construct the reduced order models, which were then employed in the online stage to evaluate the numerical solution for different time step sizes. The full set of high-fidelity snapshots was used to compare the accuracy of these reduced order solutions. In addition to comparing solution profiles, the approximation of the temporal dynamics was analyzed by comparing the root mean square error (RMSE), which is calculated separately for each component $j$ and each time step $n$ as

\begin{align}\label{rms}
    RMSE^n_j =  \sqrt{\dfrac{\sum\limits_{i=1}^N \abs*{y_{f,j,i}^n - y_{a,j,i}^n}^2}{N}}.
\end{align}
In the above definition, $y_{f,j,i}^n$ denotes the high-fidelity solution at the spatial node $i$ for component $j \in \{1,2,3\}$, $y_{a,j,i}^n$ denotes the corresponding full order solution at the node $i$, reconstructed from the ROM solution (NIROM or NPOD) by projecting back onto the high-fidelity computational mesh, and $N$ represents the number of spatial nodes in the high-dimensional mesh. 
For ease of interpretation, we focus on the primitive variables, $\vec h$, $\vec u_x$ and $\vec u_y$. That is, $y_{f,1,i}=h_{f,i}$, $y_{f,2,i}=(u_x)_{f,i}$, and $y_{f,3,i}=(u_y)_{f,i}$. 

Additionally, to compare the convergence behavior of the greedy NIROM strategies, a global measure of space-time approximation error is introduced as the space-time root mean square error (stRMSE). For any solution component, given a particular configuration of the NIROM, the stRMSE is computed as

\begin{align}\label{strms}
    stRMSE_j =  \sqrt{\dfrac{\sum\limits_{n=1}^M\sum\limits_{i=1}^N \abs*{y_{f,j,i}^n - y_{a,j,i}^n}^2}{NM}},
\end{align}
where $M$ represents the total number of time steps in the online stage.

\subsection{Comparison between NIROM and NPOD}\label{sec:results1}
\begin{omitext}
%%%-------------------------------------------------------------
\subsubsection{San Diego Bay Example}
The first numerical example involves the simulation of tide-driven flow in the San Diego Bay in California, USA (see Figure \ref{san_diego_domain}).

    \begin{figure}[htb]%{l}{0.45\textwidth}
      \begin{center}
        \includegraphics[width=0.55\columnwidth,trim=0cm 0cm .4cm .4cm, clip]{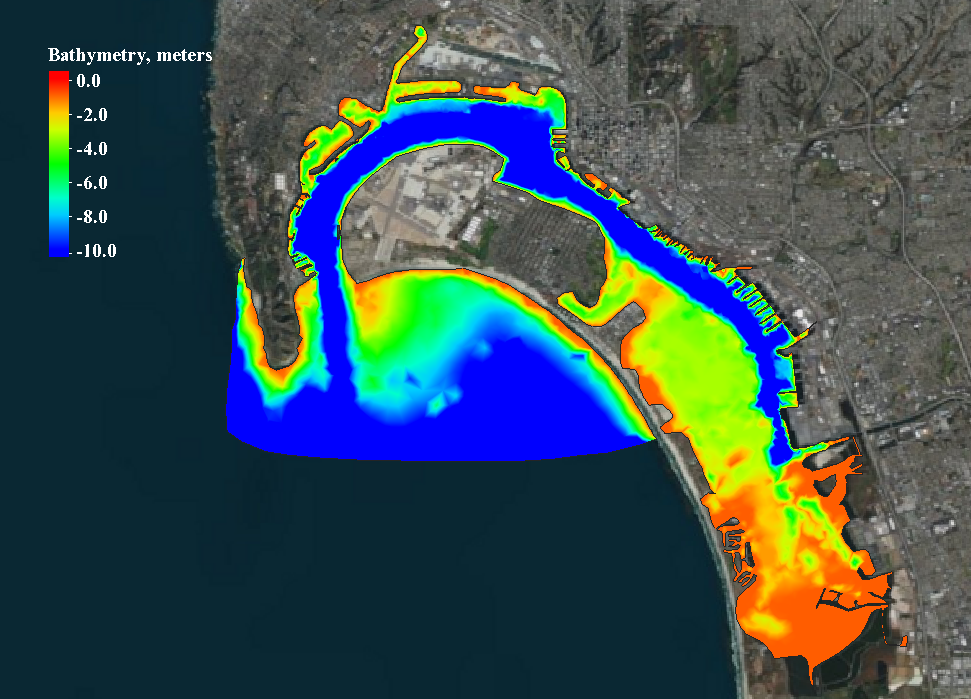}
      \end{center}
      \caption{San Diego Bay}\label{san_diego_domain}
    \end{figure}
A uniform structured triangular mesh consisting of $N = 6311$ nodes and $10999$ elements was used. The simulation was carried out until $T=50$ hours with a time step of $\Delta t = 25$ seconds. Tidal flow data was obtained from NOAA/NOS Co-Ops website at an interval of $6$ minutes. This was applied as a time series of tailwater elevation boundary data along the east-west boundary of the computational mesh at the entrance to the inner harbor, as shown in Figure \ref{san_diego_domain}. No flow boundary conditions were applied along the walls of the harbor and an initial water surface elevation for the entire domain was provided.  Moreover, this example involves viscosity $\nu = 10^{-5}$, friction $n_{mn} = 0.022$, gravity $g=9.81$, and a Coriolis latitude of $32.7$, that estimates the effect of the Coriolis force due to the Earth's rotation.

In the offline stage, a total of $7200$ high-fidelity solution snapshots were collected. A subset of $M=721$ snapshots was created by skipping over every $10$ consecutive snapshots and this subset was used to generate the optimal POD space and subsequently the first $720$ projected snapshots were used as center points to construct the RBF interpolant. The POD modes were truncated to the first $473,593,$ and $598$ modes for the primitive variables $u_1,u_2$, and $u_3$ respectively by selecting an error tolerance of $\tau = 10^{-6}$. The online simulations using the NIROM and NPOD reduced models were done with a time step of $\Delta t = 125$ and the corresponding high-fidelity snapshots were used for the error analysis.
    \begin{wraptable}[7]{R}{0.45\textwidth}
      \centering
        \begin{tabular}{c c c}
		\toprule
		$t_{HFM}$ & $t_{NIROM}$ & $t_{NPOD}$ \\ \midrule
		$21m36s$ & $6m15s$ & $5h51m23s$ \\
		& $(3.45\times)$ & $(0.06\times)$\\ \bottomrule
	    \end{tabular}
	    \caption{Computational times for the San Diego example}\label{tab:san_diego}
	\end{wraptable}

Figure \ref{fig:sandiego_cont} shows the x-velocity solutions calculated using the RBF NIROM (top row, left column) and the NPOD (bottom row, left column) at time $t=2.55\times10^4$, and the corresponding errors (right column) with respect to the high-fidelity solution. The discrete $l^2$ norm for the relative error has been computed for both the reduced models and the NPOD can be seen to perform marginally better than the NIROM. However, the computational gain for the NIROM is far superior than the NPOD model as shown in Table \ref{tab:san_diego}.

\begin{figure}[htb]
  \begin{minipage}{\textwidth}
  \centering
  \renewcommand{\tabcolsep}{0.01cm}
  \begin{tabular}{cc}  %trim=left bottom right top,
  % \footnotesize{x-velocity}&\footnotesize{error}\\
  \includegraphics[width=2.9in,trim=1cm 1cm 0cm 0cm, clip]{figures/san_diego_nirom_rbf_t204_u.pdf}&
  \includegraphics[width=2.9in,trim=1cm 1cm 0cm 0cm, clip]{figures/san_diego_nirom_err_t204_u.pdf}\\
  \includegraphics[width=2.9in,trim=1cm 1cm 0cm 0cm, clip]{figures/san_diego_npod_t-204_u.pdf}&
  \includegraphics[width=2.9in,trim=1cm 1cm 0cm 0cm, clip]{figures/san_diego_npod_err_t-204_u.pdf}\\
  \end{tabular}
  \medskip
  \caption{Solutions for the x-velocity (left column) and the corresponding errors (right column) with respect to the true solution obtained using the NIROM (top row) and the NPOD (bottom row) model at $t=2.55\times10^4$.}
  \label{fig:sandiego_cont}
  \end{minipage}
\end{figure}

Figure \ref{fig:sd_wrms} shows the weighted RMSE with time for each of the three primitive variables computed using the NPOD model and the NIROM. The local extrema in the weighted RMSE plots for the NIROM solutions are associated with the elevated and depleted flow conditions during the high and the low tide levels respectively. Figure \ref{fig:sd_modal} shows the time evolution of the time derivative of the high-fidelity solutions projected onto the POD modal space, and this function is being interpolated by the RBF interpolant. The time derivative function of the x- and y-velocity variables can be seen to coincide with the RMSE extrema locations at the high and low tide points. The extrema of the projected time derivative of depth variable are fewer in number. This explains why fewer number of POD modes ($m_1 = 473$) can acurately capture the temporal dynamics for the depth variable than that needed for the x- and y-velocity variables ($m_2 = 593$ and $m_3 = 598$ respectively). A weighted RMSE error is computed to partially normalize the effect,

\begin{align}\label{wrms}
    wRMSE^n_j =  \sqrt{\dfrac{\sum_{i=1}^N(u_{f,j,i}^n - u_{app,j,i}^n)^2}{N \sum_{i=1}^N(u_{f,j,i}^n + \text{offset})^2}},
\end{align}
where an offset = $0.0005$ is used for the velocity variables and a zero offset is used for the depth.

\begin{figure}[!ht]
     \subfloat[RMSE comparison between the NIROM and the NPOD solutions\label{fig:sd_wrms}]{%
       \includegraphics[width=0.45\columnwidth]{figures/san_diego_npod_nirom_wrms_err.pdf}
     }
     \hfill
     \subfloat[Temporal evolution of the $l^2$ norm of the modal time derivative\label{fig:sd_modal}]{%
       \includegraphics[width=0.45\columnwidth]{figures/san_diego_modal_behavior.pdf}
     }
     \caption{Error analysis for the San Diego Bay example}
     \label{fig:sd_error}
\end{figure}

In spite of the above computational artifacts, it can be seen that the approximation accuracy of the NIROM solution is comparable to the significantly more expensive NPOD solution.

\end{omitext}

%%%-------------------------------------------------------------
% \subsubsection{Kissimmee River Example}

\begin{figure}[htb] %{l}{0.65\textwidth}
    \centering
     \subfloat[Bathymetry and flow control structures\label{fig:kiss_bathy}]{%
       \includegraphics[width=0.5\columnwidth,trim=0cm 0cm 0cm 0cm, clip]{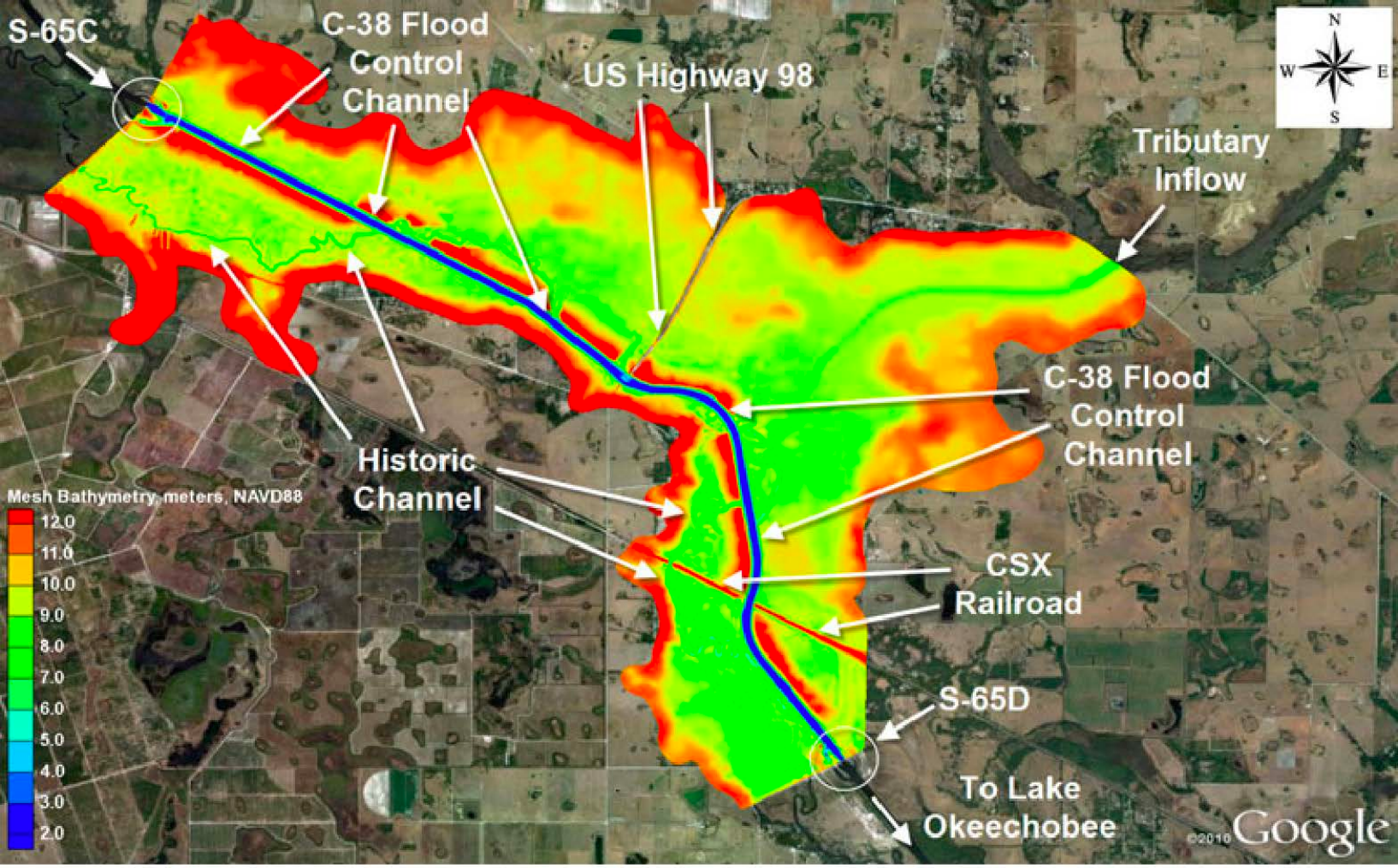}
     }
     \subfloat[Computational mesh\label{fig:kiss_mesh}]{%
       \includegraphics[width=0.49\columnwidth,trim=2cm 6cm 2cm 2cm, clip]{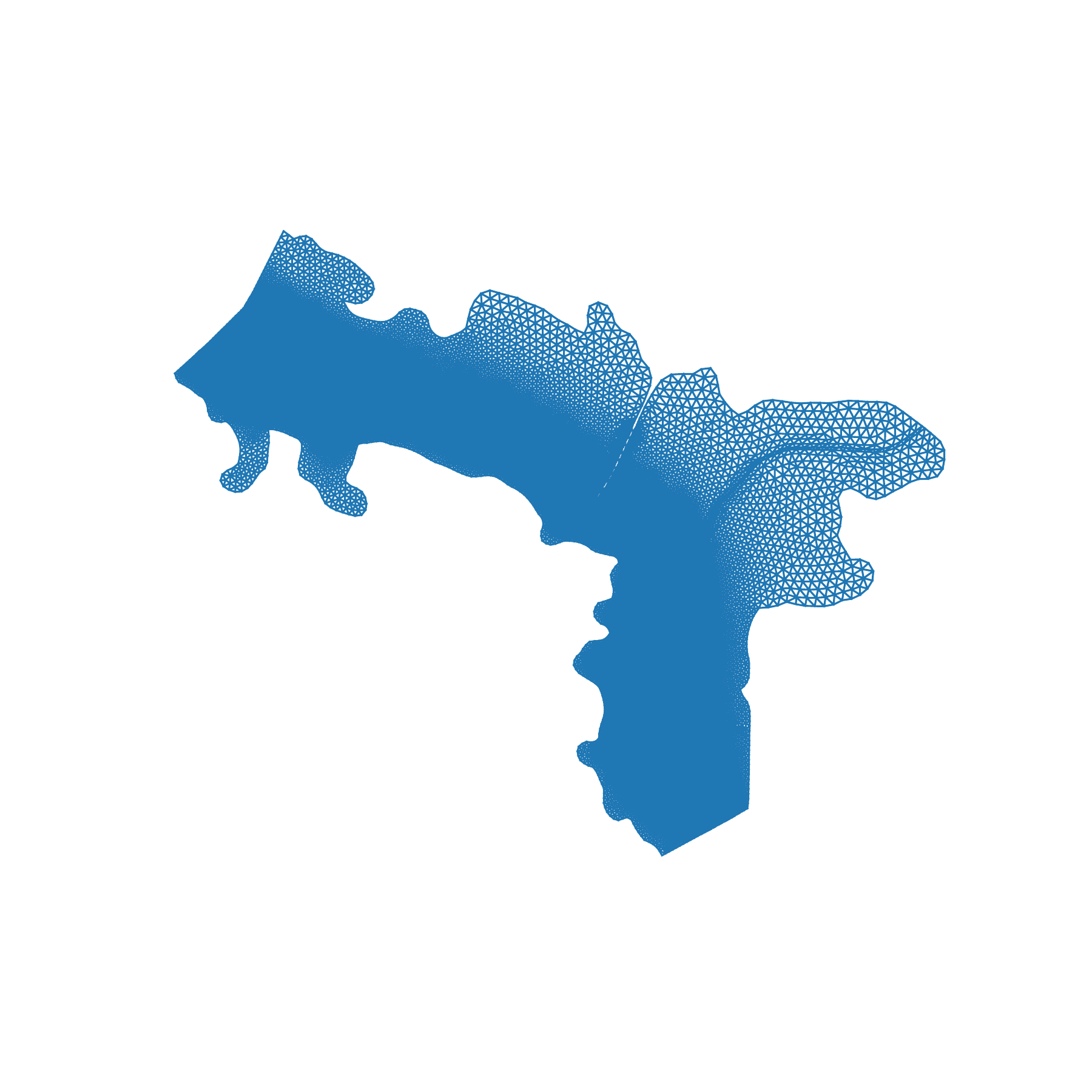}
     }
    \caption{Kissimmee River}\label{kiss_domain}
\end{figure}

The first numerical example involves the simulation of the hydrodynamic conditions along a stretch of the Kissimmee River near Fort Basinger in Florida, USA (see Fig. \ref{kiss_domain}). The numerical model was obtained from the study \cite{MSSS2013} conducted to assess the expected inundation amounts associated with two different habitat restoration alternatives. In our simulation of the existing conditions of the basin, the flow was primarily confined to the C-38 flood control drainage canal that is bounded upstream and downstream by water control structures of S-65C and S-65D, respectively (see Fig. \ref{fig:kiss_bathy}). A high-fidelity hydrodynamic modeling of the river and the surrounding floodplain involved accurate representation of the bathymetry, multiple types of friction specification for the flood control channel and the overland vegetative floodplain, as well as accurate measurement of the discharge through the control structures. Additionally, accurate modeling of the tributary inflow, hydrodynamic conditions around the bridges (U.S. Highway 98 and CSX Railroad) as well as the floodplain discharges around $10$ existing culverts along the U.S. Highway 98 causeway were necessary to determine the full range of impacts associated with the existing conditions. Details can be found in McAlpin \etal \cite{MSSS2013}.
% \newlength{\oldintextsep}
% \setlength{\oldintextsep}{\intextsep}
% \setlength\intextsep{2pt}
% \setlength{\intextsep}{0pt}

A uniform unstructured triangular mesh consisting of $N = 108816$ nodes and $216676$ elements was used. A natural inflow boundary condition was specified at S-65C along with an outflow tailwater elevation at S-65D, and another natural velocity inflow condition was specified to simulate the tributary inflow.
No flow boundary conditions were applied along the rest of the river banks and an initial water surface elevation for the entire domain was provided. Moreover, this example involves viscosity $\nu = 1.306\times 10^{-5}$, Manning's roughness coefficient $n_{mn} = 0.03$ for the channelized areas, a submerged vegetation specification for overbank areas, and gravity $g=9.81$.

\begin{wrapfigure}{R}{0.5\textwidth}
\centering
\includegraphics[width=0.5\textwidth]{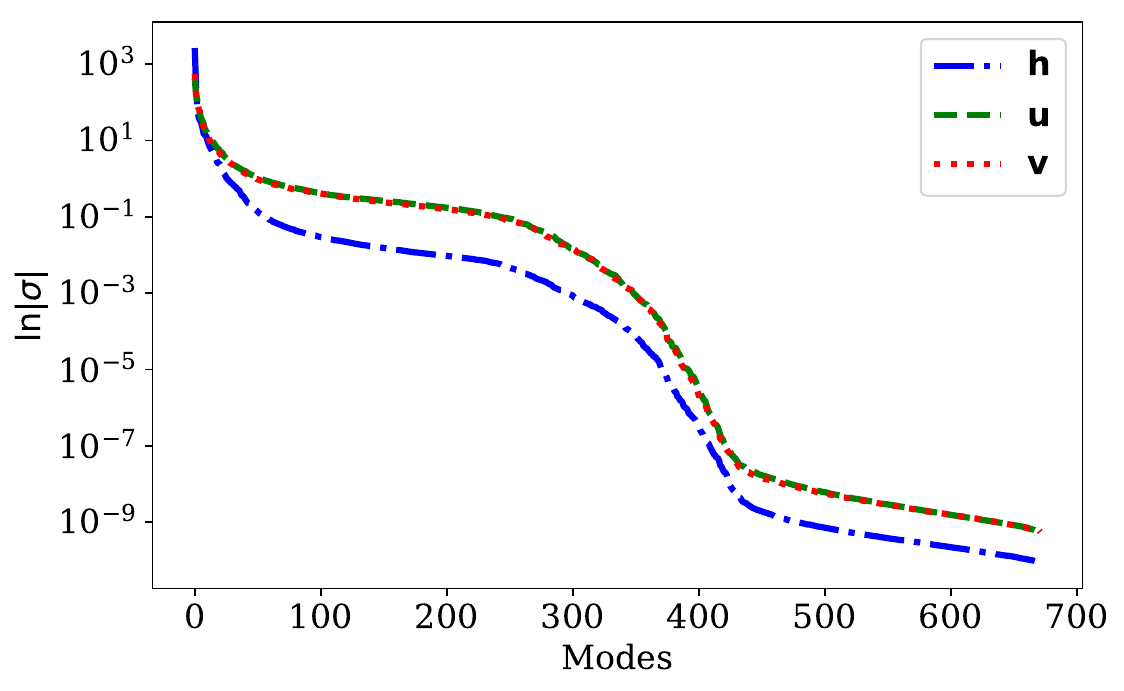}
\caption{Semi-logarithmic plot of the singular values for the Kissimmee River example}\label{fig:kiss_sing_val}
\end{wrapfigure}

The simulation was carried out until $T=4.5\times10^5$ seconds ($\approx 5$ days, $5$ hours) with a linearly increasing time step ranging from $\Delta t = 5$ seconds to $\Delta t = 500$ seconds.
In the offline stage, $2019$ high-fidelity solution snapshots were collected. 
A subset of $M=673$ snapshots, selected by uniformly skipping over every $3$ consecutive snapshots, was used as the training set to generate the optimal POD basis and to construct the RBF interpolant. The POD modes were truncated to the first $9,61,$ and $44$ modes for the primitive variables $\vec h, \vec u_x$, and $\vec u_y$ respectively by selecting an error tolerance of $\tau_{POD} = 10^{-4}$. Fig. \ref{fig:kiss_sing_val} shows the singular values for each solution component. The online simulations using the NIROM and NPOD reduced models were carried out with time steps identical to the high-fidelity simulation. The NIROM solution was computed by constructing the RBF interpolant using all the available centers, and without employing any of the greedy strategies.
  \begin{wraptable}[8]{R}{0.48\textwidth}
    \centering
      \begin{tabular}{c c c}
    		\toprule
    		$t_{HFM}$ & $t_{NIROM}$ & $t_{NPOD}$ \\ \midrule
    		$02:24:46$ & $00:02:34$ & $16:58:49$ \\
    		& $(56.40\times)$ & $(0.14\times)$\\ \bottomrule
	    \end{tabular}
	    \caption{Computational times (hh:mm:ss) for the Kissimmee River example }\label{tab:kiss}
	\end{wraptable}

\begin{figure}[ht]
    \centering
    \includegraphics[width=0.94\textwidth]{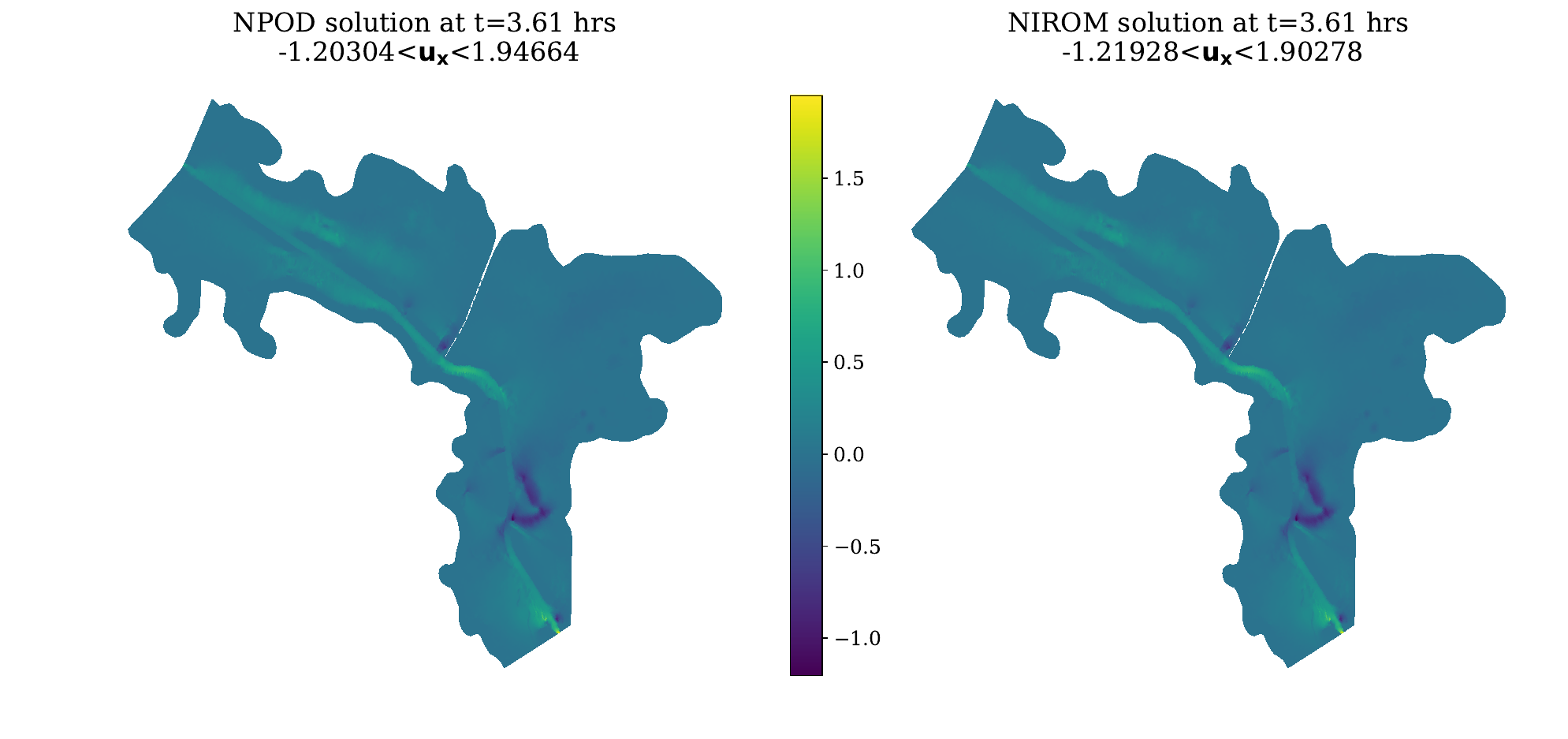}
    \caption{NPOD (left) and NIROM (right) solutions for the x-velocity variable at $t=1.2995\times10^4$ seconds $\approx 3.61$ hours.}
    \label{fig:kiss_nirom_npod_sol}
\end{figure}

\begin{figure}[H]
    \centering
    \includegraphics[width=0.94\textwidth]{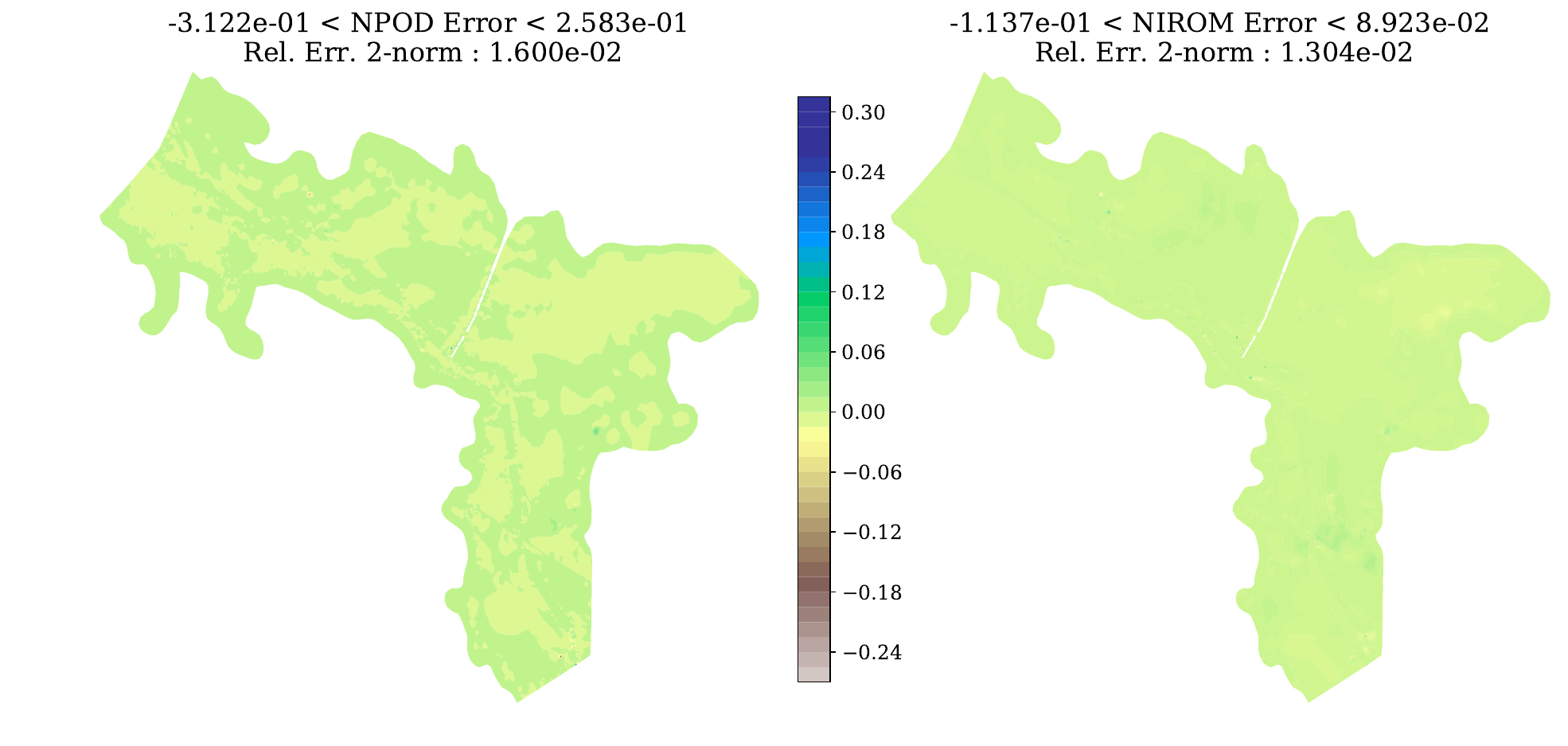}
    \caption{NPOD (left) and NIROM (right) solution errors for the x-velocity variable with respect to the true solution at $t=1.2995\times10^4$ seconds $\approx 3.6$ hours.}
    \label{fig:kiss_nirom_npod_err}
\end{figure}

Table \ref{tab:kiss} shows the CPU time for simulating the first $24$ hours of the original problem using the high-fidelity model as well as the online evaluation CPU times of the NIROM and NPOD models. The extended CPU time for the NPOD model can be attributed to the inefficiency introduced by the evaluation of the nonlinear terms in the reduced system of equations over the high-fidelity computational mesh in each time step. This is the reason why the NPOD model, in practice, is always combined with a suitable hyper-reduction scheme for improved computational efficiency. Figures \ref{fig:kiss_nirom_npod_sol} and \ref{fig:kiss_nirom_npod_err} show the x-velocity solutions and the corresponding spatial distribution of errors computed using the NPOD and the RBF NIROM methods, respectively, at time $t=3.61$ hours. The relative error $2$-norms reported in Figure \ref{fig:kiss_nirom_npod_err} were computed as the ratio of the $l^2$ norm of the absolute error to the $l^2$ norm of the true solution. The NIROM solution error can be seen to be marginally lower than the NPOD solution error, while the NIROM computation is about 56 times faster than the corresponding high-fidelity simulation.

\begin{figure}[!ht]
     \subfloat[RMSE for the x- and y-velocity solutions\label{fig:kiss_rms_uv}]{%
       \includegraphics[width=0.48\columnwidth]{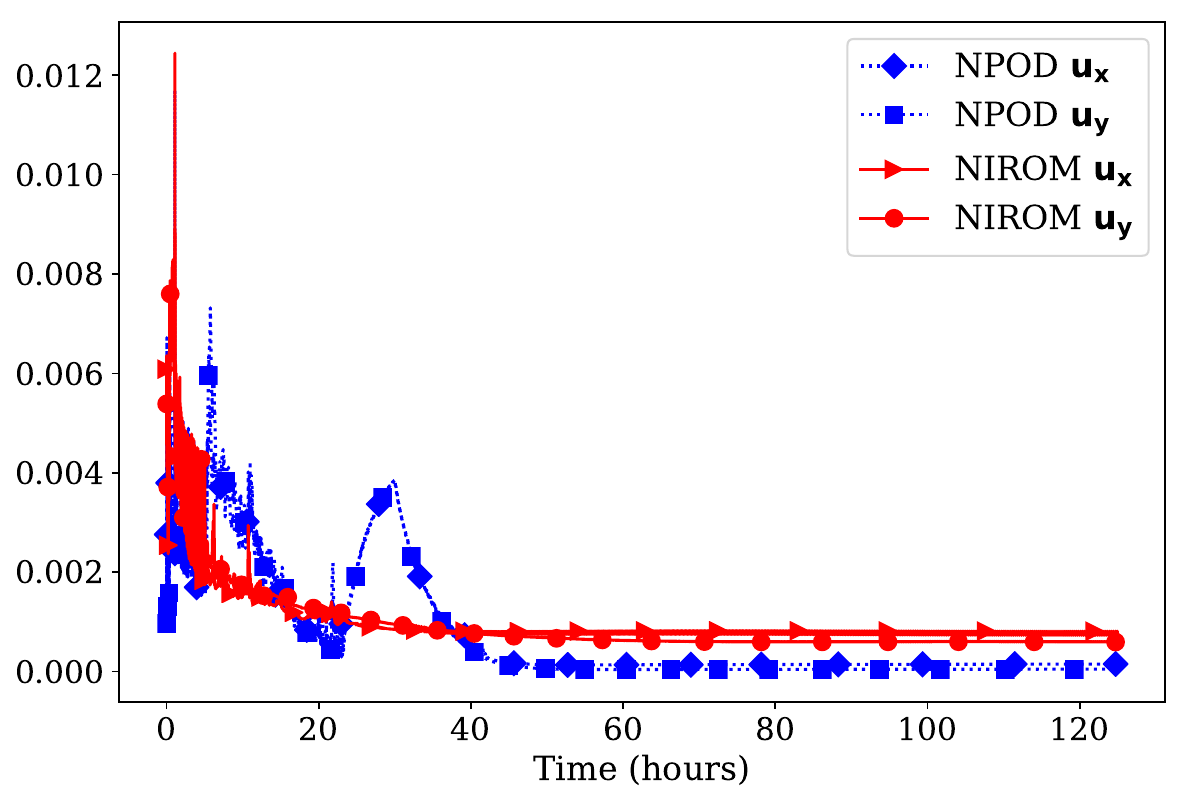}
     }
     \hfill
     \subfloat[RMSE for the depth solutions\label{fig:kiss_rms_h}]{%
       \includegraphics[width=0.48\columnwidth]{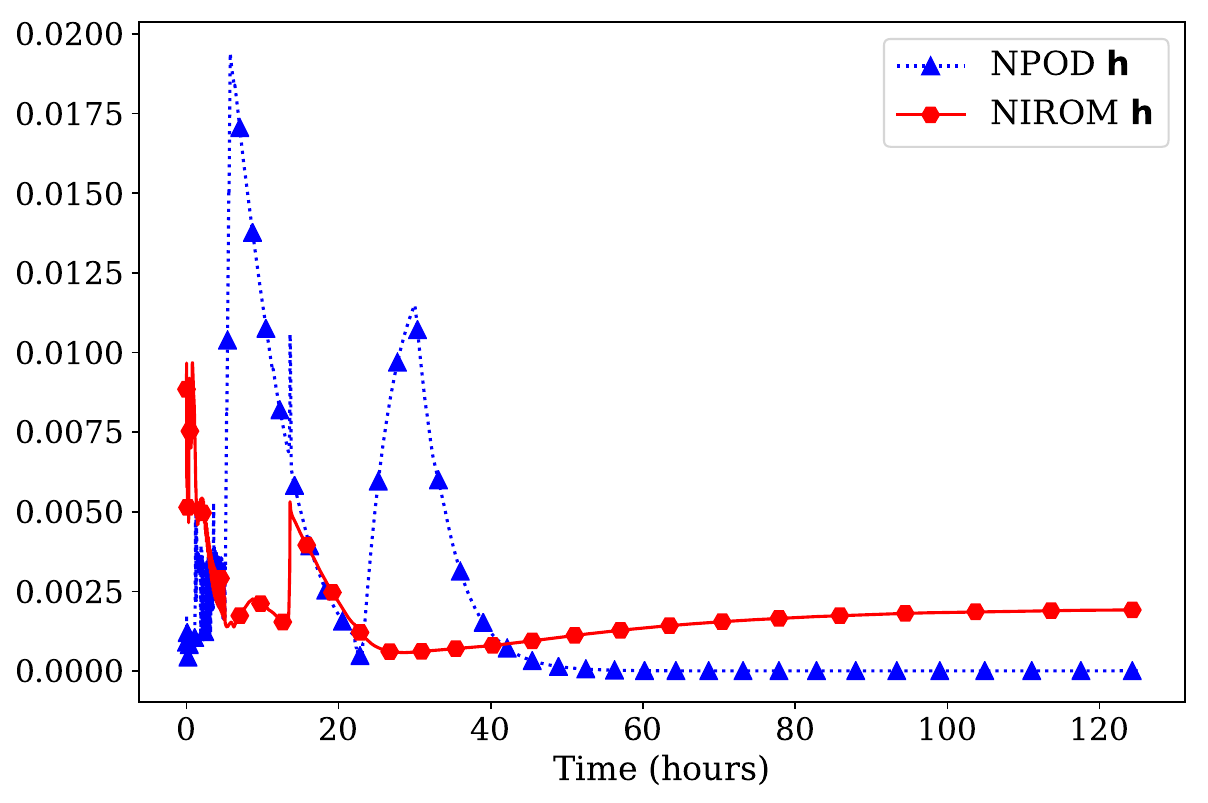}
     }
     \caption{$RMSE$ of the NIROM and NPOD reduced solutions for the Kissimmee River example}
     \label{fig:kiss_error}
\end{figure}

Figure \ref{fig:kiss_rms_uv} shows the RMSE with time for the velocity variables, $\vec u_x$ and $\vec u_y$ computed using both the NPOD and NIROM solutions while Figure \ref{fig:kiss_rms_h} shows a similar comparison for the depth variable. The temporal dynamics are non-trivial primarily at the beginning of the simulation before the flow settles down to a steady state. Hence, the local extrema in the RMSE curves for both the reduced order solutions occur at the early stages of the flow.
From the spatial distribution of the error (Figure \ref{fig:kiss_nirom_npod_err}) and temporal evolution of the RMSE (Figure \ref{fig:kiss_error}), it can be inferred that the NIROM produces a robust and accurate approximation. 

\subsection{Performance of the greedy algorithms}\label{sec:results2}
In this section, numerical results are presented to comparatively evaluate the performance of the three different greedy strategies, described in Section \ref{sec:greedy}.

%%%-------------------------------------------------------------
\subsubsection{Red River example}

    \begin{figure}[htb]
    \centering
     \subfloat[Bathymetry\label{fig:red_bathy}]{%
       \includegraphics[width=0.55\columnwidth,trim=0cm 0cm .4cm 0cm, clip]{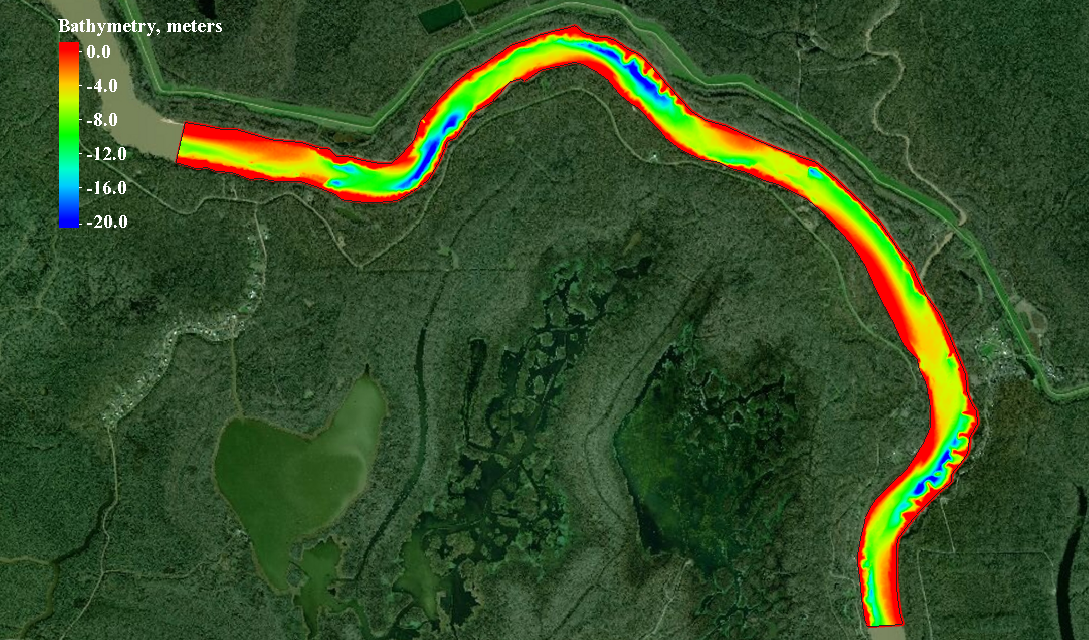}
     }
     \subfloat[Computational mesh\label{fig:red_mesh}]{%
       \includegraphics[width=0.45\columnwidth,trim=3cm 6cm 3cm 6cm, clip]{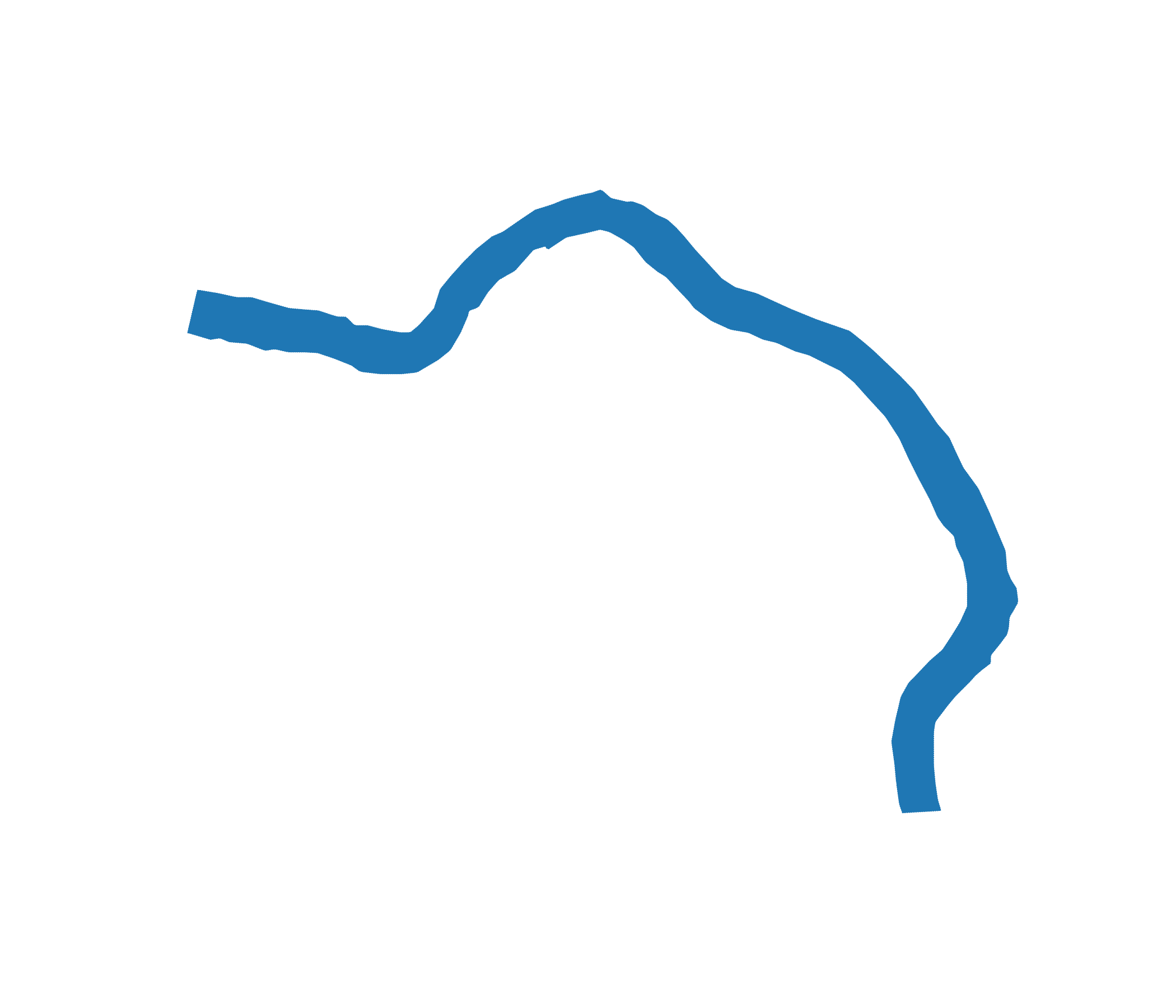}
     }
   \caption{Red River}\label{red_domain}
    \end{figure}

This example is an application of the 2D SWE to simulate riverine flow in a section of the Red River in Louisiana, USA (see Fig. \ref{red_domain}).
A uniform unstructured triangular mesh consisting of $N = 12291$ spatial nodes and $23316$ elements was used. 
%The simulation was carried out until $T=9$ hours with a time step of $\Delta t = 10$ seconds. 
A natural inflow velocity condition was specified upstream and a tailwater elevation boundary was specified downstream using hydrograph and bathymetry measurements obtained from USGS and USACE sources. No flow boundary conditions were applied along the banks of the river and an initial water surface elevation was provided for the entire fluid flow domain. Fluid viscosity $\nu = 1.139\times 10^{-6}$, and Manning's drag coefficient $n_{mn} = 0.025$ were also specified. 

\begin{wrapfigure}{R}{0.55\textwidth}
\centering
\includegraphics[width=0.54\textwidth]{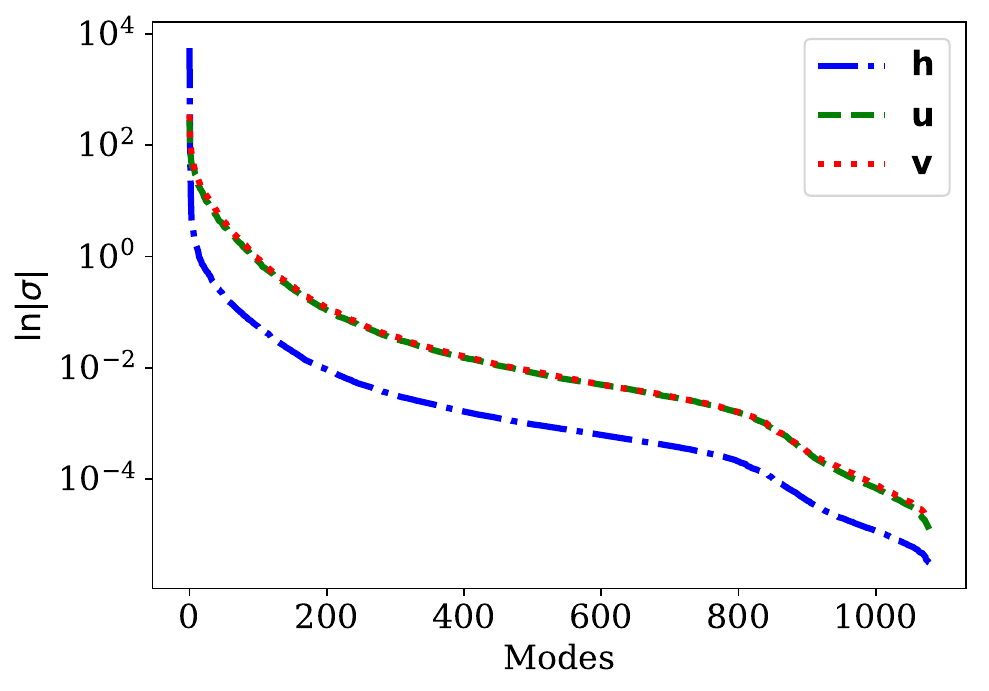}
\caption{Semi-logarithmic plot of the singular values for the Red River example}\label{fig:red_sing_val}
\end{wrapfigure}

In the offline stage, the high-fidelity simulation was run until $T=9$ hours with a time step of $\Delta t = 10$ seconds. $3141$ high-fidelity solution snapshots were collected between $t \in [0.28, 9.0]$ hours, by dropping the first $100$ snapshots to avoid the model startup effects. To compute the POD basis, $1048$ uniformly spaced, well-resolved snapshots were selected by skipping over every $3$ consecutive high-fidelity snapshots. Using a POD truncation error of $\tau_{POD} = 5 \times 10^{-6}$, only the first $3,191,$ and $179$ POD modes were selected to construct the reduced basis space for the $\vec h,\vec u_x$, and $\vec u_y$ components respectively. Figure \ref{fig:red_sing_val} shows the decay of the singular values for each solution component with respect to the number of POD modes in a semi-logarithmic scale. The online simulations using the different NIROM strategies were carried out until $t=9$ hours, with a time step of $\Delta t = 20$ seconds.

\begin{figure}[ht]
\centering
  {\includegraphics[width=0.95\columnwidth]{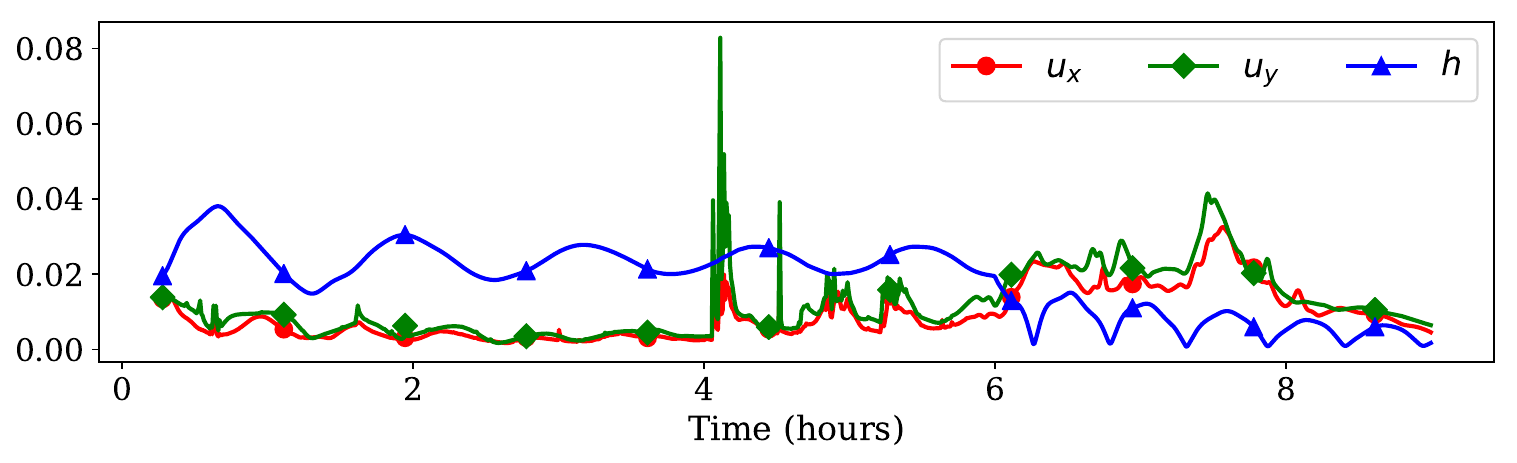}
   }
   \caption{Temporal evolution of the modal $l^2$-norm of the time derivative of the projection coefficients $\nm*{\frac{dz}{dt}(\cdot,t)}_{l^2} $ for the Red River example}
   \label{fig:red_modal_l2_norm}
\end{figure}

The $1048$ training snapshots were projected onto the POD space to generate the set $X$ of all the potential centers or data points available for the RBF interpolant. The RBF scale factor, $c$, was chosen to be $0.05$ by estimating the fill distance \cite{FZ2007} of the set of centers $X$. The \textit{p-greedy} and \textit{f-greedy} center points were selected using tolerances $\tau_p = 0.19$ and $\tau_f = 1.4\times 10^{-4}$. 
The list of modes $L$ required for the \textit{psr-greedy} iterations was determined by the following analysis. Figure \ref{fig:red_modal_l2_norm} is a visualization of the temporal dynamics of the interpolated function for each of the solution components. This is captured by computing the modal $l^2$-norm of the time derivative function, $\sum_{k=1}^{m_i}\abs*{{dz^n_{i,k}}/{dt}}^2$, for each of the projected solution components $(i=1,2,3)$. The irregular peaks in the temporal dynamics represent non-trivial changes in the overall flow patterns, potentially induced by significant variations in the inflow and the discharge boundary conditions, or by the formation of helical flows around the meandering sections of the Red River flow domain. The primary objective of the greedy algorithms is to be able to efficiently capture these flow patterns with an optimal set of data points or centers.

\begin{figure}[ht!]
 \subfloat[Modal energy computed by eq.~(\ref{eq:greedy-modes-energy}) for $\vec u_x$\label{fig:red_mode_energy_u}]{%
  \includegraphics[width=0.45\columnwidth]{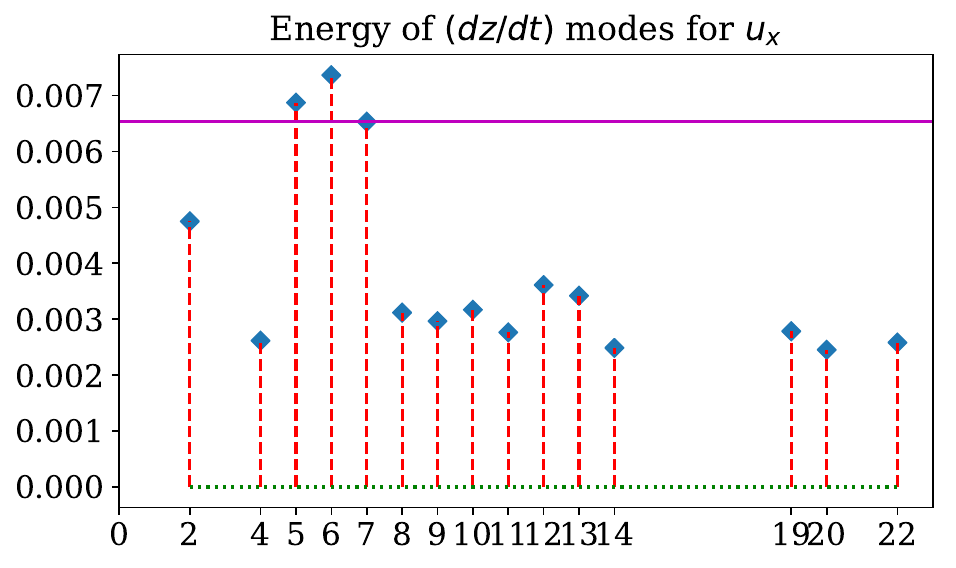}
 }\hfill
 \subfloat[Modal energy computed by eq.~(\ref{eq:greedy-modes-energy}) for $\vec u_y$\label{fig:red_mode_energy_v}]{%
  \includegraphics[width=0.45\columnwidth]{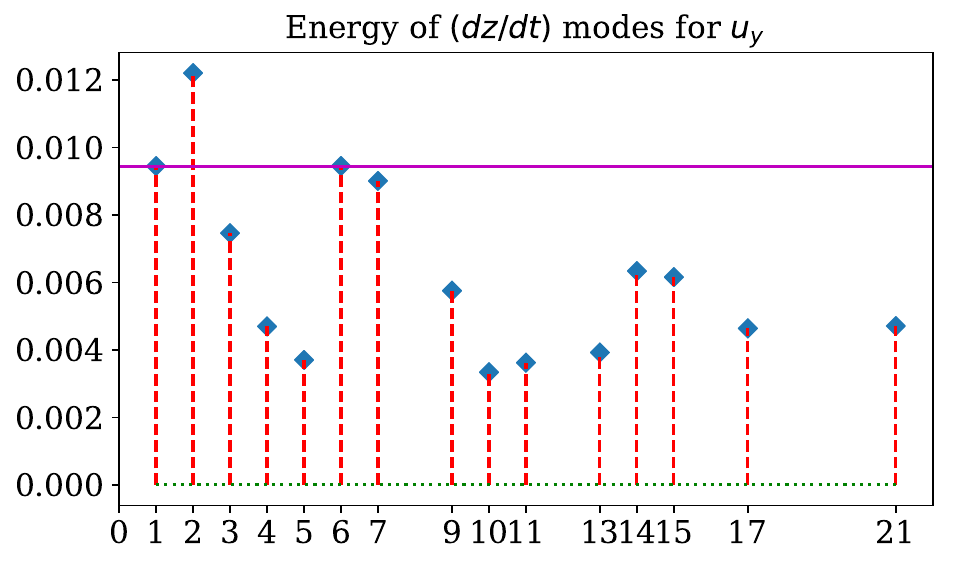}
 }
 \caption{Energy distribution across the components of the time derivative $d\bz/dt$ of the projection coefficients for the x-velocity (left) and y-velocity (right) snapshots of the Red River example}
 \label{fig:red_mode_energy}
\end{figure}

The modal energy given by eq.~(\ref{eq:greedy-modes-energy}) is distributed relatively evenly across the modes for the velocity components, as can be seen in Figure \ref{fig:red_mode_energy} where the magnitudes of the energy associated with the most dominant $16$ modes for each velocity component have been plotted. The most dominant mode ($m=0$) is left out in order to emphasize the energy distribution in the remaining modes. The y-axis represents the energy content and the x-axis denotes the respective modes. The horizontal lines denote the $25\%$ energy content barrier such that the modes that intersect the line cumulatively account for $25\%$ of the total modal energy. Specifically, the modes $\{0, 6, 5, 7\}$ for the x-velocity component, $\{ 0, 2, 6, 1\}$ for the y-velocity component, and $\{0,1\}$ for the depth are selected to populate the list $L$.

\begin{figure}[ht!]
\centering
   \subfloat[The first $8$ components ($0\leq m \leq 7$) of the time derivatives of the projection coefficients $\frac{d\bz}{dt}$ for the x-velocity snapshots. $m = (0,6,5,7)$ are selected for \textit{psr-greedy} iterations.\label{fig:red_mode_u}]{%
     \includegraphics[width=0.95\columnwidth]{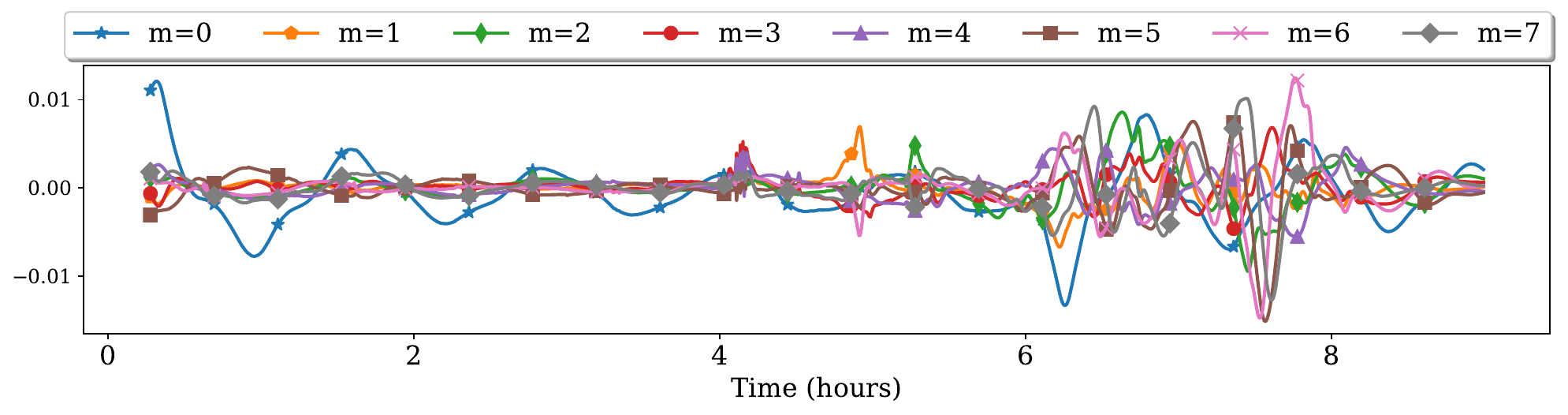}
   }\\
   \subfloat[The first $8$ components ($0\leq m \leq 7$) of the time derivatives of the projection coefficients $\frac{d\bz}{dt}$ for the y-velocity snapshots. $m = (0,2,6,1)$ are selected for \textit{psr-greedy} iterations.\label{fig:red_mode_v}]{%
     \includegraphics[width=0.95\columnwidth]{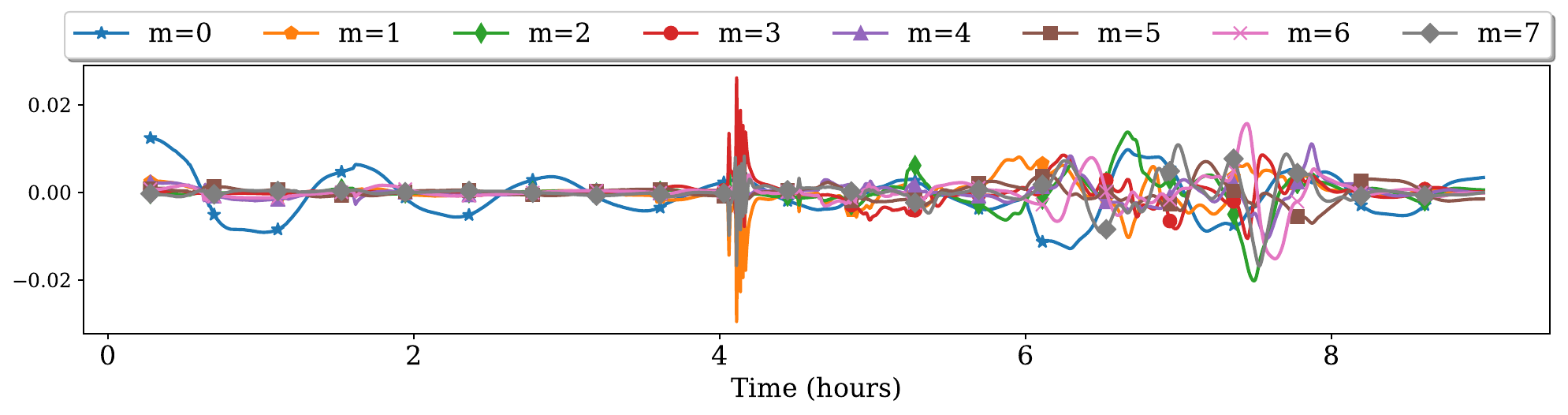}
   }\\
   \subfloat[The first $3$ components ($0\leq m \leq 2$) of the time derivatives of the projection coefficients $\frac{d\bz}{dt}$ for the depth snapshots.. $m = (0,1)$ are selected for \textit{psr-greedy} iterations.\label{fig:red_mode_h}]{%
    \includegraphics[width=0.95\columnwidth]{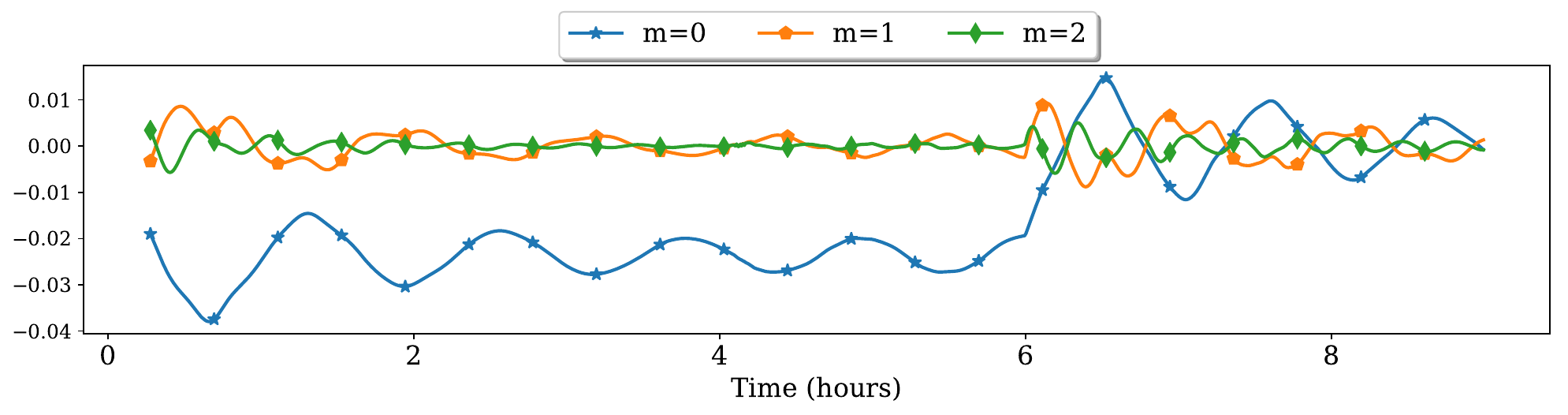}
   }
   \caption{Temporal dynamics captured by the time derivatives of the projection coefficients for the Red River example}\label{fig:red_modes}
\end{figure}

The above choice of modes for the set $L$ is further validated by studying the time derivatives of the projection coefficients as presented in Figure \ref{fig:red_modes}. It is evident that the dominant mode ($m=0$) captures most of the temporal behavior of the depth snapshots (see Figure \ref{fig:red_mode_h}). On the other hand, the dominant modes for the velocity variables describe most of the temporal behavior during the early part of the simulation time (see Figures \ref{fig:red_mode_u} and \ref{fig:red_mode_v}). The remaining selected modes for each velocity variable are required to capture the variations observed during the middle and the later parts of the simulation period. The selected modes are arranged in descending order of energy content while cycling through the solution components $\vec h, \vec u_x, \vec u_y$ such that $L = $ \{\mathlist 0, 3, 194, 1, 9, 196, 8, 10, 200, 195 \} where the $\vec h$ modes lie between $0$ to $2$, $\vec u_x$ modes range from $3$ to $193$, and $\vec u_y$ modes range from $194$ to $372$. The tolerance for the \textit{psr-greedy} algorithm was set at $\tau_{psr} = 0.12$. The choice of the greedy tolerances were dictated by the need to generate sufficiently large sets of centers for
effective comparison of the greedy NIROM solutions and for the numerical convergence studies. In practice, the greedy tolerances along with the POD truncation level $\tau_{POD}$ and the RBF scale factor $c$ act as hyperparameters for the NIROM that can be tuned to attain a desired balance of accuracy and computational efficiency.

\begin{figure}[ht!]
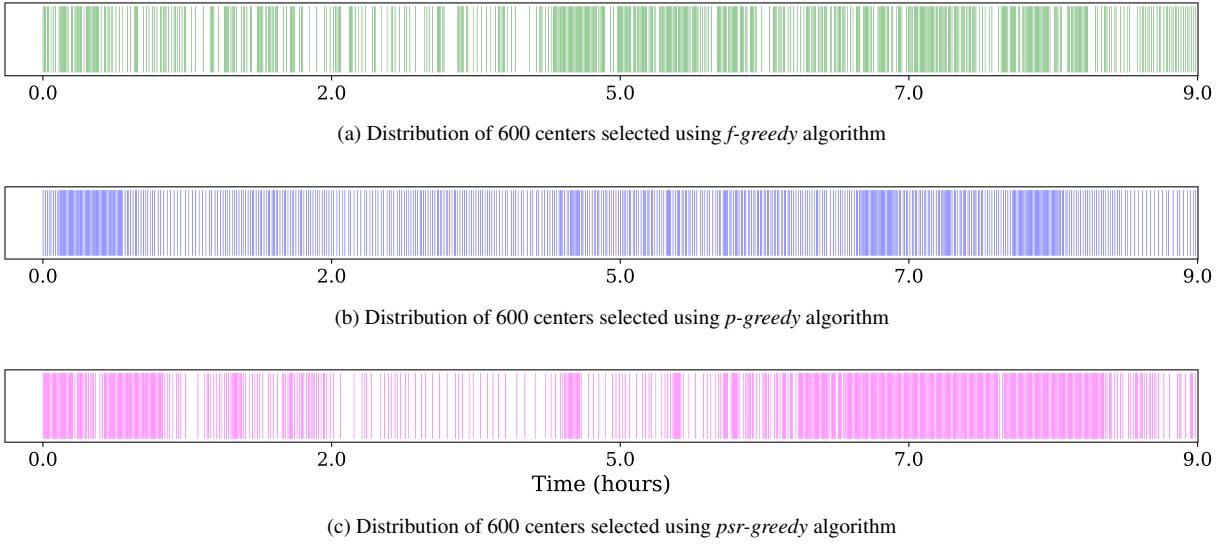

   \subfloat[Distribution of $600$ centers selected using \textit{f-greedy} algorithm\label{fig:red_centers_f}]{%
     \includegraphics[width=0.98\columnwidth]{figures/red_river_nirom_f_greedy_centers_600.png}
   }\\
   \subfloat[Distribution of $600$ centers selected using \textit{p-greedy} algorithm\label{fig:red_centers_p}]{%
     \includegraphics[width=0.98\columnwidth]{figures/red_river_nirom_p_greedy_centers_600.png}
   }\\
   \subfloat[Distribution of $600$ centers selected using \textit{psr-greedy} algorithm\label{fig:red_centers_psr}]{%
    \includegraphics[width=0.98\columnwidth]{figures/red_river_nirom_psr_greedy_centers_600.png}
   }%\\
   \caption{Comparative distribution of centers selected by the greedy algorithms for the Red River example}
   \label{fig:red_centers_greedy}
\end{figure}

Figure \ref{fig:red_centers_greedy} illustrates the temporal distribution of the first $600$ centers selected by the three greedy algorithms. It is evident that the \textit{f-greedy} (\ref{fig:red_centers_f}) centers are relatively more clustered around the locations of the local extrema of the $l^2$-norm of the time derivative, shown in Figure \ref{fig:red_modal_l2_norm}, whereas the \textit{psr-greedy} (\ref{fig:red_centers_psr}) centers tend to be distributed according to the extrema in the dominant components of the projection coefficients, shown in Figure \ref{fig:red_modes}. The \textit{p-greedy} (\ref{fig:red_centers_p}) centers are more uniformly distributed all over the time domain. 
%However, the online simulation period includes only the first two local extrema observed in Figure \ref{fig:red_modal_l2_norm}. 
Hence, the \textit{p-greedy} NIROM solution is not able to approximate all the variations in the interpolated time derivative functions as efficiently as the \textit{f-greedy} and the \textit{psr-greedy} NIROM solutions. However, the clustering effect is more pronounced for the \textit{f-greedy} centers. As a result, the \textit{psr-greedy} approach emerges as the most efficient NIROM strategy, achieving the best accuracy with the lowest number of centers among all the algorithms. 

\begin{figure}[ht]
\centering
  {\includegraphics[width=0.58\columnwidth]{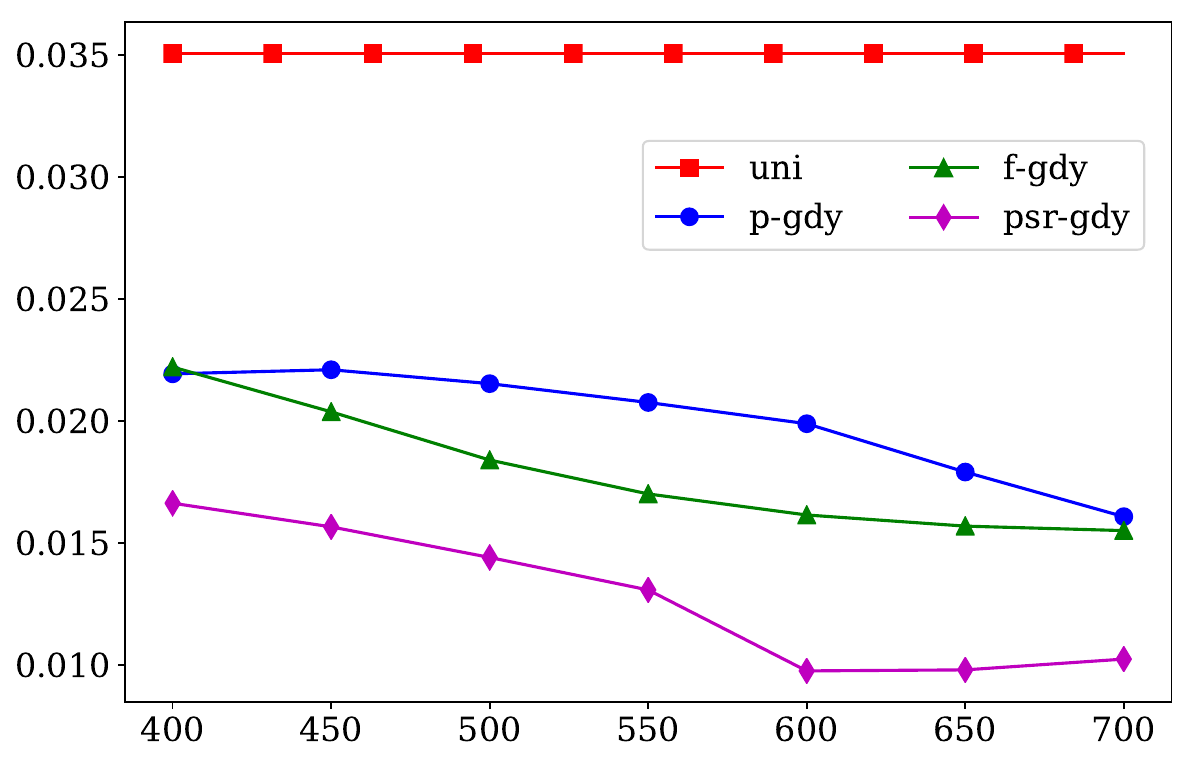}
   }
   \caption{Convergence of the $stRMSE$ norm of the greedy NIROM solutions for the x-velocity variable, with respect to the number of centers used by the RBF interpolant, for the Red River example}
   \label{fig:red_greedy_L2_L2_error}
\end{figure}

Figure \ref{fig:red_greedy_L2_L2_error} illustrates this comparative convergence behavior in terms of the $stRMSE$ norms (see eq.~(\ref{strms})) plotted with respect to the size of the RBF basis or the number of centers used by the RBF interpolant. The horizontal line represents the $stRMSE$ norm of a RBF NIROM solution in which $786$ RBF centers were selected by uniformly skipping over every $4$ snapshots from the set $X$. 
%As the \textit{p-greedy} minimizes the overall approximation error using the power function, it performs better in the beginning. 
All the greedy solutions appear to be significantly and consistently superior to the uniform NIROM solution both in terms of accuracy and efficiency, thus providing empirical evidence of the effectiveness of the greedy algorithms.
The \textit{psr-greedy} algorithm produces a more optimal set of centers than the other two greedy strategies. 
In Table \ref{tab:red_l2l2_norms}, the smallest $stRMSE$ error norms corresponding to all three components of the different greedy NIROM solutions are reported along with the respective number of centers used in each case. 

\begin{table}[ht!]
  \setlength{\tabcolsep}{5pt}
  \renewcommand{\arraystretch}{1.2}
  \centering
{\begin{tabularx}{.7\linewidth}{lcccc}
  \toprule
& \textit{uniform} $(786)$ & \textit{p-gdy} $(700)$ & \textit{f-gdy} $(700)$ & \textit{psr-gdy} $(650)$ \\ \midrule
  $\vec u_x$  & $0.0351$ & $0.0161$ & $0.0155$ & $\textsl{0.0098}$  \\ \midrule
  $\vec u_y$  & $0.0416$ & $0.0191$ & $0.0182$ & $\textsl{0.0119}$  \\ \midrule
  $\vec h$    & $0.0204$ & $0.0085$ & $0.0101$ & $\textsl{0.0063}$  \\ 
  \bottomrule
\end{tabularx}}
\caption{Lowest $stRMSE$ error norms of the greedy NIROM solutions for the Red River example}\label{tab:red_l2l2_norms}
\end{table}

{\begin{figure}[ht!]
\centering
  \includegraphics[width=0.95\columnwidth]{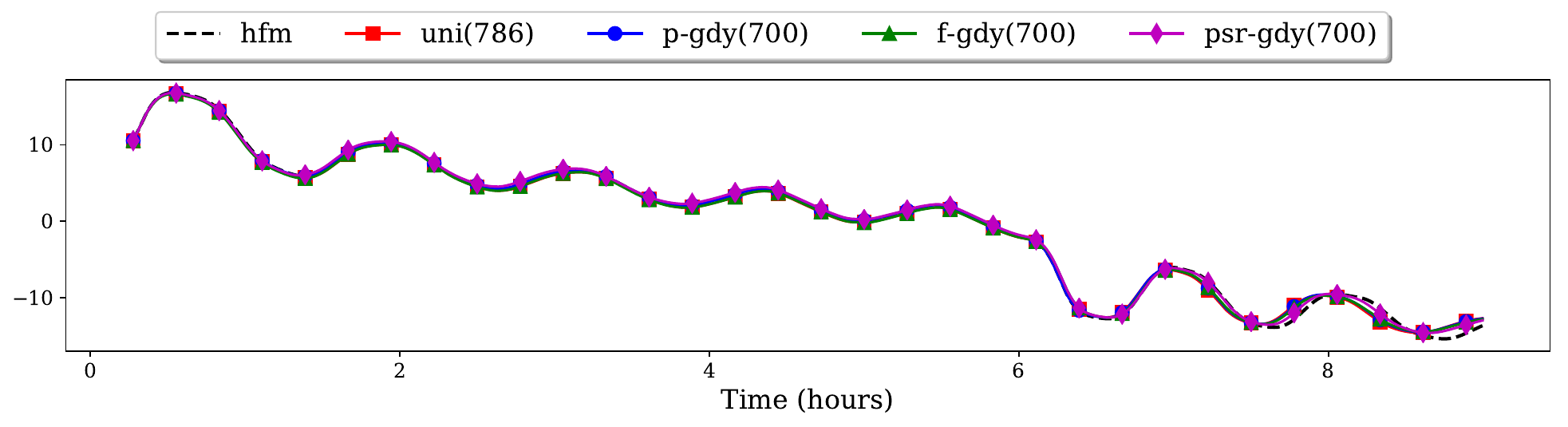}
  \caption{Comparison of the first component of reduced x-velocity solution for the Red River example}
  \label{fig:red_u_mode0}
\end{figure}}

\begin{figure}[ht!]
\centering
     \subfloat[Comparison of the $RMSE$ for the depth solutions\label{fig:red_wrms_h}]{%
       \includegraphics[width=0.95\columnwidth]{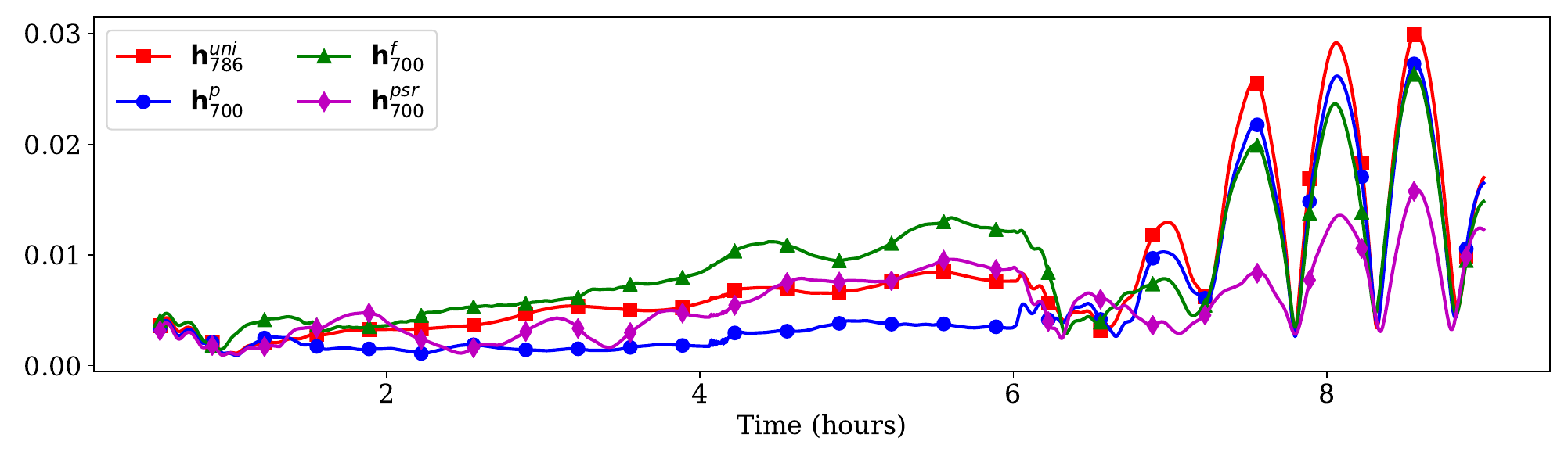}
     }\\
     \subfloat[Comparison of the $RMSE$ for the x-velocity solutions\label{fig:red_wrms_u}]{%
       \includegraphics[width=0.95\columnwidth]{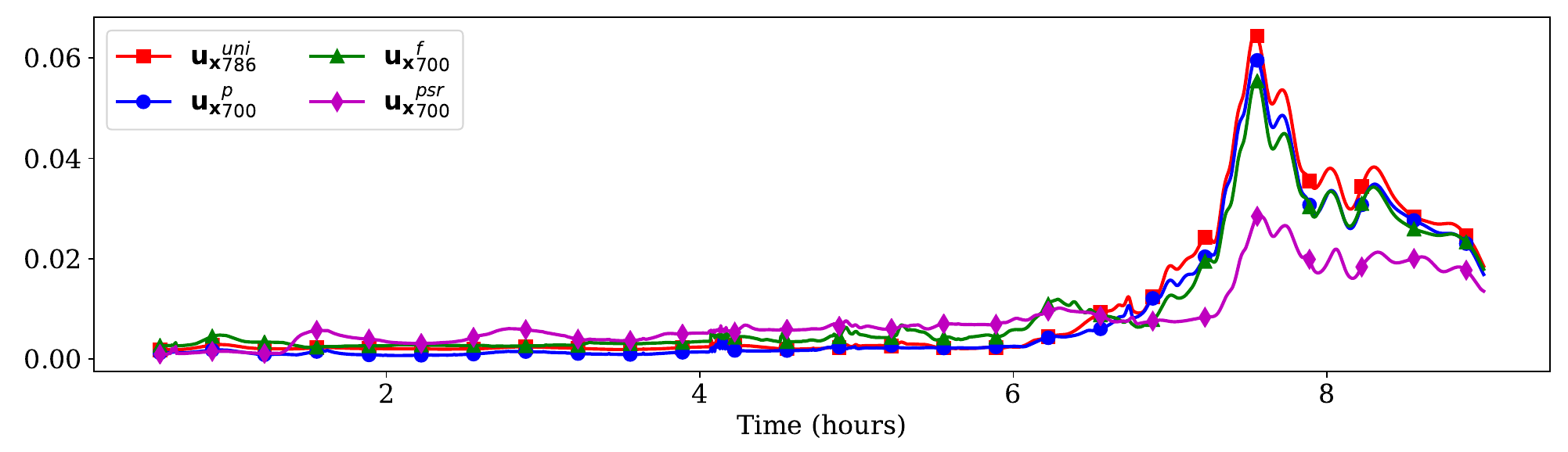}
     }
   \caption{Temporal evolution of the spatial approximation errors of the NIROM solutions for the Red River example}
   \label{fig:red_wrms}
\end{figure}
The accuracy of the greedy NIROM solutions is also studied by directly comparing the predicted coefficients in the reduced basis space as well as by comparing the RMSE in the high-fidelity, full-order space. For these comparisons, greedy NIROM solutions obtained using $700$ centers are compared with the NIROM solution obtained using $786$ uniformly-spaced centers.
Figure \ref{fig:red_u_mode0} compares the first component of the reduced order NIROM solution for the x-velocity to that obtained by projecting the high-fidelity solution on to the POD basis space (labeled as ``hfm"). Figure \ref{fig:red_wrms} shows the temporal evolution of the root mean square errors (see eq.~\ref{rms}) for the full order solution of the depth variable (Figure \ref{fig:red_wrms_h}) and the x-velocity variable (Figure \ref{fig:red_wrms_u}).

Figure \ref{fig:red_u_mode0} confirms that using comparable number of centers all the greedy RBF NIROMs are almost equally effective in approximating the dominant features of the high-fidelity solution in the reduced basis (POD) space. Figures \ref{fig:red_wrms_h} and \ref{fig:red_wrms_u} provide a closer look into how the spatially-averaged RMS errors in the high-dimensional space evolve over time. The time derivatives of the first few dominant modes, as shown in Figure \ref{fig:red_modes}, depict a slowly varying, damped, periodic behavior during the initial stage of the flow, followed by a sudden, high-frequency, irregular flow phase starting at about $t = 6$ hours. The \textit{p-greedy} NIROM solution is able to approximate the initial damped, periodic flow phase with marginally higher accuracy. However, in the region with high-frequency variations, the \textit{p-greedy} and \textit{f-greedy} algorithms fail to provide sufficient resolution to effectively capture the flow dynamics. On the other hand, the RBF centers selected by the modal residual-driven \textit{psr-greedy} NIROM are able to provide adequate resolution to capture the high-frequency, irregular variations in the solution more accurately, and thus the \textit{psr-greedy} NIROM offers a more stable and robust performance throughout the online simulation domain. The \textit{f-greedy} centers are clearly biased towards the larger local extrema in the function residual due to clustering and need to be further enriched in the under-represented locations by adopting a smaller tolerance value, $\tau_f$ in order to achieve a consistent level of accuracy. 

An argument can be made that since the modes not selected in $L$ do not inform the greedy center selection process in the \textit{psr-greedy} algorithm, the error in approximating the unselected modes should be higher than the other greedy algorithms. This is addressed in Table \ref{tab:mtRMSE_red}. For every solution component $j$, the \textit{mtRMSE} norm is computed as the $l^2$-norm of the reduced solution error over all time points $n=\{1,\ldots,M\}$ as well as over all the modes not selected in $L$ \ie $i = \{1,\ldots, m_j | i \notin L\}$ such that, 

\begin{align}\label{eq:mtrms-modal}
    mtRMSE_j =  \sqrt{\dfrac{\sum\limits_{n=1}^M\sum\limits_{\substack{i= 1,\\ i\notin L}}^{m_j} |{z_{f,j,i}^n - z_{a,j,i}^n}|^2}{Nm_j}},
\end{align}
where $z_{f,j,i}^n$ denotes the reduced representation of the high-fidelity solution and $z_{a,j,i}^n$ denotes the corresponding NIROM solution at mode $i$ and time point $n$ for component $j$. Table \ref{tab:mtRMSE_red} presents the \textit{mtRMSE} values computed using different NIROM strategies. It can be observed that even for the modes that do not instruct the center selection process, the approximation error using the \textit{psr-greedy} algorithm is significantly lower than all the other approaches.
\begin{table}[ht]
    \setlength{\tabcolsep}{5pt}
    \renewcommand{\arraystretch}{1.2}
    \centering
    \begin{tabular}{c c c c c}
        \toprule
        & uniform ($786$) & \textit{p-greedy} ($700$) & \textit{f-greedy} ($700$) & \textit{psr-greedy} ($700$) \\ \midrule
        $\vec u_x$ & $0.142$ & $0.144$ & $0.126$ & $\textsl{0.079}$\\
        $\vec u_y$ & $0.174$ & $0.177$ & $0.153$ & $\textsl{0.098}$\\
        $\vec h$   & $0.637$ & $0.599$ & $0.734$ & $\textsl{0.435}$\\ \bottomrule
        
    \end{tabular}
    \caption{\textit{mtRMSE} norms using modes not selected in $L$ for different NIROM solutions}\label{tab:mtRMSE_red}
\end{table}

%%%-------------------------------------------------------------
\subsubsection{San Diego Bay example}
The final numerical example involves the simulation of tide-driven coastal flow in the San Diego Bay in California, USA (see Figure \ref{san_diego_domain}).

    \begin{figure}[htb]%{l}{0.45\textwidth}
    \subfloat[Bathymetry\label{fig:sd_bathymetry}]{%
    \includegraphics[width=0.5\columnwidth,trim=0cm 0cm .4cm .4cm, clip]{./figures/San_diego.png}
    }\hfill
    \subfloat[Computational mesh\label{fig:sd_mesh}]{%
    \includegraphics[width=0.5\columnwidth,trim=2cm 4cm 2cm 0cm, clip]{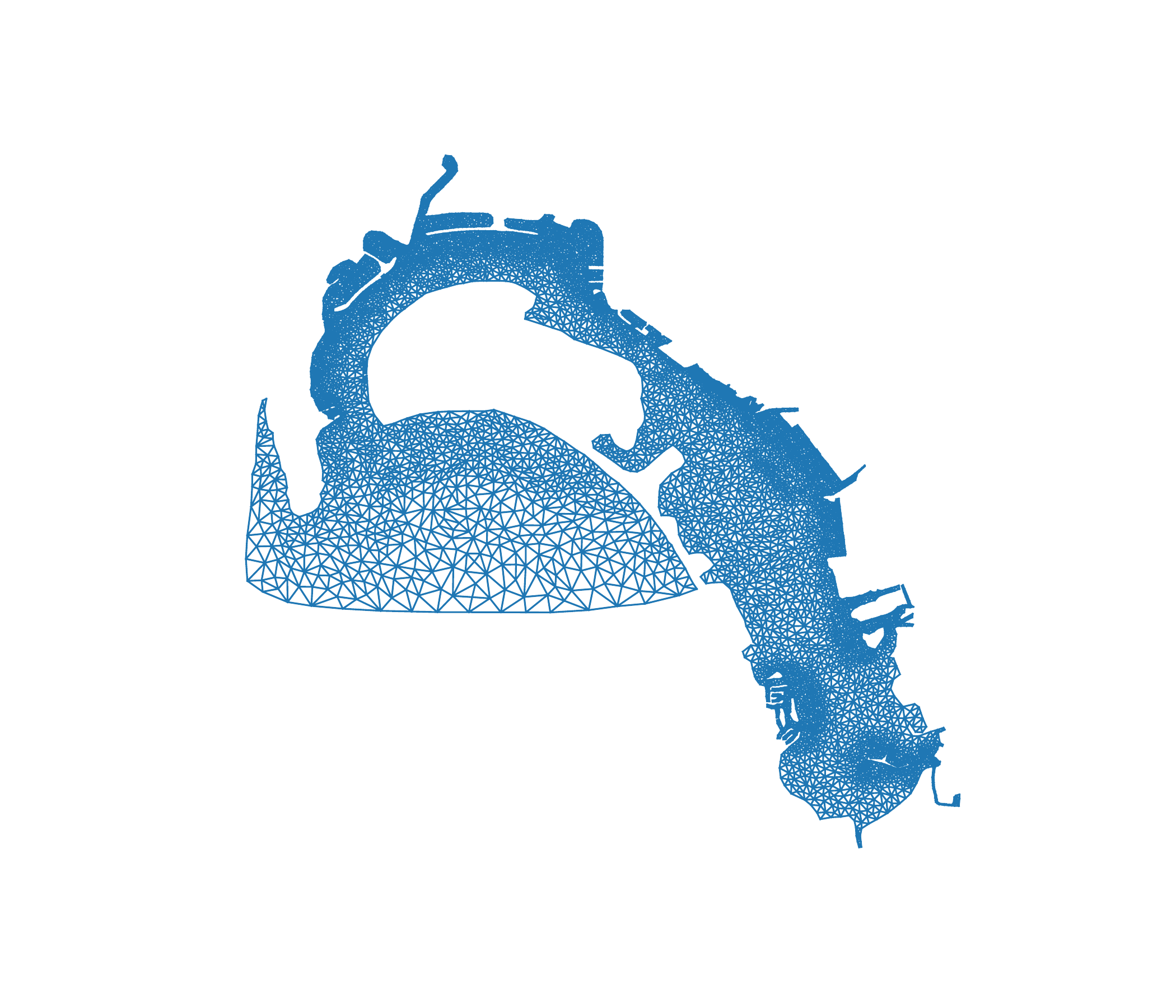}
    }
    \caption{San Diego Bay}\label{san_diego_domain}
    \end{figure}
A uniform unstructured triangular mesh consisting of $N = 6311$ nodes and $10999$ elements was used. 
Tidal flow data was obtained from NOAA/NOS Co-Ops website at an interval of $6$ minutes. This was applied as a time series of tailwater elevation boundary data along the east-west boundary of the computational mesh at the entrance to the inner harbor, as shown in Figure \ref{san_diego_domain}. No flow boundary conditions were applied along the walls of the harbor and an initial water surface elevation for the entire domain was provided. The fluid viscosity is $\nu = 10^{-5}$, Manning's friction coefficient is $n_{mn} = 0.022$, and the Coriolis latitude is $32.7$.

\begin{wrapfigure}{R}{0.55\textwidth}
\centering
\includegraphics[width=0.54\textwidth]{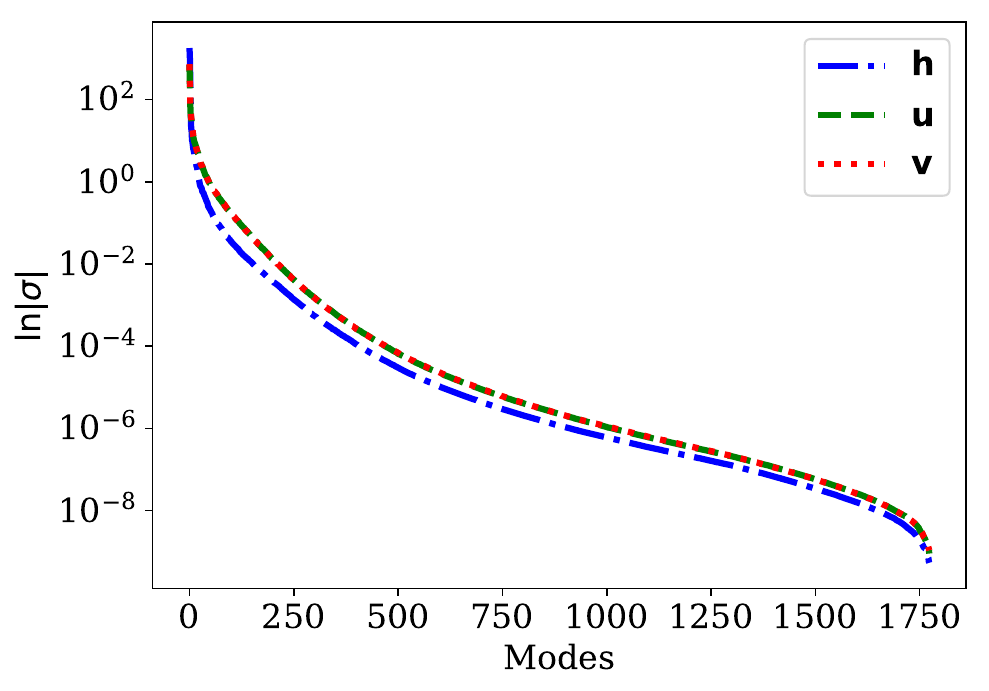}
\caption{Semi-logarithmic plot of the singular values for the San Diego bay example}\label{fig:sd_sing_val}
\end{wrapfigure}
In the offline phase, the high-fidelity model was run until $T=50$ hours with a time step of $\Delta t = 25$ seconds. Dropping the first $100$ time steps to avoid the influence of the initial model startup on the flow behavior, $7101$ offline solution snapshots were collected for $ t \in [0.69, 50]$ hours. The POD basis modes were computed using a subset of uniformly spaced $1776$ snapshots that were selected by skipping over every $4$ consecutive snapshots such that the effective time step for the offline training data was $\Delta t=100$ seconds. Using a POD truncation tolerance of $\tau_{POD} = 5\times10^{-7}$, only the first $36, 115$ and $113$ POD modes were selected for the reduced basis representation of the primitive variables $\vec h, \vec u_x$, and $\vec u_y$ respectively. Figure \ref{fig:sd_sing_val} shows the decay of the singular values for the training snapshots in a semi-logarithmic scale. The online simulations using the reduced order models were carried out for the same simulation period as the high-fidelity model and with $\Delta t = 25$ seconds.

The high-fidelity training snapshots were projected onto the POD space to generate the set $X$ containing all the available centers for the RBF interpolant. The RBF scale factor was estimated to be $c=0.05$. The \textit{p-greedy} centers were selected using a tolerance of $\tau_p = 0.33$ while the \textit{f-greedy} tolerance was $\tau_f=3.77\times 10^{-3}$. 
A list consisting of $18$ dominant modes of the solution components $\vec u_x, \vec u_y, \vec h$ was compiled by selecting the least number of modes for each component that contain $91\%$ of the total energy content for that component, \ie using $\tau_{greedy} =0.91$ in eq.~(\ref{eq:greedy-modes-energy-selection}). The residual computations for the \textit{psr-greedy} algorithms were carried out sequentially over this set of modes, $L =$ \{\mathlist{36, 151, 0, 37, 152, 1, 38, 153, 48, 154, 40, 161, 50, 164, 39, 158, 163, 157}\}. Here the modes of the projection coefficients for the depth snapshots lie between $0$ to $35$ out of which $2$ modes were selected, the x-velocity modes run from $36$ to $150$ from which $7$ were selected, and the y-velocity modes lie between $151$ to $263$ out of which $9$ were selected. 

\begin{figure}[ht!]
 \subfloat[Modal energy computed by eq.~(\ref{eq:greedy-modes-energy}) for $\vec u_x$\label{fig:sd_mode_u}]{%
  \includegraphics[width=0.45\columnwidth]{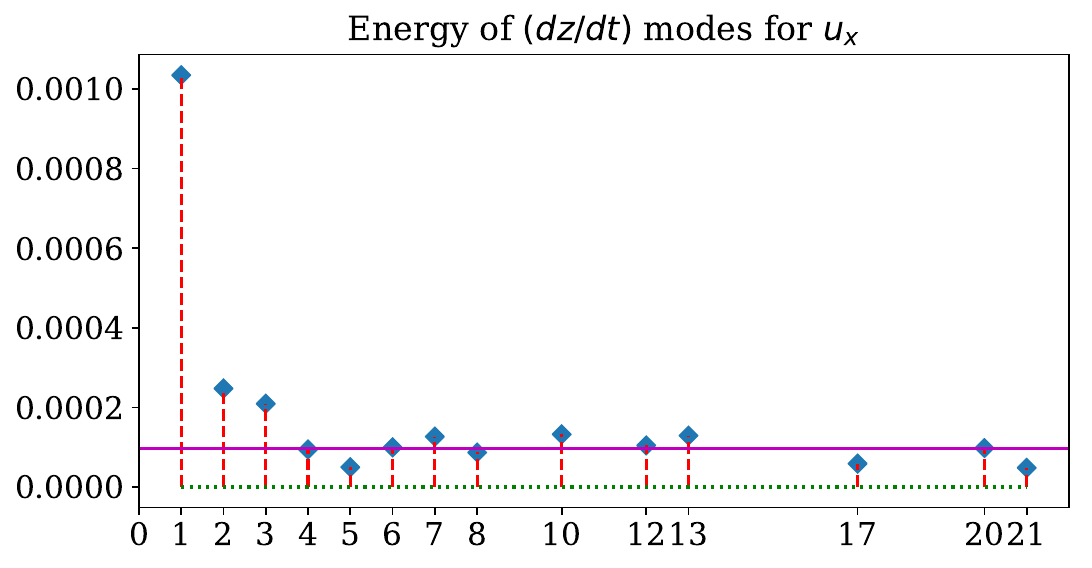}
 }\hfill
 \subfloat[Modal energy computed by eq.~(\ref{eq:greedy-modes-energy}) for $\vec h$\label{fig:sd_mode_h}]{%
  \includegraphics[width=0.45\columnwidth]{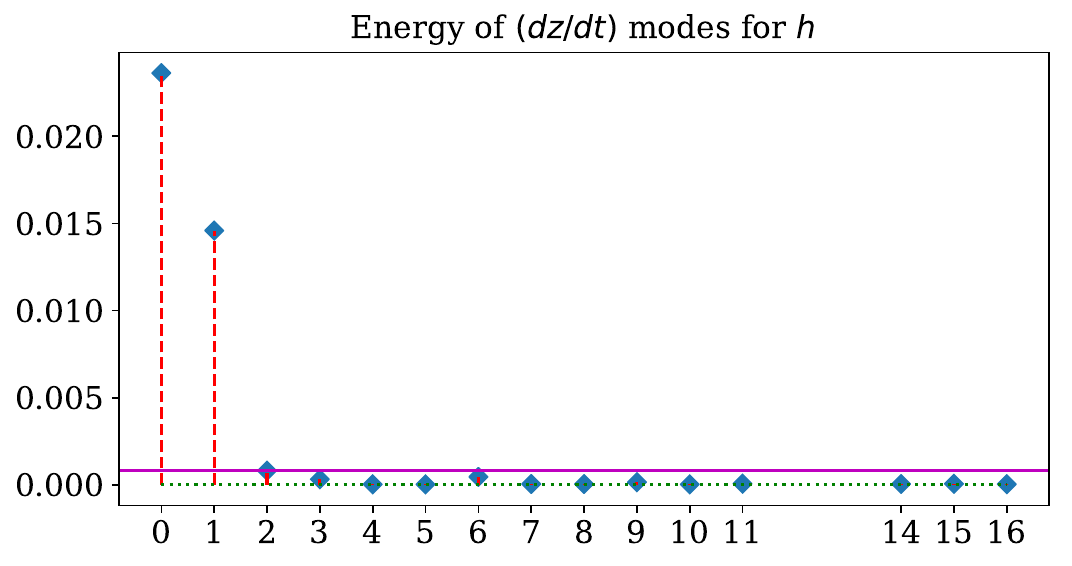}
 }
 \caption{Energy distribution across the components of the time derivative $d\bz/dt$ of the projection coefficients for the x-velocity (left) and depth (right) snapshots of the San Diego Bay example}
 \label{fig:sd_mode_energy}
\end{figure}

In Figure \ref{fig:sd_mode_energy}, the most ``energetic" $15$ modes of each solution component are plotted, where the y-axis represents the energy content and the x-axis represents the respective modes. The $91\%$ energy truncation is depicted by the horizontal line in each of the plots such that the modes corresponding to the bars that fall above the line are selected for each component. Unlike the Red River example, where modal energy was distributed fairly well across the spectrum, in this case, it can be seen that there was a sharp decay in the energy content of the most significant modes and a higher value of $\tau_{greedy} \geq 0.85$ was found to be optimal for greedy iterations.  
The \textit{psr-greedy} tolerance was set at $\tau_{psr} = 0.48$. 

\begin{figure}[ht]
\centering
    \includegraphics[width=0.98\columnwidth]{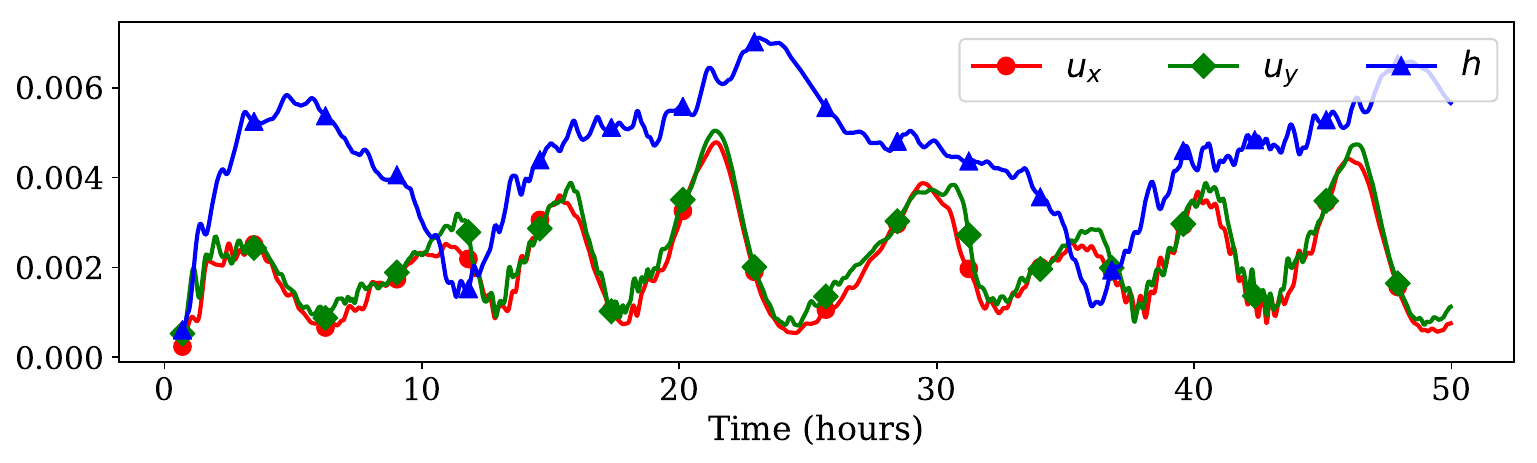}
    \caption{Temporal dynamics of the modal $l^2$-norm of the time derivative of the projection coefficients $\nm*{\frac{dz}{dt}(\cdot,t)}_{l^2} $ for the San Diego Bay example}
    \label{fig:sd_modal_l2_norm}
\end{figure}

The modal $l^2$-norms of the time derivatives of the projection coefficients for each solution component, as shown in Figure \ref{fig:sd_modal_l2_norm}, highlight the periodic nature of the tide-driven flow. The higher frequency of the variations in the $l^2$ norms of the velocity components might explain why a larger number of POD basis vectors were required for the reduced basis representation of the velocity components as compared to the depth snapshots, and also why lesser number of depth modes were required in the \textit{psr-greedy} modal residual computations.

\begin{figure}[ht!]
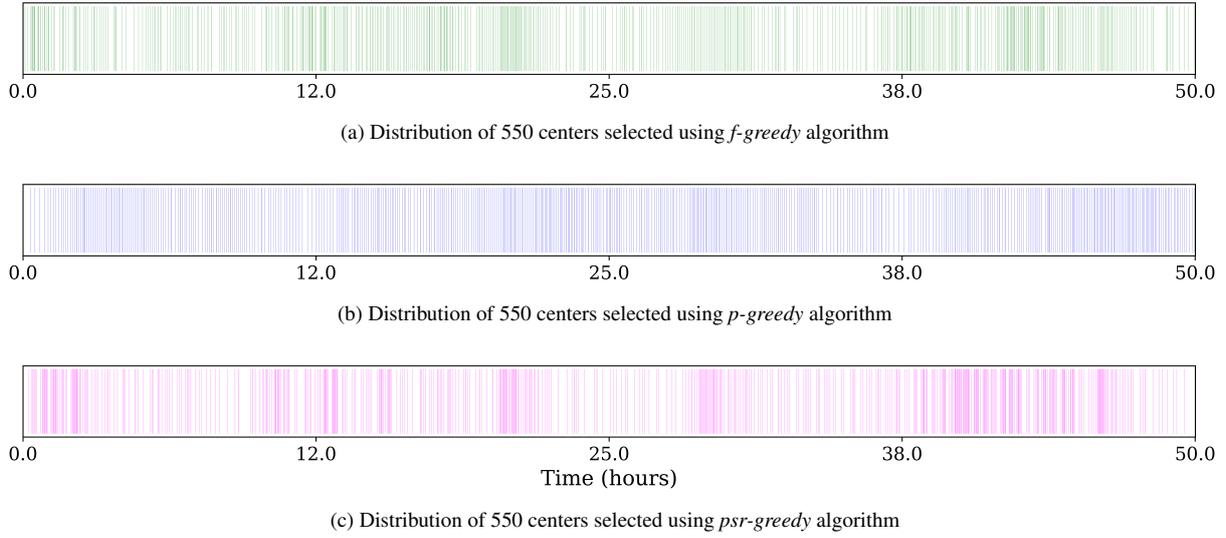

\centering
 \subfloat[Distribution of $550$ centers selected using \textit{f-greedy} algorithm\label{fig:sd_centers_f_550}]{%
   \includegraphics[width=0.98\columnwidth]{figures/sd_nirom_f_greedy_centers_550.png}
 }\\
 \subfloat[Distribution of $550$ centers selected using \textit{p-greedy} algorithm\label{fig:sd_centers_p_550}]{%
   \includegraphics[width=0.98\columnwidth]{figures/sd_nirom_p_greedy_centers_550.png}
 }\\
 \subfloat[Distribution of $550$ centers selected using \textit{psr-greedy} algorithm\label{fig:sd_centers_psr_550}]{%
  \includegraphics[width=0.98\columnwidth]{figures/sd_nirom_psr_greedy_centers_550.png}
  }
 \caption{Comparison of the distribution of centers selected by the greedy algorithms for the San Diego Bay example}
 \label{fig:sd_centers_greedy}
\end{figure}

Figure \ref{fig:sd_centers_greedy} depicts the distribution of the first $550$ centers selected by the three greedy strategies. As expected, the \textit{f-greedy} centers (Figure \ref{fig:sd_centers_f_550}) exhibit a clustering effect around the locations of the local extrema of the $l^2$-norm of the time derivative (see Figure \ref{fig:sd_modal_l2_norm}), the \textit{p-greedy} centers (Figure \ref{fig:sd_centers_p_550}) are much more uniformly distributed, and the \textit{psr-greedy} approach (Figure \ref{fig:sd_centers_psr_550}) moderates both these effects producing a less-scattered and less-clustered set of centers. However, the periodic nature of the flow with relatively lower level of small-scale variations, offers a distinct advantage to the \textit{p-greedy} and the uniform selection procedures.

\begin{figure}[ht!]
\centering
  {\includegraphics[width=0.58\columnwidth]{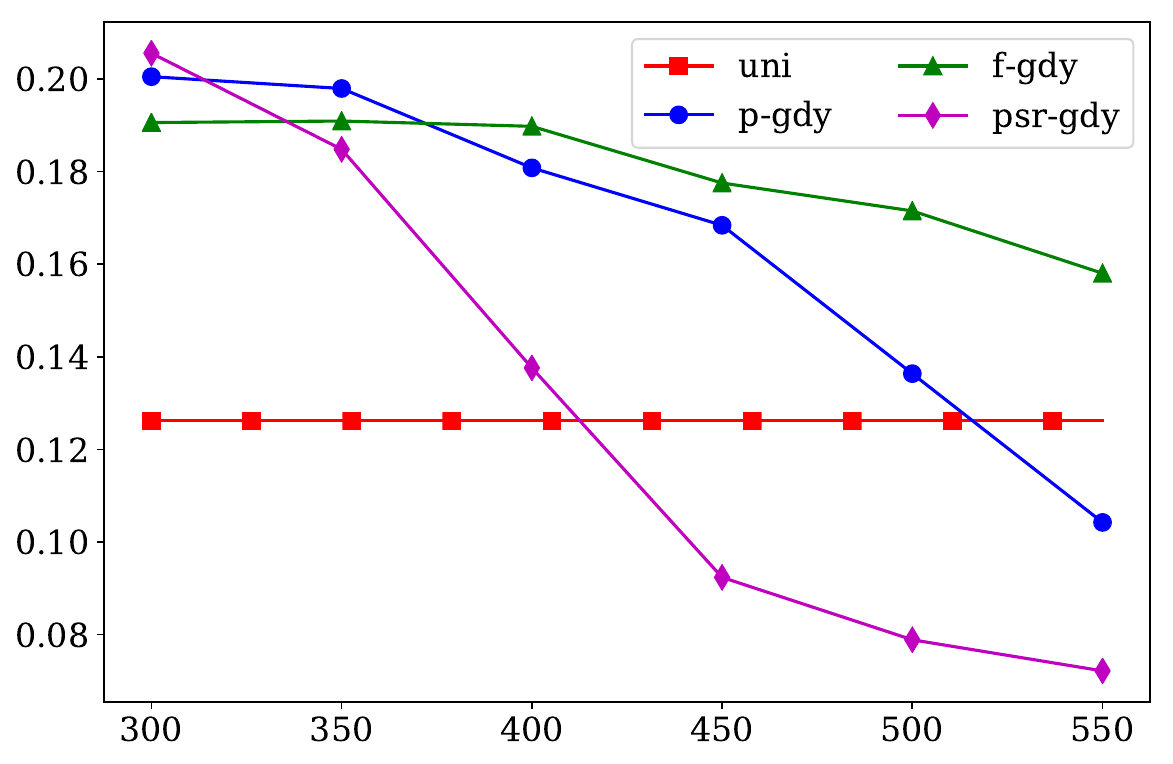}
   }
   \caption{Convergence of the $stRMSE$ norm of the greedy NIROM solutions for the x-velocity variable, with respect to the number of centers used by the RBF interpolant, for the San Diego Bay example}
   \label{fig:sd_greedy_L2_L2_error}
\end{figure}

Figure \ref{fig:sd_greedy_L2_L2_error} illustrates the convergence behavior of the different greedy algorithms in terms of the $stRMSE$ norms (see eq.~(\ref{strms})) for $\vec u_x$, where the x-axis denotes the number of centers used by the RBF interpolant. The horizontal line represents the $stRMSE$ norm of a RBF NIROM solution computed using $547$ RBF centers that were selected by uniformly skipping over every $13$ projection coefficients of the high-fidelity snapshots. The periodic nature of the flow is a more natural fit for the power function-driven sampling of centers and hence, unlike the Red River example, in this case the \textit{p-greedy} algorithm shows a more desirable improvement in the stRMSE norm with the selection of additional centers. However, the evaluation of the power function and the computation of the RBF interpolation coefficients are both dependent on the RBF scale factor, $c$ and the accuracy of the \textit{p-greedy} and the \textit{psr-greedy} NIROM solutions were found to be highly sensitive to the value of the hyperparameter $c$. In this example, $c=0.095$ was chosen for the \textit{p-greedy} algorithm and $c=0.05$ was chosen for the \textit{psr-greedy} and all other NIROM implementations.
In the \textit{psr-greedy} algorithm, the convergence is more robust and the stRMSE steadily reduces with further enrichment of centers. Moreover, the \textit{psr-greedy} selection is able to capture the small-scale features in the flow and hence the \textit{psr-greedy} NIROM offers superior accuracy than the other approaches. In Table \ref{tab:sd_l2l2_norms}, the smallest $stRMSE$ error norms corresponding to the x-velocity component of the different greedy NIROM solutions are reported along with the respective number of centers used in each case. Additionally, $stRMSE$ norms are also reported for the uniform RBF NIROM using $547$ centers. Despite the favorable nature of the flow, the uniform RBF NIROM was outperformed by the \textit{psr-greedy} NIROM. 

\begin{table}[ht!]
  \setlength{\tabcolsep}{5pt}
  \renewcommand{\arraystretch}{1.2}
  \centering
{\begin{tabularx}{.7\linewidth}{lcccc}
  \toprule
& \textit{uniform}  & \textit{p-greedy}  & \textit{f-greedy}  & \textit{psr-greedy}  \\ \midrule
  $\vec u_x$ & $0.1262$ $(547)$& $0.1042$ $(550)$& $0.1580$ $(550)$& $\textsl{0.0721}$ $(550)$ \\ \midrule
  $\vec u_y$ & $0.1347$ $(547)$& $0.1114$ $(550)$& $0.1693$ $(550)$& $\textsl{0.0782}$ $(550)$ \\ \midrule
  $\vec h$   & $0.2723$ $(547)$& $0.2087$ $(550)$& $0.2957$ $(500)$& $\textsl{0.1742}$ $(550)$ \\ 
  \bottomrule
\end{tabularx}}
\caption{Lowest $stRMSE$ error norms using each of the greedy NIROM solutions for the San Diego Bay example}\label{tab:sd_l2l2_norms}
\end{table}

\begin{figure}[ht]
\centering
    \includegraphics[width=0.48\columnwidth]{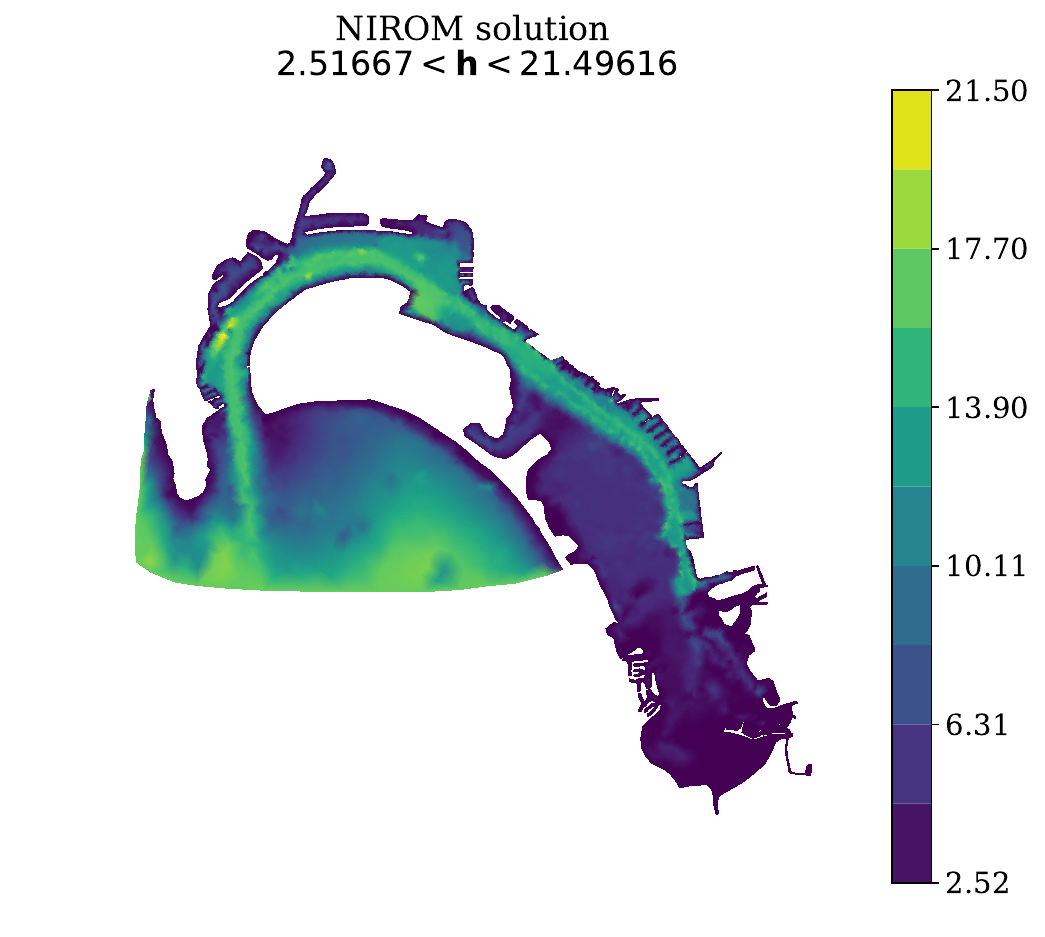}\hfill
    \includegraphics[width=0.48\columnwidth]{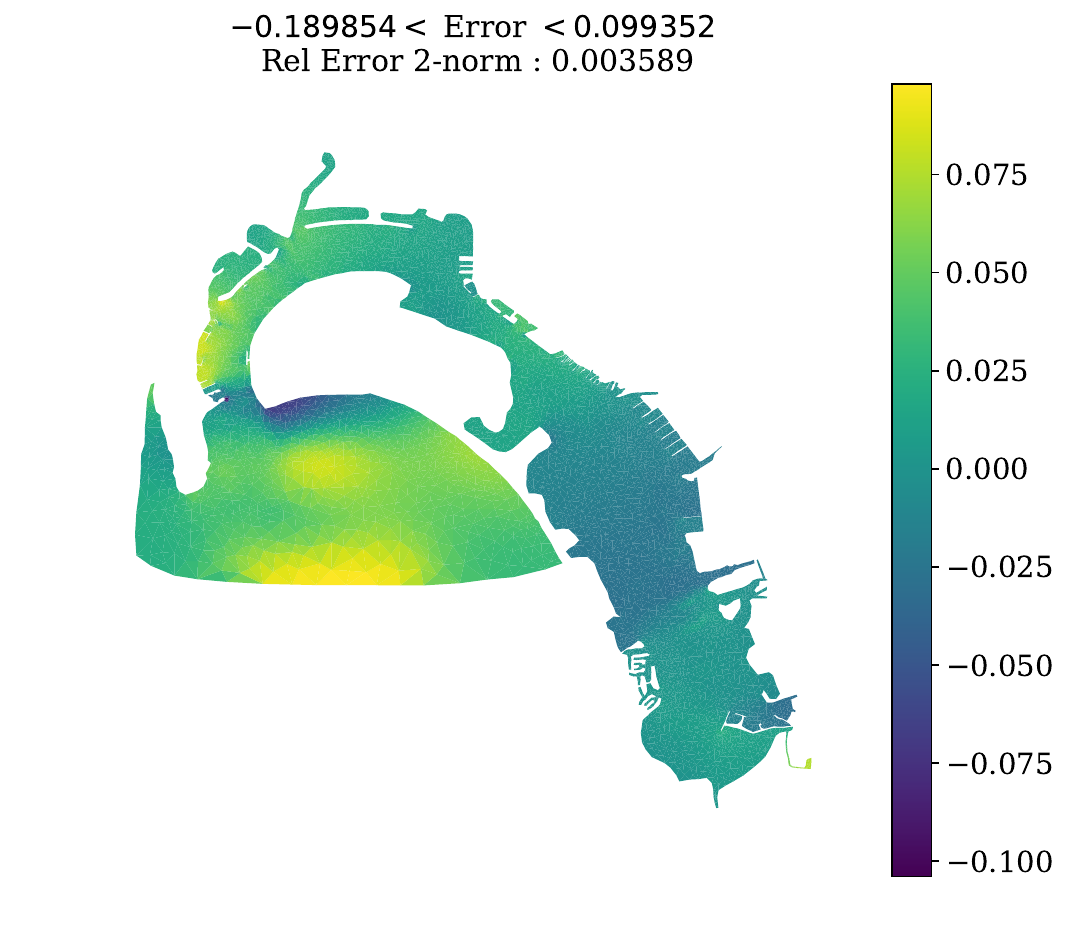}
    \caption{\textit{psr-greedy} NIROM solution for depth (left column) and the corresponding error (right column) with respect to the true solution at $t=44.24$ hours for the San Diego Bay example}
    \label{fig:sd_domain_nirom_err}
\end{figure}

Figure \ref{fig:sd_domain_nirom_err} shows the depth solution obtained by \textit{psr-greedy} NIROM at $t\approx 44.24$ hours and the corresponding error with respect to the high-fidelity solution. The localized structure of the flow features and error profile highlight the fact that in most practical applications, accurate modeling of specific regions in the flow domain is a more desirable outcome than a globally averaged and optimized model. This is where the \textit{psr-greedy} NIROM can be truly beneficial as the selection of the interpolation centers is based on the minimization of the function residuals, while being constrained by the preservation of a well-conditioned interpolation matrix.

\begin{figure}[ht!]
\centering
   \includegraphics[width=0.95\columnwidth]{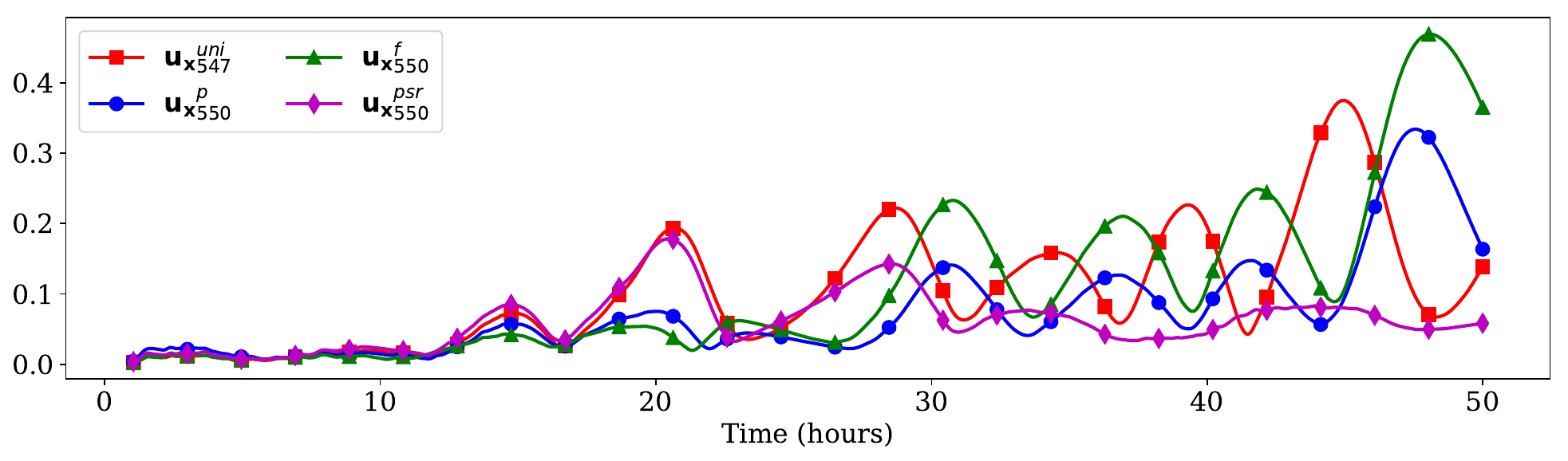}
   \caption{Temporal evolution of the $RMSE$ of the x-velocity NIROM solutions for the San Diego Bay example}
   \label{fig:sd_wrms_u}
\end{figure}

Finally, Figure \ref{fig:sd_wrms_u} shows the temporal evolution of the $RMSE$ norms of the x-velocity solution, computed using different NIROM strategies. The $RMSE$ curves show a marginal increase in the approximation error over time. This can be attributed to the modeling choice of adopting a first order discretization for the time derivatives of the projection coefficients while constructing the RBF approximation of the temporal dynamics. However, Figure \ref{fig:sd_wrms_u} shows that the accuracy of the \textit{psr-greedy} NIROM solution is relatively well preserved over time, as compared to the other NIROM solutions. These results emphasize another important feature of the \textit{psr-greedy} NIROM. Even for problems where the periodic nature of the temporal dynamics is a more natural fit for the \textit{p-greedy} algorithm, the \textit{psr-greedy} NIROM offers superior performance in terms of preserving accuracy over time while also producing a steady reduction in approximation error with the gradual enrichment of data points. Similar to the Red River example, Table \ref{tab:mtRMSE_sd} is provided to highlight the accuracy in approximating the modes that were not selected in list $L$. It can be seen again that the \textit{psr-greedy} approach produces the lowest mtRMSE arrors (see eq. \ref{eq:mtrms-modal}) among all the NIROM procedures.
\begin{table}[ht]
    \setlength{\tabcolsep}{5pt}
    \renewcommand{\arraystretch}{1.2}
    \centering
    \begin{tabular}{c c c c c}
        \toprule
        & uniform ($547$) & \textit{p-greedy} ($550$) & \textit{f-greedy} ($550$) & \textit{psr-greedy} ($550$) \\ \midrule
        $\vec u_x$ & $0.991$ & $0.772$ & $1.171$ & $\textsl{0.534}$\\
        $\vec u_y$ & $1.068$ & $0.832$ & $1.266$ & $\textsl{0.584}$\\
        $\vec h$   & $3.824$ & $2.765$ & $3.919$ & $\textsl{2.306}$\\ \bottomrule
        
    \end{tabular}
    \caption{\textit{mtRMSE} norms using modes not selected in $L$ for different NIROM solutions}\label{tab:mtRMSE_sd}
\end{table}

%\clearpage
%%%---------------------------------------------------------------
%%%---------------------------------------------------------------
\section{Discussion}\label{sec:discussions}

We briefly review some of the salient features of the greedy POD-RBF NIROM procedures and discuss some of the possible areas for future studies.

The simulations involving the greedy POD-RBF NIROMs, as reported in Section \ref{sec:results2}, were repeated with smaller tolerance levels for the POD truncation error. For instance, the San Diego Bay example was recomputed using a POD truncation error tolerance of $\tau = 10^{-10}$ such that $197, 307$ and $302$ POD modes were selected for the solution components $\vec h, \vec u_x$ and $\vec u_y$ respectively. While a marginal improvement in accuracy was observed, the relative trends in the performance of the greedy NIROMs were found to be similar to the results reported here. This provides a strong indication that the efficiency of the greedy RBF NIROM procedure for modeling the temporal dynamics is practically independent of the quality of the reduced basis representation of the high-fidelity snapshots. This property can help guide the design of optimal ROM parameters that can help strike a balance between computational reduction and desired accuracy of the ROM. 

The proposed \textit{psr-greedy} algorithm measures the dual effects of the RBF kernel as well as the function variability by defining the greedy metric to be a pointwise product of the power function and the modal residual errors. The power function acts as a variable scaling factor, enforcing the selected centers to be more widely distributed than the \textit{f-greedy} algorithm and thus reducing the clustering effect, while a suitably chosen tolerance $\tau_{psr}$ ensures that regions of nonlinearity are more adequately resolved than the \textit{p-greedy} algorithm. As a result, the \textit{psr-greedy} algorithm performed consistently well and with superior efficiency in all of the different hydrodynamic regimes considered. 

Two other greedy strategies were also examined. First, a modified \textit{f-greedy} approach was adopted in which the scalar components of the function residuals corresponding to each element of the set $L$ was greedily minimized. In other words, instead of considering the modal $l^2$-norm of the function, an approach similar to the proposed \textit{psr-greedy} method was explored. This approach yielded results that were comparable to the \textit{f-greedy} approach for an optimally tuned set of hyperparameters namely $c$, $\tau_{greedy}$, and $\tau_f$. The \textit{f-greedy} algorithm was preferred over this modified approach, as one less hyperparameter ($c$, and $\tau_f$) was required to achieve NIROM solutions with comparable accuracy.  Second, the \textit{psr-greedy} algorithm was also modified by using the modal $l^2$-norm of the function. In this case, however, the \textit{psr-greedy} algorithm, as presented here, was found to produce a more significant improvement in accuracy with enrichment of centers, and thus achieve better convergence than the modified $l^2$-norm approach.

It was observed that spatial measures of approximation error of the RBF NIROM solutions may grow with time. This is expected as the RBF interpolant is designed to approximate the discrete temporal derivative of the POD-projection coefficients. In this work, a first order accurate time discretization was adopted to compute the time derivative of the projection coefficients as well as to advance the reduced solutions in time (see eqs.~(\ref{eq:nirom-interp-cond}) and (\ref{eq:RBF-time-deriv})) which can cause the online evaluation error to accumulate over time. This effect was subdued in the Red River example due to the overall growing intensity of the underlying flow, however, this effect was more prominent in some of the greedy NIROM solutions for the periodic tidal flow simulations in the San Diego Bay example. One strategy to potentially arrest this growth would be to introduce a higher order time discretization scheme for the online model, and is a topic of ongoing study. Notably, one of the striking features of the \textit{psr-greedy} NIROM was the ability to control this error growth better than the other greedy strategies and thus, maintaining a consistent level of accuracy throughout the online simulation period. This can be attributed to the fact that the first order time discretization error would also manifest in the time derivative function that was being interpolated, and since the \textit{psr-greedy} NIROM is driven by the energetically significant components of the function residuals, it is able to counter this growth by adequately increasing the resolution of selected centers in these regions of high residual error. An alternative strategy to model and control the growth of the approximation error would be to estimate the error via evaluating appropriate response functions constructed using Gaussian process regression methods \cite{GH2019,X2019}.

In all the greedy NIROM experiments, the shape factor of the RBF kernels is estimated conservatively as the fill distance of the set of all available centers, and is kept fixed at this value for all later computations using the different greedily selected center distributions. For most of our numerical experiments $0.01 \leq c \leq 0.05$ was found to be optimal. Such a conservative estimate was adopted primarily because the optimization of the shape factor was not the objective of this work. Additionally, the Mat\'{e}rn kernel is relatively well conditioned even when using flatter shaped functions ($c \ll 1$). Hence, there was no noticeable loss in the stability of the interpolation matrix while allowing for optimal performance in terms of accuracy of the RBF NIROMs. Keeping a fixed shape factor also facilitated easier comparison between the different greedy strategies and also for studying the convergence of any given greedy strategy with the enrichment of centers. 

The efficiency of the RBF approximation may be further enhanced by increasing the flexibility of the radial kernel, so that significant changes in the rate of variability in the data as well as possible discontinuities can be represented more accurately. This may be achieved by considering Variably Scaled Kernels (VSK) (see \eg \cite{BLRS2015,DMP2019}) and Variably Scaled Discontinuous Kernels (VSDK) \cite{DeMarchi2020441}, which allow the shape factor $c$ of the radial kernel to be variable across the domain. This method has the potential for some interesting applications in building predictive ROMs and is currently under investigation.

%\clearpage
%%%---------------------------------------------------------------
%%%---------------------------------------------------------------
\section{Conclusions}\label{sec:conclusions}

In this work, we have presented different greedy strategies for defining a POD-RBF non-intrusive reduced order model (NIROM) for a stabilized finite element approximation of the shallow water equations. The POD method is adopted to generate a suitable reduced basis space from the high-fidelity snapshots, and several greedy RBF interpolation strategies are developed to accurately model the temporal dynamics. We have compared the accuracy and the performance of the NIROM with a standard nonlinear POD (NPOD) ROM. We have also compared the performance of the greedy algorithms in determining a near-optimal RBF approximation of the high-fidelity model in various challenging and realistic coastal and riverine flow problems. In our numerical experiments, we observed the following.

\begin{enumerate}[label=(\roman{*}), ref=(\roman{*})]
\item The relative ease of implementation of the POD-RBF NIROM makes it a desirable choice for model reduction of time dependent problems in "fast-replay" scenarios, especially involving large scale legacy codes where access to the underlying source code maybe limited. There was a significant reduction in computational expense in comparison to a traditional stabilized finite element high-fidelity solver while achieving a similar level of accuracy as a Galerkin-projected nonlinear POD model.
\item Three different greedy algorithms were studied as strategies of optimizing the distribution of centers for the RBF kernel. Using all the available centers \cite{XFPH2015} for interpolation often leads to redundant computational burden, which can be relieved by greedy center selection strategies, especially in the presence of nonlinear dynamics with varying temporal scales.
\item The \textit{f-greedy} algorithm directly minimizes the residual for approximating the $l^2$-norm of the function in order to select the RBF centers. This leads to a clustering problem due to oversampling of centers around regions of strong nonlinearity in the high-fidelity solution. Hence the \textit{f-greedy} algorithm usually suffers from an undesirable loss in efficiency.
\item The \textit{p-greedy} algorithm is aimed at selecting centers by minimizing the power function, and usually converges towards a fairly uniform distribution with every greedy iteration. Hence, this algorithm may suffer from under-representation in regions of strong nonlinearity. 
%Thus, an initial strong convergence rate is followed by a loss in efficiency and no significant improvement in accuracy.
\item The proposed \textit{psr-greedy} algorithm reduces the clustering effect while improving resolution in under-sampled areas to generate a well-balanced set of centers. This algorithm displayed robust convergence and superior efficiency in problems involving both coastal and riverine scales, making it the strongest candidate for obtaining an optimal distribution of centers for the RBF kernel.
\end{enumerate}
All the relevant data and codes for this study will be made available in a public repository at \url{https://github.com/erdc/podrbf_nirom} upon publication.

\section{Acknowledgments}
This research was supported in part by an appointment of the first author to the Postgraduate Research Participation Program at the U.S. Army Engineer Research and Development Center, Coastal and Hydraulics Laboratory (ERDC-CHL) administered by the Oak Ridge Institute for Science and Education through an interagency agreement between the U.S. Department of Energy and ERDC. The research contributions by the third and the fifth authors have been accomplished within Rete ITaliana di Approssimazione (RITA), partially funded by GNCS-IN$\delta$AM and through the European Union's Horizon2020 research and innovation programme ERA-PLANET, grant agreement no.~689443, via the GEOEssential project. The authors would also like to thank Corey Trahan for his valuable help in using the Adaptive Hydraulics suite (AdH) \cite{SBMT2011,SB2012,TBS2006} for the high-fidelity numerical simulation of the 2D shallow water flow examples. Permission was granted by the Chief of Engineers to publish this information.

\appendix
\section{Computational aspects of greedy algorithms}
The straightforward implementation of the \textit{p-greedy} algorithm requires the computation of an updated interpolation matrix using the Lagrange basis functions $\eta_j, j = 1,\ldots,M$, in each iterative step. As the set of selected centers is gradually populated with new centers, the corresponding interpolation matrix may become severely ill-conditioned for point distributions with small distances, thus posing severe computational challenges. Similarly, the straightforward implementation of the \textit{f-greedy} and the \textit{psr-greedy} algorithms involve using an updated set of interpolation conditions to re-compute the entire set of coefficients in each iteration, thus introducing a lot of redundant computational overhead. 

An alternative approach was proposed by Pazouki and Schaback \citep{PS2011} using adaptive calculation of Newton bases. This approach avoids the expensive computation of the interpolation matrix $\vec A$ for all selected centers in $X_k$, in each iteration. Instead, the previously computed $k$ Newton basis functions $N_1, N_2, \ldots, N_k$ are used to determine $N_{k+1}$, and subsequently the $(k+1)^{th}$ power function and the modal function residuals. This approach was also adopted in the implementation of the VKOGA algorithm \citep{WH2013} due to its computational efficiency. Here we present a modified version of the Newton basis approach to account for the modal computations of the \textit{f-greedy} and the \textit{psr-greedy} algorithms.

\theoremstyle{definition}
\begin{definition}{(Newton basis of $V_X$).}
Let $X = \{\bx_1, \ldots, \bx_{r} \} \subset \Omega$ so that $V_X = \text{span}\{\Phi(\cdot,\bx)\, : \bx \in X \}$ is an at most $r-$dimensional Hilbert space spanned by $\Phi$ over $X$. Then the Newton basis $N_1, \ldots, N_r$ of $V_X$ is obtained recursively as

\begin{align}
    & N_1(\bx) := \frac{\Phi(\bx_1,\bx)}{\sqrt{\Phi(\bx_1,\bx_1)}}. \\
    &\text{For }k=2, \ldots, r  \nonumber\\
    &N_k(\bx) := \frac{\widetilde{N_k}(\bx)}{\nm*{\widetilde{N_k}}_V},  \quad \text{where } \widetilde{N_k}(\bx) = \Phi(\bx_k, \bx) -  \sum\limits_{i=1}^{k-1} N_i(\bx_k)N_i(\bx),
\end{align}
such that $( N_i, N_j )_V = \delta_{ij}$. 
\end{definition}
The above construction of the Newton basis is equivalent to the Cholesky decomposition of the interpolation matrix $\vec A$, and hence the first $k$ basis functions $N_1, \ldots, N_k$ do not need to be recalculated when determining the $(k+1)^{th}$ basis function. This property leads to remarkable computational efficiency, as will be presented below. First we briefly review a few important properties of the Newton basis (see \citep{PS2011} for details and proofs).

\begin{lemma}{($V$-orthonormal basis property).} 
The functions $N_1,\ldots,N_r$ are, by construction, an orthonormal basis for the span of translates $\Phi(\bx_1,\cdot), \ldots, \Phi(\bx_r,\cdot)$ such that 

\begin{align}
    \left( \Phi(\bx,\cdot), N_i \right)_V = N_i(\bx), \quad \forall \bx \in \Omega.
\end{align}
\end{lemma}

\begin{lemma}{(Representations using Newton basis).} 
Let $N_1, \ldots N_r$ be the Newton basis of $V_X$. Then an approximation of $f\in V$ can be represented as

\begin{align}
    F(\bx) = \sum\limits_{i=1}^{r} (f, N_i)_V N_i(\bx).
\end{align}
Also, the power function can be computed using the Newton basis functions as 

\begin{align}
   P_X(\bx) = \Phi(\bx,\bx) - \sum_{i=1}^{r}N_i^2(\bx),
\end{align}
which leads to the useful recursive relation $P_k^2(\bx) = P_{k-1}^2(\bx) - N_k^2(\bx)$.
\end{lemma}

Using all of the above properties of the Newton basis, the greedy algorithms can be computed efficiently, as described below.
% \begin{proposition}{(Greedy algorithms using Newton basis).}
% Bla
% \end{proposition}

\begin{algorithm}[H]
  \DontPrintSemicolon
  \SetAlgoLined
	\KwResult{ $\widetilde{X}$ = Optimal set of RBF centers }
  \SetKwInOut{Input}{Input}\SetKwInOut{Output}{Output}
  \Input{$X = \{\vec z_1,\vec z_2, \ldots, \vec z_{M}\}$, $M > N_{max}$
        }
  \BlankLine
	For $k=1$ set: \newline
	$\vec z_1 = \argmax_{\vec z_i \in X} \{ \Phi(\vec z_i, \vec z_i) \}, \quad X_1 \leftarrow \{\vec z_1\}$ \newline
    $N_1(\vec z) = \dfrac{\Phi(\vec z_1, \vec z)}{\sqrt{\Phi(\vec z_1, \vec z_1)}} $ \;
  
   \Do{$\max\limits_{\vec z_i \in X\setminus X_k} P_{X_k}(\vec z_i) > \tau_p $ \& $k < N_{max}$}
    {
      Compute $P_{X_k}^2(\vec z_i) =  \Phi(\vec z_i, \vec z_i) - \sum_{l=1}^k N_l^2(\vec z_i), \quad \forall \vec z_i \in X\setminus X_k $ \;
      Set $\vec z_{k+1} = \argmax\limits_{\vec z_i \in X\setminus X_k} \{ P_{X_k}(\vec z_i) \} $\;
      $X_{k+1} \leftarrow X_k \cup \{ \vec z_{k+1}\}$ \;
      $N_{k+1}(\vec z_i) = \dfrac{\Phi(\vec z_{k+1}, \vec z_i ) - \sum_{l=1}^k N_l(\vec z_{k+1})N_l(\vec z_{i})}{P_{X_k}(\vec z_{k+1})} \quad \forall \vec z_i \in X\setminus X_{k+1}$ \;
      Set $k \leftarrow k+1$ \;
    }

	\caption{\textit{p-greedy} algorithm using Newton basis}\label{alg:p-greedy-Newton}
\end{algorithm}

In the \textit{f-greedy} algorithm (see Algorithm \ref{alg:f-greedy-Newton}), we define $g(\bz) = \nm*{\vec{f}(\bz)}_2$ and the function residual at the $k^{th}$ iteration as $\xi^k(\bz) = g(\bz) - F^k(\bz)$ where $F^k$ is the approximation for $g(\bz)$ at the $k^{th}$ iteration.

\begin{algorithm}[ht]
  \DontPrintSemicolon
  \SetAlgoLined
	\KwResult{ $\widetilde{X}$ = Optimal set of RBF centers }
  \SetKwInOut{Input}{Input}\SetKwInOut{Output}{Output}
  \Input{$X = \{\vec z_1,\vec z_2, \ldots, \vec z_{M}\}$, $M > N_{max}$}
  \BlankLine
	For $k=1$ set: \newline
	$\vec z_1 = \argmax_{\vec z_i \in X} \{g(\vec z_k)\}      $ \newline
    $X_1 \leftarrow \{\vec z_1\} $ \newline
    $c_1 = \dfrac{g(\vec z_1)}{\sqrt{\Phi(\vec z_1, \vec z_1)}} $ \newline
    $N_1(\vec z) = \dfrac{\Phi(\vec z_1, \vec z)}{\sqrt{\Phi(\vec z_1, \vec z_1)}} $ \;

    \Do{$\max\limits_{\vec z_i \in X\setminus X_k} \abs*{\xi^{k}(\vec z_i)} > \tau_f $ \& $k < N_{max}$}
    {
      %Compute $P_{X_k}^2(\vec z_i) =  \Phi(\vec z_i, \vec z_i) - \sum_{l=1}^k N_l^2(\vec z_i), \quad \forall \vec z_i \in X\setminus X_k $ \;
      Compute $\xi^{k}(\vec z_i) = g(\vec z_i) - \sum_{l=1}^k c_l N_l(\vec z_i) \quad \forall \vec z_i \in X\setminus X_k $\;
      Set $\vec z_{k+1} = \argmax\limits_{\vec z_i \in X\setminus X_k} \{ \abs*{\xi^{k}(\vec z_i)} \} $ \;
      $X_{k+1} \leftarrow X_k \cup \{ \vec z_{k+1}\}$ \;
      $c_{k+1} = \dfrac{g(\vec z_{k+1}) - \sum_{l=1}^k c_l N_l(\vec z_{k+1})}{\left(\Phi(\vec z_{k+1}, \vec z_{k+1}) - \sum_{l=1}^k N_l^2 (\vec z_{k+1}) \right)^{1/2}}$\;
      $N_{k+1}(\vec z_i) = \dfrac{\Phi(\vec z_{k+1}, \vec z_i ) - \sum_{l=1}^k N_l(\vec z_{k+1})N_l(\vec z_{i})}{\left(\Phi(\vec z_{k+1}, \vec z_{k+1}) - \sum_{l=1}^k N_l^2(\vec z_{k+1}) \right)^{1/2}} \quad \forall \vec z_i \in X\setminus X_{k+1}$ \;
      Set $k \leftarrow k+1$ \;
    }

	\caption{\textit{f-greedy} algorithm using Newton basis}\label{alg:f-greedy-Newton}
\end{algorithm}

For the \textit{psr-greedy} algorithm (see Algorithm \ref{alg:greedy-Newton}) we define the $k^{th}$ function residual vector as $\vec{\xi}^k(\bz) = \vec{f}(\bz) - \vec{F}^k(\bz)$, where $\vec{f}(\bz) = [f_1(\bz), f_2(\bz), \ldots, f_q(\bz)]^T$, $\vec{F}^k(\bz)$ is the $k^{th}$ approximation for $\vec{f}(\bz)$, and $q = \abs{L}$.

\begin{algorithm}[ht]
  \DontPrintSemicolon
  \SetAlgoLined
	\KwResult{ $\widetilde{X}$ = Optimal set of RBF centers }
  \SetKwInOut{Input}{Input}\SetKwInOut{Output}{Output}
  \Input{$X = \{\vec z_1,\vec z_2, \ldots, \vec z_{M}\}$, $M > N_{max}$ \;
        $L = \{j \,|\,  j \leq m_i^j\}, \, |L| \ll m$ : List of selected significant modes }
  \BlankLine
	For $k=1$ set: \newline
	$\vec z_1 = \argmax_{\vec z_i \in X} 
       \{\Phi(\vec z_i, \vec z_i)\, \abs*{f_j(\vec z_k)} : j \text{ is the first mode in } L $ \newline
    $X_1 \leftarrow \{\vec z_1\} $ \newline
    $\vec{c}_1 = \dfrac{\vec{f}(\vec z_1)}{\sqrt{\Phi(\vec z_1, \vec z_1)}} $ \newline
    $N_1(\vec z) = \dfrac{\Phi(\vec z_1, \vec z)}{\sqrt{\Phi(\vec z_1, \vec z_1)}} $ \;
  \For{$j$ in $L$}
  {
    \Do{$\max\limits_{\vec z_i \in X\setminus X_k} \{P_{X_k}(\vec z_i) \, \abs*{\xi_j^{k}(\vec z_i)} \} > \tau_{psr}$ \& $k < N_{max}$}
    {
      Compute $P_{X_k}^2(\vec z_i) =  \Phi(\vec z_i, \vec z_i) - \sum_{l=1}^k N_l^2(\vec z_i), \quad \forall \vec z_i \in X\setminus X_k $ \;
      Compute $\xi_j^{k}(\vec z_i) = f_j(\vec z_i) - \sum_{l=1}^k c_{l,j} N_l(\vec z_i) \quad \forall \vec z_i \in X\setminus X_k $\;
      Set $\vec z_{k+1} = \argmax\limits_{\vec z_i \in X\setminus X_k}
            \left\{P_{X_k}(\vec z_i) \, \abs*{\xi_j^{k}(\vec z_i)} \right\} $ \;
      $X_{k+1} \leftarrow X_k \cup \{ \vec z_{k+1}\}$ \;
      $\vec{c}_{k+1} = \dfrac{\vec{f}(\vec z_{k+1}) - \sum_{l=1}^k \vec{c}_l N_l(\vec z_{k+1})}{\left(\Phi(\vec z_{k+1}, \vec z_{k+1}) - \sum_{l=1}^k N_l^2 (\vec z_{k+1}) \right)^{1/2}}$\;
      $N_{k+1}(\vec z_i) = \dfrac{\Phi(\vec z_{k+1}, \vec z_i ) - \sum_{l=1}^k N_l(\vec z_{k+1})N_l(\vec z_{i})}{P_{X_k}(\vec z_{k+1})} \quad \forall \vec z_i \in X\setminus X_{k+1}$ \;
      Set $k \leftarrow k+1$ \;
    }
  }
	\caption{\textit{psr-greedy} algorithm using Newton basis}\label{alg:greedy-Newton}
\end{algorithm}

%% References

\clearpage
\small
\bibliographystyle{elsarticle-num}
\bibliography{mybib.bib}

\end{document}